\documentclass[
pre,aps,twocolumn,amsmath,amssymb,floats,superscriptaddress,nofootinbib,10pt,showkeys
 reprint,
 amsmath,amssymb,
 aps,
]{revtex4-2}

\usepackage{graphicx}
\usepackage{dcolumn}
\usepackage{bm}
\usepackage{hyperref}
\usepackage{siunitx}

\usepackage{titletoc}
\titlecontents{lsection}   
  [0em]                    
  {}                       
  {\makebox[4.5em][l]{\thecontentslabel}\enspace}  
  {}                       
  {\titlerule*[0.5pc]{.}\contentspage}             
\titlecontents{lsubsection}
  [4.5em]
  {}
  {\makebox[3em][l]{\thecontentslabel}\enspace}
  {}
  {\titlerule*[0.5pc]{.}\contentspage}

\usepackage[normalem]{ulem}

\DeclareMathOperator\erf{Erf}

\pdfpageattr {/Group << /S /Transparency /I true /CS /DeviceRGB>>}

\usepackage{chngcntr}
\let\normalsection\section
\begin{document}

\title{Experimentally accessible measurement of irreversibility in stochastic systems\\ by categorizing single-molecule displacements.}

\author{Alvaro Lanza}\affiliation{Department of Physics, King's College
London, London WC2R 2LS, United Kingdom}
\author{Inés Martínez-Martín}\affiliation{Department of Physics, King's College
London, London WC2R 2LS, United Kingdom}
\author{Rafael Tapia-Rojo}
\affiliation{Department of Physics, King's College
London, London WC2R 2LS, United Kingdom}
\affiliation{Centre for the Physical Sciences of Life, King's College
London, London WC2R 2LS, United Kingdom}

\author{Stefano Bo}
\affiliation{Department of Physics, King's College
London, London WC2R 2LS, United Kingdom}
\affiliation{Centre for the Physical Sciences of Life, King's College
London, London WC2R 2LS, United Kingdom}

\date{\today}

\begin{abstract}
Quantifying the irreversibility and dissipation of non-equilibrium processes is crucial to understanding their behavior, assessing their possible capabilities, and characterizing their efficiency.
We introduce a physical quantity that quantifies the irreversibility of stochastic Langevin systems from the observation of individual molecules' displacements. 
Categorizing these displacements into a few groups based on their initial and final position allows us to measure irreversibility precisely without the need to know the forces and magnitude of the fluctuations acting on the system.
For short times, our model-free estimate of irreversibility is related to entropy production by a conditional fluctuation theorem.  For short times and in general for stationary protocols, our estimate provides a lower bound to the average entropy production.
We validate the method on single-molecule force spectroscopy experiments of proteins subject to force ramps. We show that irreversibility is sensitive to detailed features of the energy landscape underlying the protein folding dynamics and suggest how our methods can be employed to unveil key properties of protein folding processes.
 \end{abstract}

\maketitle

 \section{\label{sec: intro}Introduction}

Life is out of equilibrium, sustained by the supply of energy, which is dissipated in irreversible processes.
Such energy supplies lift the
  strict constraints imposed by equilibrium, 
  and allow life to exhibit its astonishing variety.
The further from equilibrium a system is driven, the richer its behavior can be, with strong fluxes, large irreversibility~\cite{peliti2021stochastic,Seifert_2025}, enhanced sensing~\cite{lan2012energy,skoge2013chemical,sartori2014thermodynamic,govern2014energy,bo2015thermodynamic,mehta2012energetic}, better polymer copying~\cite{hopfield1974kinetic,murugan2012speed,sartori2013kinetic,ouldridge2017thermodynamics,sartori2015thermodynamics}, faster and more accurate reactions to the environment, and many other biologically relevant features~\cite{yang2021physical,Brown19}. 
To properly characterize a system, it is fundamental to assess how far from equilibrium it operates.
This allows us to compare the costs (in terms of dissipation, irreversibility, and energy expenditures) to the benefits (in terms of enhanced performances) and quantitatively assess trade-offs~\cite{lan2012energy,Pietzonka16,Gnesotto18}.
Recent fundamental results, such as the Thermodynamic Uncertainty Relation (TUR), have contributed to this analysis, unveiling that dissipation bounds precision~\cite{Barato15}, fluctuations~\cite{gingrich2016dissipation}, coherence and correlations of these systems~\cite{ohga2023thermodynamic,santolin2025dissipation,ptaszynski2025dissipation}.

Modern single-particle tracking techniques allow us to probe the dynamics of a wide range of systems at small scales~\cite{Greenleaf07, Leake13, Hansen18, Bustamante21,muenker_accessing_2024, tapia-rojo_single-molecule_2024, Mackay2024}, revealing a fluctuating world. Such stochastic thermal fluctuations make it challenging to detect if the system is out of equilibrium and to quantify how far. Stochastic thermodynamics has provided a solid theoretical framework to define and quantify irreversibility and entropy in stochastic systems, revealing the laws governing their statistics, such as fluctuation theorems~\cite{Seifert12,peliti2021stochastic,Seifert_2025,Jarzynski1997NonequilibriumDifferences,crooks1999entropy,Seifert05}.
Applying this framework to experiments with limited information on the forces at play is far from trivial and has led to the development of many methods for measuring irreversibility in Markovian stochastic processes~\cite{Battle16,Pietzonka16,martinez2019inferring, Dechant18, Koyuk20, Manikandan20, VanVu20, Roldan21, Otsubo22, Lee23, Di_Terlizzi24, Degunther24, Singh24,Harunari22,van2022thermodynamic,hyeon_physical_2017,van_vu_thermodynamic_2020,dieball_direct_2023,chatzittofi2024thermodynamic,di2025force}. 
Many such methods rely on the thermodynamic uncertainty relation, which provides a lower bound to the entropy that is produced to achieve a certain accuracy~\cite{Barato15,gingrich2016dissipation,Pietzonka17,Dechant18, Koyuk20}
and only requires the mean and variance of observable currents. 
A recent work independent of the TUR obtained a bound to entropy in systems with time-dependent driving, using only these first two moments and the diffusion coefficient~\cite{Singh24}.
Machine-learning techniques~\cite{Manikandan20, Otsubo22} and related methods~\cite{VanVu20, Lee23} have also been developed, and also estimators from coarse-graining experimentally inaccessible degrees of freedom of full trajectories
~\cite{Battle16,Roldan10,Li19, Roldan21, Degunther24,bao2025}. Despite this remarkable progress, applications to experimental systems remain challenging and therefore limited. While each method has its capabilities and constraints, the main roadblocks include obtaining sufficient data, knowledge of the forces acting on the system and applicability to cases with time-dependent protocols.\\

Aiming for experimental applicability, we present a technique to calculate the irreversibility of Langevin systems based on the measurement of particles' displacements at finite time intervals, which are experimentally accessible in subcellular systems.
The key insight behind the method is to compensate for the finite temporal resolution by categorizing the observed displacements into classes, which depend on their starting and final positions.
The method requires no knowledge of the forces acting on the system or the diffusion coefficient, and we show it to
provide a lower bound to entropy production for stationary systems or when the system can be observed with a high time resolution.
For systems that can be measured at high frequency, the method also captures the space-dependent profile of local dissipation.\\

To validate the method, we perform single-molecule magnetic tweezers experiments on the mechanosensing protein talin, which we unfold and refold out of equilibrium using force ramps. 
By comparing wild-type and mutant protein domains with simulations, we show how quantifying irreversibility unveils subtle properties of the protein free-energy landscape.\\

We start the paper by briefly introducing the basic concepts of stochastic thermodynamics in Section~\ref{sec: lang}.
In Section~\ref{sec: method} we introduce our method, which we apply to various case studies in Section~\ref{sec: case}. We close by discussing the results in Sections~\ref{sec: discussion} and~\ref{sec: conclusion}. 

\section{Stochastic thermodynamics of Langevin systems}\label{sec: lang} 
We consider systems described by the overdamped Langevin equation
\begin{align}\label{eq: langevin}
    \dot x_t = \frac{F}{\gamma}+\sqrt{2D}\xi_t
\end{align}
where, $x_t$ is the degree of freedom we are tracking (for example, the position of a single particle), $F$ is the deterministic force acting on it (which can depend on position and time), $\xi_t$ is a Gaussian white noise satisfying $\langle\xi_t\xi_{t'}\rangle=\delta(t-t')$  and the fluctuation-dissipation theorem links the diffusion $D$ to friction $\gamma$ via the temperature $T$, $D=k_BT/\gamma$.
The entropy produced along a trajectory $\mathbf{x}$ can be defined in terms of its irreversibility as
the log-ratio 
\begin{align}
    \Delta S_{tot}(\mathbf{x}) \equiv k_B \ln{\frac{\mathcal{P}(\mathbf{x})}{\mathcal{P}^R(\widetilde{\mathbf{x}})}} \label{eq: entropy-onepath}
\end{align}
where $\mathcal{P}(\mathbf{x})$ is the probability of observing the trajectory $\mathbf{x}$ forward in time and $\mathcal{P}^R(\widetilde{\mathbf{x}})$ is the probability of observing the time reversed one~\cite{Seifert05}. Since the space of trajectories grows exponentially with their length, it is practically impossible to sample it efficiently, and non-trivial coarse-graining procedures are required~\cite{Gomez-Marin08,Roldan10,gladrow2021experimental,kappler2020stochastic}. 
Taking the average of Eq.~\eqref{eq: entropy-onepath} shows that the average entropy production can be written as a
 Kullback-Leibler Divergence ($D_{KL}$)
\begin{align}\label{eq: pathEP} 
    \langle \Delta S_{tot} \rangle = k_B D_{KL}\left(\mathcal{P}(\mathbf{x}) ||\mathcal{P}^R(\widetilde{\mathbf{x}})\right)\,,
\end{align}
where the $D_{KL}$ for two probability distributions $p$ and $p'$ is defined as $D_{KL}(p||p')\equiv \int dx\, p\ln(p/p')$. 
This expression agrees with the second law of thermodynamics since Kullback-Leibler divergences are never negative.
A key result of stochastic thermodynamics is that this definition of entropy coincides with the thermodynamic entropy defined by studying heat exchanges with reservoirs
\begin{align}\label{eq: thermoEnt}
    \Delta S_{tot}(\mathbf{x})=&  \underbrace{k_B\ln{\frac{P_0(x_0)}{P_t(x_t)}}}_{\Delta S_{sys}} +\underbrace{\frac{1}{T}\int F(x_t)\circ  dx_t}_{\Delta S_{env}}\,,
\end{align} 
which has contributions $\Delta S_{sys}$ due to changes in the entropy of the system and $\Delta S_{env}$ from dissipation into the environment. 
The average rate of entropy production can be written as~\cite{Seifert05} 
$\langle \dot{S}_{tot}\rangle=\int dx\,\dot{\sigma}(x)$, where
\begin{align}\label{eq: sigmadot}
    \dot{\sigma}(x)\equiv k_B\frac{\nu^2(x)}{D}P(x)\,,
\end{align}
with $\nu(x)$ the local mean velocity, related to the probability flux as $\nu(x)=j(x)/P(x)$.
$\dot{\sigma}(x)$ corresponds to the average rate at which entropy is generated conditioned on being at position $x$ times the probability density of being there.
\section{Measuring irreversibility}\label{sec: method}
\subsection{Classes of displacements and their distributions.} 
We consider a system that is observed at discrete time intervals $\Delta t$, in which it moves from the initial position $x'\equiv x_t$ to $x\equiv x_{t+\Delta t}$, and define the displacement 
$\ell=x-x'$.
We start by considering the case of stationary systems so that there is no dependence on the overall time $t$. We discuss the generalization to time-dependent dynamics in Section~\ref{sec: time_dep}.

\begin{figure}
\includegraphics[width=0.45\textwidth]{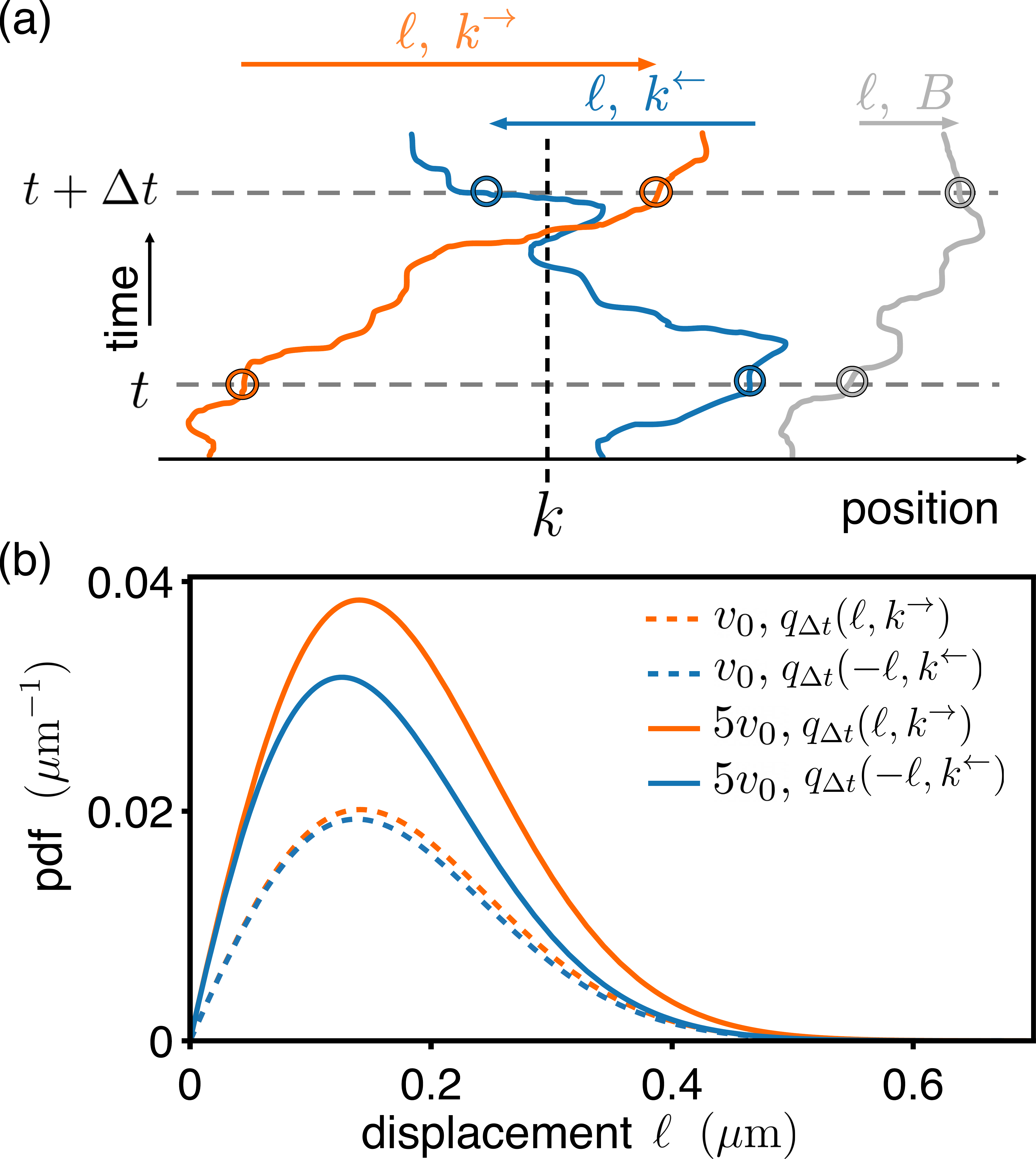}
\caption{\label{fig: fig1} (a) Illustration of displacements $\ell$ of three trajectories in the time window $\Delta t$,  categorized into displacement classes with respect to the boundary at $k$. Trajectories of class $k^\rightarrow$ cross $k$ left to right, $k^\leftarrow$ right to left and $B$ remain on the same side. (b) Sample displacement distributions at $k=0$ for the square potential on a ring [shown in Fig.~\ref{fig: fig2}(a)], at different driving speeds with $v_0=0.5~\si{\micro\meter\per\second}$. At equilibrium $q(\ell,k^\rightarrow)=q(-\ell,k^\leftarrow)$.}
\end{figure}

In many systems, such as biomolecular condensates~\cite{Hyman2014}, protein and DNA (un)folding~\cite{woodside_reconstructing_2014, thorneywork_direct_2020}, cell membrane domains~\cite{Sakamoto23membrane} and broader diffusion in the cytoplasm~\cite{Cherstvy14}, the diffusive motion takes place in physically distinct environments.
For such heterogeneous diffusion, be it with respect to an inhomogeneous diffusion coefficient and/or complex energy landscapes, it is natural to group displacements on the basis of where they take place, which is a key novelty of our approach. We categorize displacements based on where the initial and final positions are with respect to reference boundaries.
These displacement distributions were studied experimentally in Ref.~\cite{McSwiggen19}, and shown to contain relevant information about the diffusive nature of motion in biomolecular condensates and their equilibrium properties~\cite{Bo21}.
In the example depicted in Fig.~\ref{fig: fig1}(a), we draw a boundary at $x=k$ and partition all possible displacements into three classes, forming the set $\mathcal{C}_3=\{k^\rightarrow,\, k^\leftarrow,\,B\}$.
The last class is composed of displacements where the initial and final points are on the same side of the boundary (bulk, $B$), like the gray trajectory in Fig.~\ref{fig: fig1}(a). 
The remaining two classes are for
displacements where the initial and final points are on opposite sides of $k$, where $k^\rightarrow$ and $k^\leftarrow$ denote displacements that cross $k$ from left to right and right to left (orange and blue trajectories in Fig.~\ref{fig: fig1}(a), respectively.) 
Let us consider as an example, class $k^\rightarrow$, where $x'<k$ and $x>k$.
The statistics of such displacements are described by the joint probability density of observing a displacement $\ell$ and of the displacement belonging to class $k^\rightarrow$
\begin{align}
    q_{\Delta t}(\ell, k^\rightarrow)
    \equiv \int\limits_{-\infty}^k dx' \int\limits_{k}^{\infty} dx \, \delta(\ell-(x-x'))P_{\Delta t}(x | x')P(x') \label{eq: qCrossR}
\end{align}
where $P(x')$ is the probability of finding the particle at $x'$, $P_{\Delta t}(x|x')$ is the propagator, \textit{i.e.}, the probability of finding the particle in position $x$ after a time $\Delta t$, given that it started at $x'$, and $\delta$ denotes the Dirac delta function.
We also define the marginal probability of observing a displacement of class $k^\rightarrow$
\begin{align}
    p_{\Delta t}(k^\rightarrow) \equiv \int_{-\infty}^k dx' \int^{\infty}_k dx \, P_{\Delta t}(x | x')P(x'). \label{eq: pCrossR}
\end{align}
For a generic class of displacements $C$, the probability density is defined as
\begin{align}
    q_{\Delta t}(\ell, C) \equiv \iint\limits_{\Omega_C} dx' 
    \, dx \, \delta\left(\ell-(x-x')\right)P_{\Delta t}(x | x')P(x')\,, \label{eq: qdefn}
\end{align}
where the integral is over the domain $\Omega_C$, which defines the class, and 
\begin{align}
    p_{\Delta t}(C) \equiv \iint\limits_{\Omega_C} dx' 
    \, dx\, P_{\Delta t}(x | x')P(x')\,. \label{eq: pdefn}
\end{align}
By normalization we have that $\sum_{C\in\mathcal{C}}\int d\ell\, q_{\Delta t}(\ell,C)=\sum_{C\in\mathcal{C}}\, p_{\Delta t}(C)=1$.\\

\subsection{Displacements and irreversibility}
Displacements and their classes contain information about time reversibility. 
For example, for stationary systems, the time reversal 
of a displacement $\ell$ of class $k^\rightarrow$ is a displacement  $-\ell$ of class $k^\leftarrow$ (and vice versa), while time reversal of bulk displacements still belongs to the same class. In general, we denote by $\tilde{C}$ the class associated with the time reversal of a displacement of class $C$.
Formally, they are found by exchanging initial and final positions, $x'$ and $x$. For the example of Fig.~\ref{fig: fig1}(a), the time-reversed classes of $k^\rightarrow,\, k^\leftarrow,\,B$  correspond to $k^\leftarrow,k^\rightarrow,B$, respectively.
For a general partition, including $n$ classes of displacements, $\mathcal{C}_n$, we define the following measure of irreversibility
\begin{align}
\Sigma_{\Delta t}
(\mathcal{C}_n) \equiv k_B\sum_{C\in\mathcal{C}_n} \int d\ell \, q_{\Delta t}(\ell, C) \ln{\frac{q_{\Delta t}(\ell, C)}{q_{\Delta t}(-\ell, \tilde{C})}}\,,  \label{eq: dklndef}
\end{align}
which is the Kullback-Leibler divergence of the mixed (discrete in $C$, continuous in $\ell$) joint probabilities between time-forward and reversed displacements over all classes, and is the key contribution of this work.
This measure of irreversibility is based on coarser measurements that focus on the initial and final point $x'$ and $x$ in a displacement taking a time $\Delta t$ and ignore the intermediate positions visited by the paths. Therefore, it is easier to estimate from experimental data than the full path probabilities with Eq.~\eqref{eq: pathEP}.

\subsection{Irreversibility and entropy}
At equilibrium, the detailed balance condition $P_{\Delta t}(x | x')P(x') = P_{\Delta t}(x' | x)P(x)$ ensures that $q_{\Delta t}(\ell, C)=q_{\Delta t}(-\ell, \tilde{C})$, resulting in  $\Sigma_{\Delta t} =0 $, as seen, for a special case, in Ref.~\cite{Bo21}.  This suggests a relation between the coarse measure of irreversibility $\Sigma_{\Delta t}$ and entropy production defined in Eq.~\eqref{eq: entropy-onepath}, which we explore in this section.

In the limit of vanishing $\Delta t$, entropy production can be expressed in terms of the breaking of the detailed balance condition and Eq.~\eqref{eq: entropy-onepath}
can be written as 
\begin{align}
  \lim_{\Delta t\to 0}  \Delta S_{tot}(x | x')=k_B\lim_{\Delta t\to 0}\ln\frac{P_{\Delta t}(x|x')P(x')}{P_{\Delta t}(x'|x)P(x)}\,, 
\end{align}
where $\Delta S_{tot}(x | x')$ denotes the entropy produced in a trajectory starting at position $x'$ and arriving at position $x$. 
As shown in Appendix~\ref{app: ft}, for small $\Delta t$, by exchanging the initial and final points, $x'$ and $x$ the probability distribution of a displacement $q_{\Delta t}(\ell, C)$ can be expressed in terms of the same integrals appearing in the definition of the time-reversed displacement $q_{\Delta t}(-\ell, \tilde{C})$, weighted by the exponential of the entropy produced in the displacement.
One can then obtain that the ratio of the displacement distribution and its time reversal obeys the conditional fluctuation theorem
\begin{align}\label{eq: ft}
   \lim_{\Delta t\to 0} \frac{q_{\Delta t}(-\ell, \tilde{C})}{q_{\Delta t}(\ell, C)}=\lim_{\Delta t\to 0}
    \langle e^{-\Delta S_{tot}/k_B}|\ell, C\rangle\,
\end{align}
where the right-hand side is the conditional average of the exponential of minus the entropy that is produced in a displacement, provided that this displacement is $\ell$ and of class $C$. 
For a one-dimensional stationary system, this holds for all $\Delta t$, as shown in~\ref{supp: prop} of the Supplementary Material~\cite{supplementary}, and in a similar result~\cite{hartich_emergent_2021} based on splitting probabilities. 
A more general fluctuation theorem of this kind was obtained in Ref.~\cite{Wimsatt25} by considering a broader definition of classes, not defined only in terms of the initial and final point.
Equation~\eqref{eq: pathEP} expresses the average entropy production in terms of a Kullback-Leibler divergence of the path probabilities. From an information-theoretic viewpoint, this implies that if instead of observing the full trajectories in detail, we only have access to coarse-grained measurements, we will typically obtain a lower bound to the entropy production~\cite{Gomez-Marin08,Roldan10}.
This is a consequence of the fact that projecting or marginalizing two distributions cannot increase their Kullback-Leibler divergence, which is known as the data-processing inequality and proven by the log-sum inequality~\cite{cover2012elements}\footnote{ 
For non-negative $a_i$ and $b_i$, the log-sum inequality states $\sum a_i\log (a_i/b_i)\ge a\log (a/b)$, where $a=\sum_i a_i$ and $b=\sum_i b_i$.}.
The displacements we are considering
record the initial and final point ($x'$ and $x$, respectively) but ignore the intermediate points visited in the time interval $\Delta t$.\footnote{This projection, combined with Eq.~\eqref{eq: pathEP}, provides a lower bound to entropy production $\langle\Delta S_{tot}\rangle
    \geq k_BD_{KL}(P_{\Delta t}(x',x) ||P_{\Delta t}(x,x'))$. This bound is saturated for one-dimensional stationary processes or for generic processes in the small $\Delta t$ limit (see ~\ref{supp: prop}).}.
Furthermore, the displacements only retain $\ell=x-x'$ and the type of displacement (\textit{i.e.}, less information than $x$ and $x'$). Because of this coarse-graining, comparing the displacements to their time reversal provides the lower bound 
 \begin{align}\label{eq: bound}
    \Sigma_{\Delta t}(\mathcal{C}_n) \leq \langle \Delta S_{tot} \rangle\, \qquad
     \text{(steady-state)}\;,
\end{align}
as shown explicitly in Appendix~\ref{app: proofBound}.
We remark that this bound holds for stationary dynamics and discuss the time-dependent case in ~\ref{sec: time_dep}.

\subsubsection{A hierarchy of bounds}
Intuitively, categorizing the displacements in many different classes has the potential of achieving a tighter bound to entropy production.
This intuition can be made precise by the data processing inequality. Let us consider a partition including $n$ classes $\mathcal{C}_n$. If two of these classes are merged, we obtain a partition with $n-1$ classes, which we call $\mathcal{C}'_{n-1}$\footnote{Note that this merging procedure is not unique, and different classes $\mathcal{C}'_{n-1}$ exist.}.
We can repeat the procedure until we are left with a single class, including all the possible displacements, which we denote as $\mathcal{I}$, for which the time reversal simply corresponds to flipping the sign of $\ell$.
For these partitions, we have the hierarchy of bounds
\begin{align}\label{eq: hierarchy}
    0 \leq  \Sigma_{\Delta t}(\mathcal{I}) \leq \,... \leq  \Sigma_{\Delta t}(\mathcal{C}'_{n-1}) \leq  \Sigma_{\Delta t}(\mathcal{C}_n) \leq \langle \Delta S_{tot} \rangle.
\end{align}

\subsubsection{Time-dependent dynamics.}\label{sec: time_dep}
For time-dependent dynamics, the probabilities of observing displacements evolve in time. 
It is then necessary to generalize our approach to take into account the time $t$ when the displacements are observed and consider the time-dependent
 displacement probability distribution $q_{\Delta t, t}(\ell, C)$, with the added time in the subscript $t$ denoting the starting time of the displacement.
To assess irreversibility, we compare displacements starting at time $t-\Delta t$ and ending at $t$, to displacements starting at time $t$ and ending at time $t+\Delta t$.
This guarantees that the starting time of the reversed displacements corresponds to the ending time of the forward displacement, as detailed in Appendix~\ref{app: td}.
Our definition of irreversibility, then, generalizes to the time-dependent case as
\begin{align}
    \Sigma_{\Delta t, t}(\mathcal{C}_n) \equiv k_B\sum_{C\in\mathcal{C}_n} \int d\ell \, q_{\Delta t, t-\Delta t}(\ell, C) \ln{\frac{q_{\Delta t, t-\Delta t}(\ell, C)}{q_{\Delta t,t}(-\ell, \tilde{C})}}.\label{eq: irr_transient}
\end{align}
Systems where the applied forces do not vary in time relax towards a steady state. For such relaxation dynamics, $\Sigma_{\Delta t, t}$ provides a lower bound to the average entropy produced in the $[t-\Delta t,t]$ time window $\langle\Delta S_{tot}\rangle^t_{t-\Delta t}$
\begin{align}
    \Sigma_{\Delta t, t}(\mathcal{C}_n) &\leq \langle\Delta S_{tot}\rangle^t_{t-\Delta t}\, \qquad\text{(relaxation)} \,.
\end{align}
If the applied forces follow a time-dependent protocol, Eq.~\eqref{eq: irr_transient} still provides a well-defined measure of irreversibility.  However, this measure provides a lower bound to the average entropy production only in the limit of a short time window $\Delta t\rightarrow0$, for which the time reversal of the protocol plays a negligible role (see Appendix~\ref{supp: shift})
\begin{align}
    \lim_{\Delta t\rightarrow0}\Sigma_{\Delta t, t}(\mathcal{C}_n) &\leq \langle\Delta S_{tot}\rangle^t_{t-\Delta t}\, \qquad\text{(time-dep. driving)}\,. \label{eq: time-dep bound}
\end{align}

\subsection{Local dissipation}
Let us now focus on the case where we have a single boundary at $k$ as in Fig.~\ref{fig: fig1}(a). Disregarding the length of the displacements, the probability of a displacement crossing the boundary [defined in Eq.~\eqref{eq: pCrossR}], provides an alternative measurement of irreversibility. For a steady state, it reads 
\begin{align}
    \Sigma_{\Delta t}^{trans}(k) \equiv k_B[ p_{\Delta t}(k^\rightarrow)-p_{\Delta t}(k^\leftarrow) ]\ln{\frac{p_{\Delta t}(k^\rightarrow)}{p_{\Delta t}(k^\leftarrow)}}\,, \label{eq: dkltrans}
    \end{align}
    while the definition for the time-dependent case is discussed in Appendix~\ref{app: td}.
    Since it disregards the length of the displacements, this measure is coarser compared to the one defined in Eq.~\eqref{eq: dklndef}. As a consequence, it is easier to estimate from experimental data, but it captures less irreversibility, since
       $ \Sigma_{\Delta t}^{trans}(k) \leq \Sigma_{\Delta t}(\mathcal{C}_3)$. 
In the limit of short time intervals $\Delta t$, $\Sigma_{\Delta t}^{trans}$ provides an estimate of the dissipation locally occurring around $k$. Defining the diffusive lengthscale $\ell_0=\sqrt{\pi D\Delta t}$, we have (see~\ref{supp: transExpand} for the derivation) 
\begin{align}
\frac{\Sigma_{\Delta t}^{trans}(k)}{\Delta t} = \ell_0\dot{\sigma}(k) + \mathcal{O}(\Delta t^{3/2}), \label{eq: transExpand}
\end{align}
where the density of dissipation rate $\dot{\sigma}(x)$ is defined in Eq.~\eqref{eq: sigmadot}.
\begin{figure*}[t]
    \includegraphics[width=\textwidth]{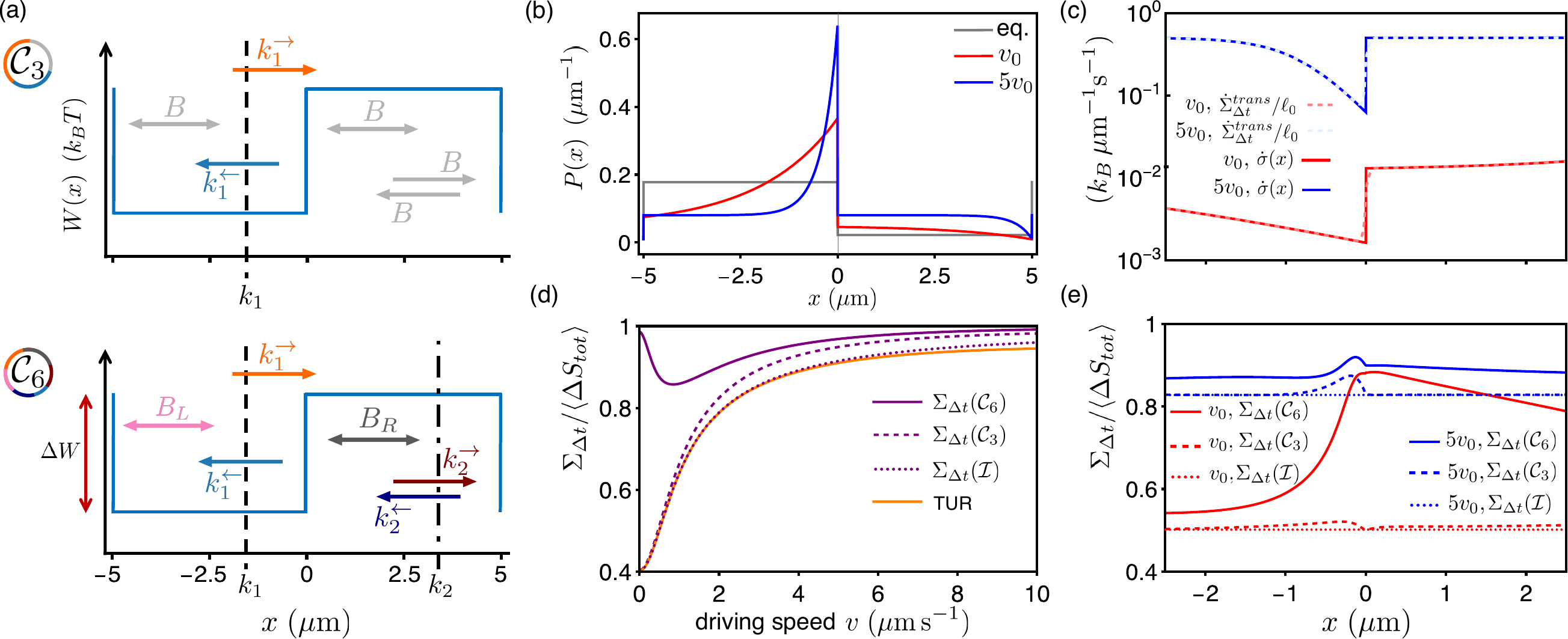}
    \caption{(a) Profile of square potential $W(x)$ and different classes of displacements: top $\mathcal{C}_3(k_1)$ and bottom $\mathcal{C}_6(k_1,k_2)$. (b) Steady-state probability distributions for a particle in a square potential on a periodic topology, subject to different constant drivings $v$ with $v_0=0.5~\si{\micro\meter\per\second}$. For the equilibrium (eq.) distribution, $v=0$. (c) Local dissipation: the solid lines give the density of entropy production rate $\dot{\sigma}$ defined in Eq.~\eqref{eq: sigmadot} 
    for near ($v=0.5$) and far ($v=2.5$) from equilibrium driving. The dashed curves are the local irreversibility estimates obtained from $\dot{\Sigma}_{\Delta t}^{trans}=\Sigma_{\Delta t}^{trans}/\Delta t$ [see Eq.~\eqref{eq: dkltrans} and~\eqref{eq: transExpand}], as a function of the position of the boundary $k_1=x$.
    (d) Fraction of irreversibility captured
     ($\Sigma_{\Delta t}/\langle\Delta S_{tot}\rangle$) for three types of displacement partitions $\mathcal{I}$ (purple dotted line), $\mathcal{C}_3(k_1=-0.1)$ (purple dashed line), $\mathcal{C}_6(k_1=-0.1,k_2=\pm5)$ (purple solid line) and TUR (orange line) as a function of the nonequilibrium driving $v$. (e) $\Sigma_{\Delta t}/\langle\Delta S_{tot}\rangle$ for different partitions as a function of the boundary location $k_1$ for far and near equilibrium driving (blue and red lines, respectively), with $k_2=\pm 5$ fixed. For all drivings considered in the plots above, $D = 1$~\si{\micro\meter\squared\per\second}, $\gamma=1\, k_BT$~\si{\second\per\micro\meter\squared}, $\Delta W = \ln{8}\; k_BT$, and irreversibilities are measured with $\Delta t=10$~\si{\milli \second}. 
    }
    \label{fig: fig2}
\end{figure*}
\\

\section{Case studies}\label{sec: case}
\subsection{Nonequilibrium steady states}
\subsubsection{Driven particle on a ring}
As a first example, we consider a particle on a ring subject to a constant force $F=v\gamma$.
The system reaches a non-equilibrium steady-state (NESS) with a constant (in time and space) probability distribution and a constant probability flux $j\propto v$. This leads to a constant entropy production rate [see~Eq.~\eqref{eq: sigmadot}] and in a time $\Delta t$, there is an average entropy production $  \langle \Delta S_{tot} \rangle=k_Bv^2\Delta t/D$.
The displacement distributions can be calculated analytically, because the propagators are Gaussian and the steady-state distribution is space-independent. 
Since the system is uniform in space, it is not necessary to partition the displacements and $\Sigma_{\Delta t}(\mathcal{I})=k_Bv^2\Delta t/D$ coincides with total entropy produced, with $\Sigma_{\Delta t}(\mathcal{I})=\Sigma_{\Delta t}(\mathcal{C}),\, \forall \mathcal{C}$. For this example, a suitably defined TUR bound [see Eq.~\eqref{eq: TUR}], and methods based on milestoning~\cite{hartich_emergent_2021,Blom24}, and other coarse-grainings~\cite{Gomez-Marin08,Roldan21} also capture the full dissipation.
\subsubsection{Driven particle in a square potential on a ring}\label{sec: squarepot}
As a second, richer example, we consider a particle on a ring inside a square potential well, as shown in Fig.~\ref{fig: fig2}(a). 
The well extends from $x=-L$ to $x=0$, and has height $\Delta W$: $W(x)=\Delta W[\Theta(x)+\Theta(-x-L)]$, where $\Theta(x)=1$ for $x>0$ and $0$ otherwise. We implement the ring by imposing periodic boundary conditions at $x=-L$   and $x=L$. In addition to the potential, the particle is subject to a constant non-conservative force $\gamma v$ so that in Eq.~\eqref{eq: langevin} $F=-\nabla W + \gamma v$.
This force creates a non-equilibrium steady-state, and pushes the particle on one side of the well, accumulating a peak in probability distribution which decays exponentially within the well (over a length $D/v$) and exhibits a sharp drop upon leaving the well. We report the probability distributions for different amounts of driving $v$ in Fig.~\ref{fig: fig2}(b) and give the explicit expressions in~\ref{supp: pSS_NESS}. 
For larger external forces, the peak within the well becomes narrower and the profile away from the peak becomes flatter [compare the blue and red curves in Fig.~\ref{fig: fig2}(b)].
The propagators for this model can be computed analytically~\cite{Uhl21}, and integrating them provides explicit expressions for the displacement distributions, which we report in~\ref{supp: pSS_NESS} and illustrate in Fig.~\ref{fig: fig1}(b) for $q(\ell,k^\rightarrow)$ and $q(-\ell,k^\leftarrow)$ at $k=0$.
As expected, the difference between a displacement and its time reverse increases as the system is driven further from equilibrium.
To obtain $\Sigma_{\Delta t}$, we integrate these displacement distributions numerically.
In this example, since the system is not homogeneous in space, considering more classes of displacements refines the estimated irreversibility.
We compare three ways of partitioning the possible displacements: a single class, $\mathcal{I}$, the three classes defined with respect to the crossing of a boundary at $x=k_1$,  $\mathcal{C}_3(k_1)=\{k_1^\rightarrow,\, k_1^\leftarrow,\,B\}$, and 
a more refined partition that considers a second boundary at $x=k_2$, with six classes of displacements $\mathcal{C}_6(k_1,k_2)=\{k_1^\rightarrow,k_1^\leftarrow,B_L,B_R,k_2^\rightarrow,k_2^\leftarrow\}$~\footnote{We consider time scales such that displacements crossing both boundaries have negligible probabilities, \textit{i.e.} $k_2\pm k_1\gg\sqrt{D\Delta t}$, $k_2\pm k_1\gg v\Delta t$ }, as illustrated in Fig.~\ref{fig: fig2}(a).
As predicted by the hierarchy of bounds Eq.~\eqref{eq: hierarchy}, considering more classes of displacements allows us to greatly improve the estimate, capturing more irreversibility, as shown in Fig.~\ref{fig: fig2}(d) and (e). 
How much irreversibility each estimator captures depends on how far from equilibrium the system is driven [see Fig.~\ref{fig: fig2}(d)] and on where we place the boundary $k_1$ [see Fig.~\ref{fig: fig2}(e)].
The first interesting finding is that, when we do not distinguish between classes of displacements, $\Sigma_{\Delta t}(\mathcal{I})$ tracks closely the dissipation measured via the TUR. 
Figure~\ref{fig: fig2}(d) shows how the estimators approach one another and the average entropy production as the system is driven further from equilibrium. This is likely due to the fact that the system becomes ``more homogeneous'' (more similar to the case without the potential) as the driving increases, making it less important to distinguish different classes of displacements. 
For systems near equilibrium, distinguishing the displacements can greatly improve 
the irreversibility estimate.
Interestingly, $\Sigma_{\Delta t}(\mathcal{C}_6(k_1,k_2))$ displays a non-monotonic dependence on the external driving, capturing accurately the dissipation as $v\rightarrow0$.
A reason for this is that the homogeneity of the system is also non-monotonic with $v$. For instance, the steady-state probability depicted in Fig.~\ref{fig: fig2}(b) shows how, away from the boundary, the profile is flat at equilibrium (grey curve), displaying a marked gradient at intermediate driving (red curve)
 and approaching flatness again for higher driving (blue curve). 
 Far from equilibrium, the plateaus inside and outside the potential are at the same value (blue curve), while at equilibrium they are at different levels (grey curve).
 This suggests that, near equilibrium, it may be advantageous to consider separately the displacements taking place inside or outside the well.
 This is achieved by the partition $\mathcal{C}_6$ with the boundaries located near the potential barriers $k_1\simeq 0$, $k_2\simeq L$.
To gather further insight, in Fig.~\ref{fig: fig2}(e) we show the fraction of the average entropy production $\langle \Delta S_{tot} \rangle$ that is captured by $\Sigma_{\Delta t}$, as a function of where we set the boundary $k_1$ ($k_2$ is kept at $x=L$). As a reference, not grouping displacements into separate classes, $\Sigma_{\Delta t}(\mathcal{I})$ captures only $50\%$ of $\langle \Delta S_{tot} \rangle$ near equilibrium  ($v=0.5$~\si{\micro\meter\per\second}) and up to $83\%$ further from equilibrium ($v=2.5$~\si{\micro\meter\per\second}).
As anticipated, adding a boundary at $x=k_1$ provides an improvement when the boundary is drawn in the vicinity of the potential barrier that the force is pushing towards, $k\simeq 0$. Further away, the improvement is slight.
 $\Sigma_{\Delta t}(\mathcal{C}_6(k_1,k_2))$ provides a significant increase, especially for the near equilibrium case, where, placing a boundary at $k_1\simeq 0$, nearly doubles the amount of dissipation captured. There is also a stronger dependence on where the boundary is drawn. The main contribution to the increase comes from splitting the two bulk classes $B_R$ and $B_L$, where the displacements are markedly different (indeed, the probability of crossing a boundary scales with $\sqrt{\Delta t}$ so for the short times we are considering, there are relatively few crossings, so that crossing displacements in classes $\{k_1^\rightarrow,k_1^\leftarrow,k_2^\rightarrow,k_2^\leftarrow\}$  play a minor role). In Fig.~\ref{fig: 6displacements} we plot the six displacement distributions of $\Sigma_{\Delta t}^{(6)}(\mathcal{C}_6)$ for the near and far from equilibrium cases at their optimal $k_1$.
 
 Note that, for a particle on a one-dimensional ring, it is possible to recover the full entropy production using discrete transition statistics~\cite{Harunari22, van2022thermodynamic, hartich_emergent_2021}, which can be obtained through the coarse-graining procedure known as milestoning~\cite{Blom24}. In an empirical setting, the data requirements of this procedure are different from $\Sigma_{\Delta t}(\mathcal{C}_6)$. Depending on the situation, $\Sigma_{\Delta t}(\mathcal{C}_6)$ can return more accurate estimates of entropy production as we illustrate in~\ref{app: milestoning} with simulated data for the settings described in Fig.~\ref{fig: fig2}, discussed in Section~\ref{sec: tur_mile}.\\

Figure~\ref{fig: fig2}(c)
shows that the local dissipation can be estimated accurately with Eq.~\eqref{eq: transExpand}, counting displacements crossing a boundary located at $k_1$ via $\Sigma_{\Delta t}^{trans}(k_1)$ for times up to $\Delta t=10$~\si{\milli \second}, which corresponds to a lengthscale $\ell_0=0.18$ \si{\micro \meter}.

\subsection{Double-well potential under force ramps.}\label{sec: sim}
\begin{figure*}[t]
    \includegraphics[width=\textwidth]{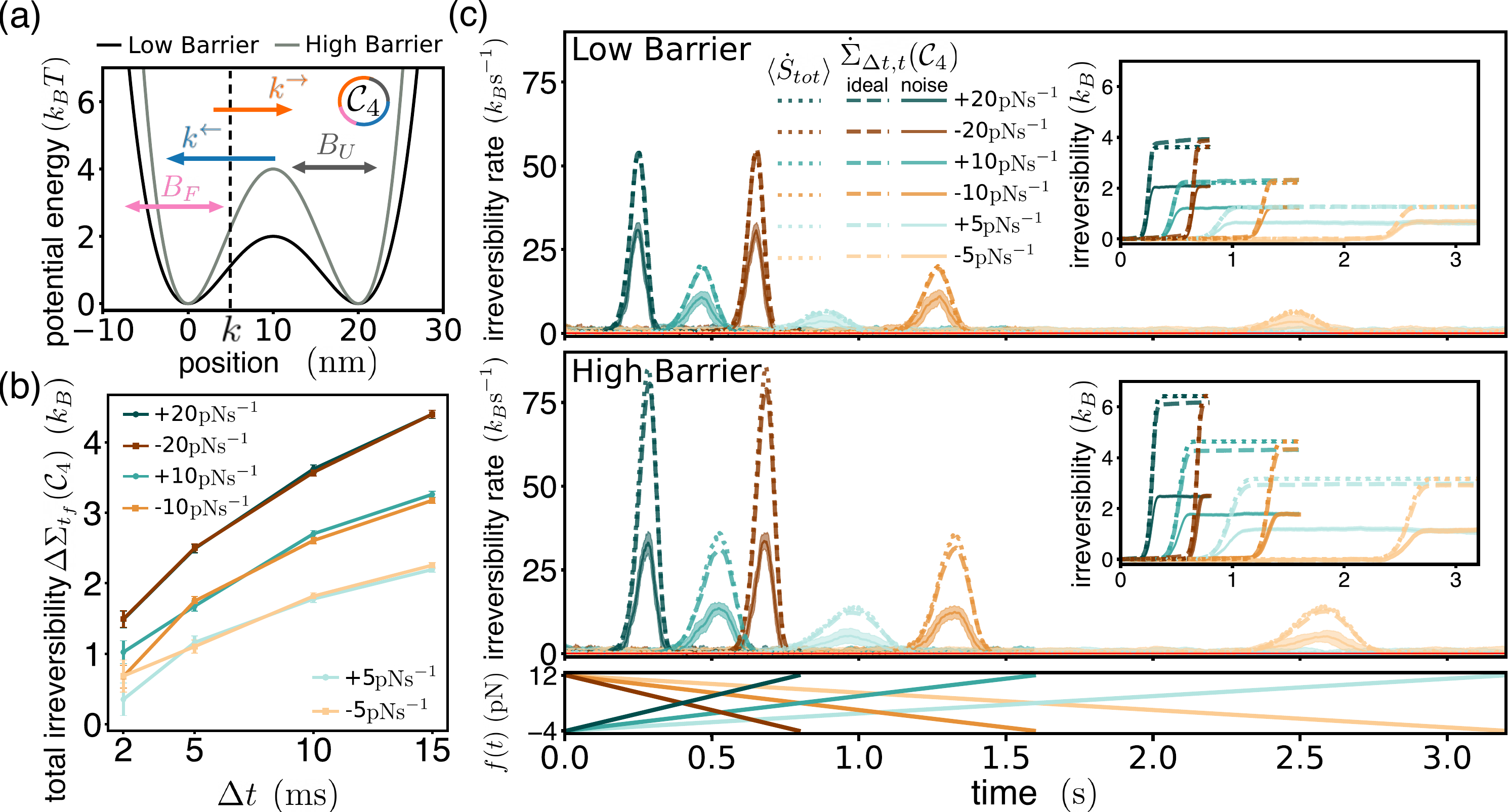}
    \caption{
    (a) Symmetric double-well potential 
    with a high ($\Delta U=4k_BT$) and low ($\Delta U=2k_BT$) barrier and $x_M=L=10\si{\nano\meter}$. Arrows indicate the types of displacements measured in the $\mathcal{C}_4$ classification used to calculate $\Sigma_{\Delta t,t}(\mathcal{C}_4)$. The boundary used to define the displacements is located at $k$. 
    (b) Influence of displacement time $\Delta t$ on the total measured irreversibility $\Delta \Sigma_{t_f}(\mathcal{C}_4)$ in each force ramp, for the high barrier case with measurement noise. (c) Entropy production rate $\langle\dot S_{tot}\rangle$ (dotted lines), computed via Eq.~\eqref{eq: thermoEnt}, and irreversibility rate $\dot\Sigma_{\Delta t,t}$ for different pulling rates for low and high potential barriers, respectively.
    Solid curves: running average of the irreversibility rate estimate $\dot\Sigma_{\Delta t,t}(\mathcal{C}_4)$, measured every millisecond with $\Delta t=5~\si{\milli \second}$, using 5000 trajectories and adding measurement noise. The running averages are taken over 70, 35, 18 points for the $|r|=5,\,10,\,20$ curves, respectively, while the shaded error regions are taken from the running standard deviation over the same windows. Dashed curves: $\dot\Sigma_{\Delta t,t}(\mathcal{C}_4)$ but estimated from 50000 trajectories without measurement noise. 
    Inset: Cumulative time integral of measured irreversibility $\Delta\Sigma_t(\mathcal{C}_4)$ and entropy $\langle \Delta S_{tot}\rangle$. Bottom panel: trace of pulling force curves with time, for the different ramp rates as color-coded above.
    $k_BT=4.11~\si{\pico\newton\nano\meter}$ and $D=3000~\si{\nano\meter\squared\per\second}$. 
    }
    \label{fig: fig3}
\end{figure*}
After discussing stationary cases, we now consider a scenario subject to a time-dependent driving. In this case, the time-dependent irreversibility $\Sigma_{\Delta t, t}$ as defined in Eq.~\eqref{eq: irr_transient} provides a bound to
entropy production only in the vanishing $\Delta t$ limit [Eq.~\eqref{eq: time-dep bound}].
In this section, we show that highly informative measurements of irreversibility can be obtained despite this limitation.
\\

We consider a particle in a symmetric double-well potential $W(x)=\Delta U[(x-x_M)^2/L^2-1]^2$,
where $x_M$ is the location of the potential barrier, which is of height $\Delta U$. $L$ is the distance between the maximum and either of the two minima. Double-well potentials are ubiquitous as a model of bistability in physics and have been successfully employed to model protein and RNA (un)folding~\cite{neupane_protein_2016}, chemical reactions~\cite{Kramers40}, etc. In anticipation of our experimental analysis of the thermodynamics of proteins under force ramps in Section~\ref{sec: exp}, we tilt the potential by an external time-dependent force $f(t)=f_{\rm{0}}+rt$ linearly with rate $r$, so that the force in Eq.~\eqref{eq: langevin} is $F(x,t)=-\nabla W(x)+f(t)$. We consider two potentials, characterized by relatively low and high barrier heights as sketched in Fig.~\ref{fig: fig3}(a), with parameters also motivated by the experimental regime.
For initial and final forces $f_{\rm{0}}=-4$~\si{\pico\newton} and $f_{\rm{f}}=12$~\si{\pico\newton}, we explore three different pulling rates $r$: 5 \si{\pico\newton\per\second}, 10 \si{\pico\newton\per\second} and 20 \si{\pico\newton\per\second}, where the total duration of each experiment is $t_f=(f_{\rm{f}}-f_{\rm{0}})/r$. We also look at the opposite protocol, with $f_{\rm{0}}=12$~\si{\pico\newton} and $f_{\rm{f}}=-4$~\si{\pico\newton} and the same, but negative rates. Trajectories are always initialized from the equilibrium distribution at a tilt of $f_{\rm{0}}$.
To estimate irreversibility in this scenario, we identify 4 types of displacements by placing the measurement boundary $k$ between the two wells $\mathcal{C}_4=\{k^\rightarrow,k^\leftarrow,B_F,B_U\}$ as in Fig.~\ref{fig: fig3}(a), where $k^\rightarrow$ and $k^\leftarrow$ displacements represent crossing from the left to the right well (unfolding for a protein model) and viceversa (folding for a protein model). Class $B_F$ ($B_U$) includes displacements that stay in the left (right) well, representing fluctuations in the folded (unfolded) state. 
In agreement with the hierarchy of bounds (Eq.~\eqref{eq: hierarchy}), we found this $\mathcal{C}_4$ classification captures more irreversibility than the $\mathcal{C}_3$ with one bulk term, as shown in~\ref{sec: symmC3}.  We fix $k$ throughout each ramp (one can potentially vary it at each measurement time), but place it at different locations for each of the six ramp types we explore, choosing the location that records the most irreversibility in total. We show how $\Sigma_{\Delta t}$ depends on the location of $k$ in~\ref{sec: movek}. 
The dashed lines in Fig.~\ref{fig: fig3}(c) show the numerical evaluation of the dissipation rate $\dot\Sigma_{\Delta t, t} (\mathcal{C}_4)=\Sigma_{\Delta t, t} (\mathcal{C}_4)/\Delta t$, which tracks closely the average total entropy production $\langle \Delta S_{tot}\rangle$
reported as the dotted lines (for the low barrier case, the lines are barely distinguishable). The peaks of the dissipation estimate (and of the entropy production rate) correspond to the occurrence of transitions across wells, as they agree with the transition probability curves shown in~\ref{sec: symmC3}. This differs from the NESS case studied above, as the crossing $(k^\rightarrow,k^\leftarrow)$ terms are now main contributors to $\Sigma_{\Delta t}(\mathcal{C}_4)$.
The locations of these peaks in time are different between the unfolding (positive $r$) and refolding (negative $r$) ramps, despite the potential well being symmetric. This is simply because the protocols start at forces that are not equidistant with respect to the coexistence force of $f=0$. As expected, faster ramps are more irreversible and accompanied by higher dissipation, visible in the height of the peaks in Fig.~\ref{fig: fig3}(c) and in the total irreversibility 
$\Delta\Sigma_t=\int_0^tdt'\,\dot\Sigma_{\Delta t, t'}$
shown in the insets (we measure every millisecond so $dt'=1~\si{\milli\second}$).
 We also see that barrier hopping is more reversible for the potential with the lower barrier, as testified both by entropy production and by our measure of irreversibility.

In an experiment, only a finite number of ramps can be recorded, which complicates the empirical estimation of the Kullback-Leibler divergences.
This is not a trivial task for continuous distributions of small data sets~\cite{Bonachela08, Harunari22}, as simply `plugging-in' binned probabilities to the base $D_{KL}$ formula raises systematic biases, and is strongly dependent on the chosen binning. In~\ref{supp: NN}, we discuss our implementation of the nearest-neighbors algorithm for a continuous probability $D_{KL}$~\cite{Wang09} to the case of the mixed discrete-continuous $\Sigma_{\Delta t}(\mathcal{C}_n)$. 
To test the robustness of our method, we consider the case of 5000 ramps simulated via Brownian dynamics.
We additionally explore the effect of measurement noise by adding to the recorded simulated positions a Gaussian noise with variance $9$~\si{\nano\meter\squared}. This value is motivated by the instrumental noise of our magnetic tweezer setup (see~\ref{supp: setup}).
The resulting curves are shown as full lines in
Fig.\ref{fig: fig3}(c), and indicate that the empirical measure still captures very well the location of the dissipation peaks.
The height and the width of the peaks are smaller because measurement noise influences in the same way the forward and the time-reversed displacements, making their distribution more similar.
However, the rescaling is consistent across conditions, testifying that our method can be used to compare irreversibility in different conditions for experimentally relevant settings. The error shading from our estimator in Fig.~\ref{fig: fig3}(c) denotes the running standard deviation.
In Section~\ref{app: noise}, we show that the addition of noise is the
principal factor responsible for the rescaling, while the reduction of the sample size plays a minor role.

In Fig.~\ref{fig: fig3}(b), we explore how varying the time over which we compute irreversibility affects the estimate on simulated data, using this noisy dataset. We see that larger $\Delta t$ gives rise to higher dissipation estimates, but that the overall ranking that sees faster ramps dissipating more 
and both positive and negative ramps dissipating similarly, is conserved among different times (\textit{i.e.} the lines corresponding to different ramp rates do not cross). This suggests that the estimated irreversibility yields robust information under varying extents of non-equilibrium drivings, and that it can be used to systematically compare which system is more dissipative. The dependence on $\Delta t$, however, suggests keeping the same $\Delta t$ when comparing different scenarios.\footnote{What influences the estimate is how $\Delta t$ compares with the time scales of the system, one of which is set by the pulling rate, another by the typical oscillation time scales near the potential bottom, and another one by the transition rates across the barrier. This suggests caution when comparing widely different systems.} 
Large $\Delta t$ can provide irreversibility estimates that exceed entropy production, as already evident by the noiseless estimate at $r=\pm20~\si{\pico\newton\per\second}$ in the low barrier inset of Fig.~\ref{fig: fig3}(c). 
We choose $\Delta t =5$ ms in an effort to balance the tradeoff between the advantage of larger timesteps, which provide a sufficient number of transitions across $k$, and the requirement of short timesteps to closely monitor dissipation. 
For this, the timesteps should be much shorter than the typical timescales of the system, which include the timescales of the protocol, the intrawell relaxation times, and the unfolding/refolding ones.
\subsection{Experimental protein unfolding and refolding under force ramps}\label{sec: exp}
\begin{figure*}[t]
    \includegraphics[width=\textwidth]{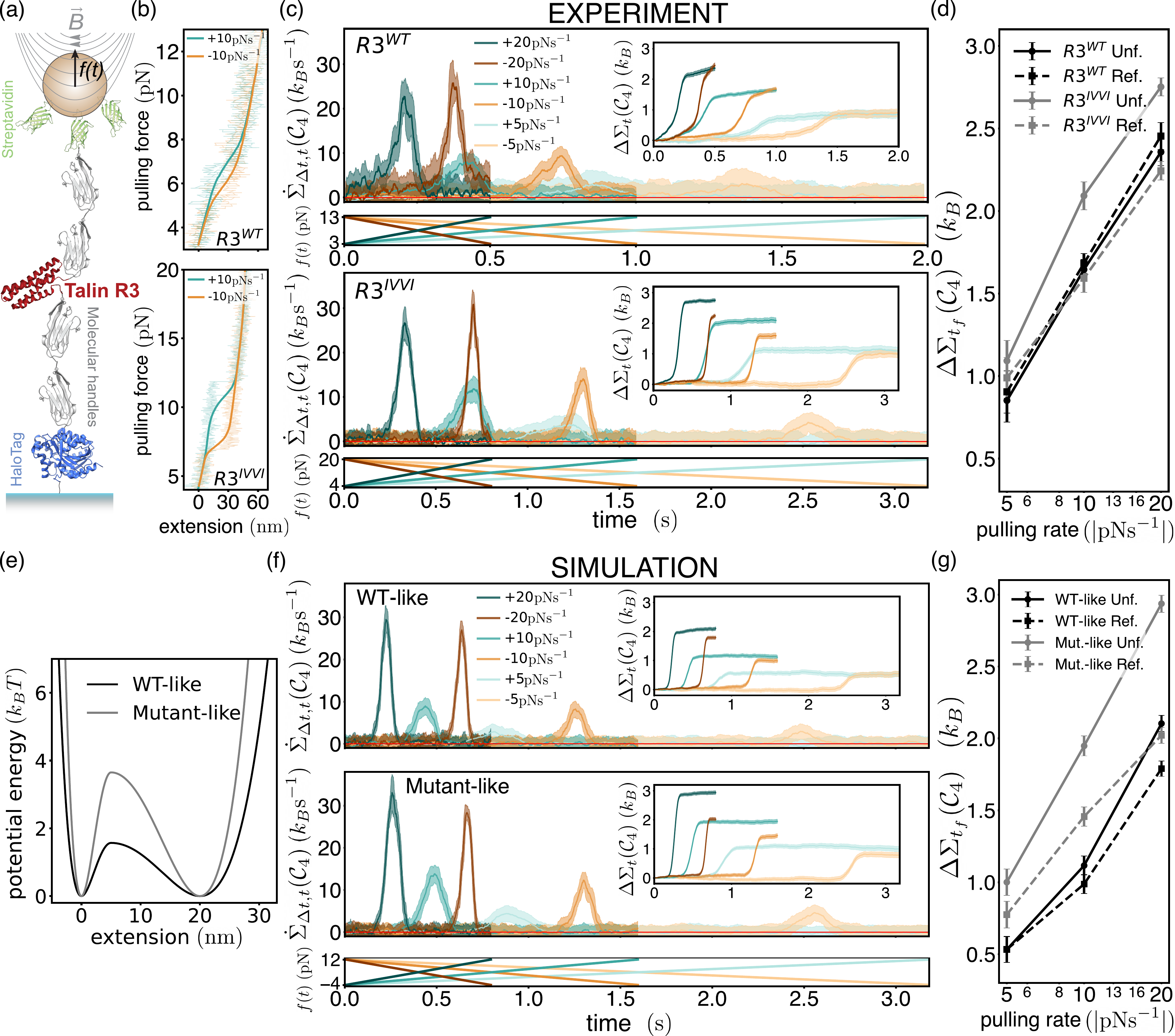}
    \caption{(a) Schematic of experimental magnetic tweezer setup. (b) Thin traces: example trajectories of the Talin R3 extension as the pulling force is linearly increased (blue) and decreased (orange) at $r=\pm10$~\si{\pico\newton\per\second}, for the wild-type R3$^{\rm{WT}}$ (top) and mutant R3$^{\rm{IVVI}}$ (bottom). The thick curves show the respective mean extension over all of the experiments.\\ (c) Irreversibility rate $\dot\Sigma_{\Delta t,t}(\mathcal{C}_4)$ measurements from experiments on the R3$^{\rm{WT}}$ (top) and mutant R3$^{\rm{IVVI}}$ (bottom) protein domains at different pulling rates. Measuring every millisecond with $\Delta t=5~\si{\milli\second}$, we show the running average over 70, 35, 18 points for the $|r|=5,\,10,\,20$ curves, respectively. Inset: Cumulative time integral $\Delta\Sigma_t(\mathcal{C}_4)$ of measured irreversibility. Lower panels: pulling forces for each ramp rate as color-coded above. (d) Total irreversibility at the end of each protocol as a function of unfolding (positive) and refolding (negative) pulling rates. 
     The shown data is obtained from 1682, 5000, 1441 trajectories for $|r|=5,\,10,\,20$ on the R3$^{\rm{WT}}$ and 3933, 3500, 5322 on the R3$^{\rm{IVVI}}$, respectively.
    (e)
Simulations are performed on the shown asymmetric potentials,
with $D=3000$~\si{\nano\meter\squared\per\second}, $k_BT=4.11~\si{\pico\newton\nano\meter}$.
The potential with the lower energy barrier mimics the R3$^{\rm{WT}}$ protein and the high barrier the mutant R3$^{\rm{IVVI}}$ construct.\\ (f) As in (c) for 5000 simulated trajectories, with added noise on the recorded positions. (g) As in (d) for the simulated data.
    }
    \label{fig: fig4}
\end{figure*}
To test our measure of irreversibility on an experimental two-state system, we use single-molecule magnetic tweezers on a mechanosensitive protein. We study the non-equilibrium protein unfolding and refolding dynamics upon cyclic linear force ramp protocols, using the setup presented in Ref.~\cite{tapia-rojo_single-molecule_2024}, and shown in Fig.~\ref{fig: fig4}(a). As a model protein system, we choose talin R3, a force-sensing protein domain at focal adhesions showcasing a low mechanical stability,  unfolding at forces of just $\sim$5 pN. Engineered mutations in its four-threonine belt residue (R3$^{{\rm{IVVI}}}$) increase its mechanical stability to $\sim$9 pN while keeping its physiological function intact~\cite{goult2013riam,tapia-rojo_talin_2020}. Both the R3 wild-type R3$^{\rm{WT}}$ and the R3$^{{\rm{IVVI}}}$ mutant have been previously characterized as classic two-state folders whose native folding dynamics under force are well-described by a one-dimensional double-well energy landscape along the end-to-end pulling coordinate, similar to Fig.~\ref{fig: fig3}(a)~\cite{Tapia24, tapia-rojo_single-molecule_2024}.\footnote{Over long timescales ($\sim10^4~\si{\second}$), R3$^{{\rm{IVVI}}}$ has been shown to populate reversible misfolded states, therefore departing from the simple two-state folding scenario. However, these dynamics are not relevant to the fast non-equilibrium experiments of this work.} We confirm this by conducting constant force experiments and using the committor and transition path tests for one-dimensional diffusion~\cite{neupane_transition-path_2015, neupane_protein_2016}, as shown in~\ref{app: committor}. The results indicate that, indeed, protein extension is a suitable projection of the protein dynamics, following a one-dimensional diffusion in a bistable landscape.
Our measure of irreversibility and the alternative ones we discuss in Section~\ref{sec: discussion} are based on this one-dimensional diffusive description. As usual, dissipation in faster, unobservable degrees of freedom cannot be excluded \cite{esposito2012stochastic,bo2017multiple}.

As magnetic tweezers enable direct control of the pulling force, we subject a single R3$^{\rm{WT/IVVI}}$ domain to a non-equilibrium pulling force protocol. As in the simulations of Section~\ref{sec: sim}, we increase the force linearly between two values $f_{\rm{0}}$ and $f_{\rm{f}}$ for a time $t_{\rm{f}}$. We make sure that the protein is equilibrated in the folded state at $f_{\rm{0}}$ ($f_{\rm{0}}$=3 pN for R3$^{{\rm{WT}}}$ and $f_{\rm{0}}$=4 pN for R3$^{{\rm{IVVI}}}$) before starting the ramp.
Once the unfolding ramp is completed we keep the force constant to allow the protein to relax and in the unfolded state at  $f_{\rm{f}}$ ($f_{\rm{f}}$=13 pN for R3$^{\rm{WT}}$ and $f_{\rm{f}}$=20 pN for R3$^{{\rm{IVVI}}}$). We then perform the refolding ramp, linearly decreasing the force back to $f_{\rm{0}}$. The high stability of magnetic tweezers enables us to probe the (un)folding dynamics of a single protein by cycling this force protocol thousands of times (see Fig.~\ref{fig: ramps_SI} for the protocol and some examples of the recorded trajectories). Additionally, and in order to modulate the degree of irreversibility in our experiment, we probe three different pulling rates for each protein, $r$= $\pm5$, $\pm10$, and $\pm20$~\si{\pico\newton\per\second}.
The dynamics of the system upon a force cycle are characterized by force extension traces as shown in Fig.~\ref{fig: fig4}(b), displaying the typical hysteresis loops between the unfolding (green) and refolding (orange) force ramps due to the non-equilibrium nature of the system (see Fig.~\ref{fig: exp_extension} for force-extension curves at other pulling rates). The area enclosed in the loop represents the total work dissipated in an unfolding and refolding cycle. As expected, the average dissipated work increases with the pulling rate, and the R3$^{\rm{WT}}$ protein shows lower dissipation than R3$^{{\rm{IVVI}}}$, likely due to its lower mechanical stability.

To compute the irreversibility over the non-equilibrium unfolding/refolding ramps, we identify the same four classes of displacements  $\mathcal{C}_4$ as discussed for the simulations [Fig.~\ref{fig: fig3}(a)], and compute  $\Sigma_{\Delta t, t} (\mathcal{C}_4)$ as per Eq.~\eqref{eq: irr_transient} using the estimator discussed in~\ref{supp: NN}. The magnetic tweezers setup allows for a measurement frame-rate of $\sim1~\si{\per\milli\second}$, so we calculate $\Sigma_{\Delta t,t} (\mathcal{C}_4)$ every $1~\si{\milli\second}$, and, as in the simulations, the accumulated dissipation is calculated as 
$\Delta\Sigma_t=\int_0^tdt'\,\dot\Sigma_{\Delta t,t'}$
with $dt'=1~\si{\milli\second}$. Guided by the simulation results in Section~\ref{sec: sim} (Fig.~\ref{fig: fig3}(b)), we choose $\Delta t=5~\si{\milli\second}$ for all the ramps. 
In Fig.~\ref{fig: dtSigma_exp} we show how different choices of $\Delta t$ influence the irreversibility estimation in experiments. 
The hysteresis loops in
Fig.~\ref{fig: fig4}(b) 
show that unfolding and refolding events occur stochastically at different forces, which leads to different extension changes following the polymer elasticity of an unfolded protein.
We therefore choose to place the boundary $k$ at different locations for the unfolding and refolding ramps, as discussed in~\ref{sec: movek}.
For instance, at $r=\pm 10~\si{\pico\newton\per\second}$ we find the best choice of displacement partitions $k=24$~\si{\nano\meter} and $19$~\si{\nano\meter} for the unfolding and refolding ramps of R3$^{{\rm{IVVI}}}$, respectively. For the wild-type domain, we place them at $k=30$~\si{\nano\meter} and $28$~\si{\nano\meter}. 

As in the simulated scenarios, the peaks of irreversibility (Fig.~\ref{fig: fig4}(c)) take place when most unfolding and refolding events occur, shown by the survival probabilities in Fig.~\ref{fig: exp_escapeTexp}. As a second finding, we confirm the expected trend that faster ramps are more irreversible, as manifested by larger total irreversibility $\Delta\Sigma_{t_{\rm{f}}}$ in Fig.~\ref{fig: fig4}(d).
For the wild type, unfolding and refolding ramps are similarly irreversible as in the simulations of Section~\ref{sec: sim}.
Surprisingly, for the R3$^{{\rm{IVVI}}}$ mutant, unfolding is systematically more irreversible than refolding.
The lower panel in Fig.~\ref{fig: fig4}(c) shows that this difference is due to the 
peaks of irreversibility being broader for the unfolding events (mostly visible for the $r$= $10$~\si{\pico\newton\per\second} pulling rate). This effect is not predicted by the symmetric simulations of Section~\ref{sec: sim}, therefore calling for the development of a richer model to describe the folding dynamics of talin R3.

Previous analytical studies based on simplified free-energy landscapes obeying Bell-Evans kinetics and experiments on DNA hairpins have highlighted that 
the distance between each of the wells and the barrier governs the irreversibility of the folding and unfolding processes~\cite{manosas2009dynamic}. 
We therefore hypothesize that the difference we observe may also be linked to asymmetries in the free-energy landscape. Evidence from long equilibrium experiments conducted close to the coexistence force, suggests an asymmetric folding landscape, where 
the unfolded well is broader than the folded one and thus further from the barrier~\cite{Tapia24}. This hypothesis is further supported by the different force-sensitivity in the folding and unfolding rates (Fig.~\ref{fig: constant_force}). We therefore set out to test whether an asymmetric potential with a narrower folded well recapitulates the experimental observation that unfolding is more irreversible than refolding, by performing numerical simulations.

We choose the asymmetric potential wells to be $W(x)=\Delta U[(a(x-x_M)/L)^2-1]^2$, where the asymmetry factor $a=1$ for $x<L$ and $a=1/3$ for $x\geq L$.
Since the mutant is more mechanically stable, we
model it with a higher potential barrier ($\Delta U=3.65k_BT$) compared to the wild type ($\Delta U=1.56k_BT$), as  shown in Fig.~\ref{fig: fig4}(e), and discussed in~\ref{supp: expAsymmetry}.
The simulations in an asymmetric potential show that unfolding is more irreversible than refolding for the high-barrier (mutant-like) potential, while this difference vanishes for the low-barrier (wild-type-like) case, as reported in Fig.~\ref{fig: fig4}(g).
This recapitulates the experimental observation. Furthermore, 
Fig.~\ref{fig: fig4}(f) shows that the difference in irreversibility for the mutant-like simulations stems from the wider peaks in the irreversibility profile of the unfolding ramps, as observed experimentally.
 Figure~\ref{fig: fig4}(d) and~\ref{fig: fig4}(g) show that for the low-barrier (wild-type-like) potential, the irreversibility is very similar for unfolding and refolding, as seen in the experiments.
The simulations also confirm the qualitative trend that (as for the symmetric case) the peaks of irreversibility of the mutant are higher than those of the wild type.
The difference in the total irreversibility is more pronounced in the simulations than in the experiments [Figs.~\ref{fig: fig4}(d) and (g)], mostly owing to the wild-type experiment featuring some dissipation also away from the folding and refolding transitions [see $\Delta \Sigma_t$ gradients at start and end of protocols in top of Fig.~\ref{fig: fig4}(c) inset]. These gradients are not due to early or late folding or unfolding events, which are very rare for these times, as can be seen in the survival probability shown in Fig.~\ref{fig: exp_escapeTexp}. This is confirmed by the fact that the main contribution to irreversibility is given by displacements in the class associated with the folded/unfolded states. 
This irreversibility may arise from the non-equilibrium stretching of the polymer.

 \section{Discussion}\label{sec: discussion}
 \subsection{Summary of the results}
We have introduced an experimentally accessible, model-free, direct measure of irreversibility based on classifying displacements according to their location and comparing them with suitably defined time-reversal displacements. 
The method is experimentally feasible because it does not attempt to sample the overwhelmingly large space of trajectories and focuses on the initial and final points of displacements. 
The risk of overlooking dissipation in this potentially lossy approach is mitigated by the fact that the method distinguishes displacements that take place in different locations.
We have discussed the relation between the irreversibility estimated by our method and entropy production, to which it can provide a lower bound for stationary dynamics, for systems relaxing towards a steady state, and, in general, in the short-time limit.
By studying a forced particle in a potential on a ring, we have shown how the method captures more irreversibility than a naive TUR and progressively improves as we increase the number of classes in which we categorize the displacement.
We have also shown how, by monitoring the system at relatively high (but experimentally feasible) frequency, the method can identify the spatial profile of dissipation, as shown in Fig.~\ref{fig: fig2}(c).
We have showcased the method's potential in time-dependent nonequilibrium systems and demonstrated how it can be applied to study single-molecule force-spectroscopy experiments. This demonstrates that the applicability of the method is not restricted to ideal situations but extends to experiments where measurement noise and a limited number of trajectories impact the study.

 \subsection{Relation to existing measures of irreversibility}
 A direct comparison between the measure of irreversibility we are presenting and existing methods in the literature is not straightforward.
 The main difficulty lies in the scarcity of model-free methods that can be applied to time-dependent continuous systems (such as driven Langevin systems) without requiring non-trivial processing of the data.
 This is easier for stationary systems, where the TUR can be directly applied, as we discuss below. 
 For time-dependent systems, however, this is challenging. Applications of the TUR to systems under time-dependent force protocols require observing the response of the system to slight changes in the driving at every time point~\cite{Koyuk20}, or knowledge of the spatial derivatives of the full force~\cite{van_vu_thermodynamic_2020} (and thus an error-free estimation of the energy landscape). \\
 An alternative approach to characterize irreversibility involves slicing the time-dependent trajectory into short intervals and applying the TUR on each of the intervals. The machine-learning method of~\cite{Otsubo22} and the estimator of~\cite{Lee23} optimize this approach and show accurate applicability to an RNA unfolding protocol. These methods come at the cost of decreasing interpretability and require non-trivial hyperparameter and architecture optimization. We discuss the performances of~\cite{Otsubo22} and of a short-time displacements TUR on our experimental data in Section~\ref{sec: comp_exp} and~\ref{sec: otsubo}.
 \\
 In addition, the recent method presented in~\cite{Degunther24} is suitable for studying linear force ramps of a protein, but it requires both the forward and backward protocols. Additionally, it needs sampling of two continuous random variables, the times of starting and completing (un)folding transitions in both protocols, which are binned in two dimensions, which makes it rather data-intensive. We present its implementation for our experimental data in~\ref{sec: snippet} and summarize its performance in Section~\ref{sec: comp_exp}.
 \\
 The method of Ref.~\cite{Singh24} is accessible and not data-intensive, but requires knowing the diffusion coefficient. This is challenging to estimate for systems evolving in the complex energy landscapes (such as proteins) measured at finite time resolution~\cite{ragwitz2001indispensable}, and made additionally difficult by measurement noise~\cite{martin2002apparent}. 
In~\ref{sec: proesmans}, we compare our method to that of Ref.~\cite{Singh24} for simulations in which $D$ is known a priori. We find that, while the trend of higher dissipation at higher driving is also inferred, our method better captures the relative difference in irreversibility between low and high barrier systems and between folding and refolding ramps of the same, as shown in Fig.~\ref{fig: proesmans}.
For the experimental studies, due to the challenges in measuring the diffusion coefficient,
we assume a reference space constant diffusion coefficient in both protein constructs. This prevents us from comparing the magnitude of the estimators but allows us to study their profiles in time and trends across experimental conditions. While the trends are roughly consistent, the method 
of Ref.~\cite{Singh24} does not capture that unfolding is more irreversible than folding in the R3$^{{\rm{IVVI}}}$ construct at any pulling rate.
 
 \subsubsection{Relation to the thermodynamic uncertainty relation (TUR) and milestoning for stationary systems.}\label{sec: tur_mile}
 For stationary systems, it is possible to directly compare our method to estimates made by the TUR using the same type of displacement statistics.
The TUR provides a lower bound to entropy production by considering the mean and fluctuations of a time-integrated current.
The displacements we are considering can be analyzed with a TUR approach since $\ell$ is an integrated current
over the time window $\Delta t$,  $\ell=\int_{t=0}^{t=\Delta t} dx_t$ satisfying the $\ell\rightarrow-\ell$ odd-parity under time reversal. For such a current, the TUR gives a bound of the entropy production by the mean net displacement $\langle \ell\rangle$ and its variance as:
\begin{align} \label{eq: TUR}
    2k_B\frac{\langle\ell\rangle^2}{\langle\ell^2\rangle-\langle\ell\rangle^2} \leq \langle\Delta S_{tot}\rangle \,.
\end{align}
This type of analysis does not take advantage of categorizing the displacements based on where they take place. In this respect, it can be compared to our method for the partitioning where all displacements belong to the same class $\mathcal{I}$.
Our case studies confirm that the two measures yield similar estimates of irreversibility, as shown in Fig.~\ref{fig: fig2}(d). 
For the cases we considered, categorizing displacements based on where they occur returns a better estimate of irreversibility, making them preferable to the TUR, especially for cases where the dynamics vary considerably in space.
When the system is driven far from equilibrium, all estimators capture more of the irreversibility (as noted for the TUR in a sinusoidal potential in~\cite{hyeon_physical_2017}), but our method still outperforms the TUR.
It would be an interesting future direction to investigate the relation between the multi-dimensional TUR and our method based on categorizing displacements~\cite{Lee23}.

We also compare the estimate provided by $\Sigma_{\Delta t}(\mathcal{C}_6)$ to that of the milestoning method of Ref.~\cite{Blom24} on simulated data from the non-equilibrium steady state in a potential illustrated in Fig.~\ref{fig: fig2}, as detailed in \ref{app: milestoning}. For a particle on a one-dimensional ring, milestoning is known to recover the full entropy production rate for long trajectories sampled at short time intervals. However, for finite trajectory length and time intervals, $\Sigma_{\Delta t}(\mathcal{C}_6)$ can capture a higher fraction of entropy production, and milestoning may also overestimate entropy production.
The comparison in Fig.~\ref{fig: milestoning} shows that, far from equilibrium, $\Sigma_{\Delta t}(\mathcal{C}_6)$ provides a more reliable estimate of entropy production for a finite amount of data, with faster convergence. 
Increasing the number of milestones provides a faster convergence, but due to the finite sampling time $\Delta t$, results in an overestimation of the entropy production for longer trajectories as shown in Fig.~\ref{fig: milestoning_dt}. 
Closer to equilibrium, milestoning converges faster than $\Sigma_{\Delta t}(\mathcal{C}_6)$, but is still prone to overestimations. 
Increasing the sampling time $\Delta t$ improves the convergence of $\Sigma_{\Delta t}(\mathcal{C}_6)$ and makes the overestimation of milestoning more severe.
These results suggest the advantage of using our method for short trajectories far from equilibrium and for trajectories that cannot be sampled at high frequency.

\subsubsection{Comparison of methods to estimate dissipation in protein pulling experiments.}\label{sec: comp_exp}

  Our method does not involve the difficult exponential average of Jarzynski-like estimators, which are typically biased~\cite{gore2003bias,palassini2011improving} and is more robust to the presence of outliers. Additionally, unlike estimators using Crooks' relation~\cite{crooks1999entropy}, it does not require performing experiments with both the time-forward and time-backward protocols.
  However, since our experimental protocol contains both the forward and backward protocols, we are able to estimate the dissipated work using Crooks' relation and cross-validate our results. The dissipated work computed via Crooks' relation is directly linked to entropy production when the protocol starts and ends at equilibrium.
  The trend that wild-type talin R3 is more reversible than the IVVI mutant is confirmed, as is the fact that, for the mutant, unfolding is more irreversible than refolding.
  We report these measurements in Fig.~\ref{fig: crooks} and~\ref{fig: cycle}.
  
  We stress that, for systems driven by time-dependent protocols, $\Sigma_{\Delta t,t}$ is a lower bound to entropy production only for vanishing $\Delta t$.
  However, the presence of measurement noise decreases the irreversibility estimated by $\Sigma_{\Delta t,t}$ and typically mitigates the risk of overestimating entropy production, as shown in Fig.~\ref{fig: fig3} and~\ref{fig: SI_asym}.
   One should, anyway, recall that the exact value of $\Delta \Sigma_{t}$ depends on the chosen $\Delta t$, as shown in Fig.~\ref{fig: fig3}(b) and~\ref{fig: dtSigma_exp}. As a consequence, the relative amount of irreversibility captured by the method can depend on the magnitude of $\Delta t$ relative to the relevant time scales.
  In the experiments we performed, for $\Delta t=5$~ms, we estimate the fraction relative to the dissipated work to be in the range between 14 and 36\%, as shown in Fig.~\ref{fig: cycle}(b), where the higher values are obtained for faster pulling rates and the wild-type, which are characterized by faster dynamics (resulting in relatively larger $\Delta t$). Nonetheless, $\Delta \Sigma_{t}$ provides a principled and systematic measurement of irreversibility, as confirmed by the high degree of correlation with the dissipated work measured by Crooks' relation shown in Fig.~\ref{fig: cycle}(a).

 Having access to both the forward and reverse protocol allows us to implement the method of Ref.~\cite{Degunther24}. We find that, while the estimates of the wild-type are fairly robust, the results for the IVVI mutant are crucially sensitive to the specific implementation of the method, depending strongly on how the discrete states are chosen and the histogram binning (see Fig.~\ref{fig: snippetEnt}). As suggested in Ref.~\cite{Degunther24}, when data is limited, it is advisable to bin the histogram coarsely. We find that with this choice, and a suitable selection of the states (which is non-trivial without knowing the ground truth), the correct irreversibility trends can be identified [see Fig.~\ref{fig: snippetEnt}(d)].
 
   Similarly to our approach, one could use other estimates devised for stationary or short-time dynamics and apply them to time-dependent systems measured over small time intervals. We compare our experimental estimates of irreversibility $\Sigma_{\Delta t, t}$, to the TUR taken at short-time displacements, and the machine-learning technique~\cite{Otsubo22} for time-dependent entropy production. This is more extensively discussed in~\ref{sec: otsubo}.
  Both techniques produce similar results to $\Sigma_{\Delta t, t}$ on the wild-type, but do not recover all the expected trends in the more irreversible mutant, as shown in Fig.~\ref{fig: otsubo} in the SM. For instance, the method of Ref.~\cite{Otsubo22} with the default architecture and hyperparameter choice does not reproduce the increase of dissipation with higher pulling rates, finding a higher dissipation for the intermediate ramp at 10~pN/s than for the fast one at 20~pN/s. Furthermore, both alternative methods typically find a lower dissipation for the mutant than for the wild-type, which is in contrast with the results from $\Sigma_{\Delta t, t}$, and Crooks' relation. 
  The short-time TUR also erroneously predicts that unfolding dissipates less than refolding. The estimates of $\Sigma_{\Delta t, t}$ are in line with physical expectations (faster ramps dissipate more), with numerical simulations (higher irreversibility for unfolding and for more stable proteins), and confirmed by independent estimates via Crooks' relation. The discrepancies of the short-time TUR, and the methods from Refs.~\cite{Otsubo22,Singh24} suggest  $\Sigma_{\Delta t, t}$ is more reliable
at estimating irreversibility in these experiments.

\subsection{Applications to biomolecular folding}\label{sec: app_biomolecular}
 Stochastic thermodynamics and its application to biomolecular folding have a rich history, dating back to Jarzynski and Crooks' seminal work~\cite{Jarzynski1997NonequilibriumDifferences,crooks1999entropy}. The application of fluctuation theorems to nonequilibrium single-molecule rupture experiments has enabled the recovery of folding free energies in general with good agreement with bulk biochemical data~\cite{Collin05}, and, more recently, of other thermodynamic quantities such as enthalpy or heat capacity changes~\cite{rico2022molten}. 
These methodologies require starting from an equilibrium case, and therefore have mostly been employed to study thermodynamic differences between the folded and unfolded states.
Our measurement of irreversibility naturally handles nonequilibrium scenarios, providing detailed profiles of how irreversibility evolves in time. This offers new insight into the molecular mechanisms that drive transitions between the folded and unfolded states.
 
  By precisely quantifying irreversibility, we have demonstrated that it is possible to distinguish thermodynamic differences between the folding mechanisms of the R3 wild-type and the widely studied R3$^{{\rm{IVVI}}}$ mutant. Strikingly, while the mechanically fragile wild-type variant exhibits nearly symmetric unfolding and refolding transitions, R3$^{{\rm{IVVI}}}$ shows higher irreversibility for unfolding compared to refolding. We traced back this asymmetry to features of the underlying energy landscape, in which the unfolding barrier lies closer to the folded basin, producing distinct curvatures for the folded and unfolded wells. These results agree with the observations made for DNA hairpins~\cite{manosas2009dynamic,alemany2017force}. 
  
 From the perspective of protein folding, talin R3 shows a rather unique behavior, hallmarked by its exquisite force-sensitivity both for unfolding and refolding that underpins its biological function as a threshold for cellular mechanotransduction~\cite{elosegui2016mechanical}. By contrast, most classically studied proteins are mechanically rigid, unfolding at high forces with only weak force-dependence in their unfolding rates, while refolding at very low forces. 
  This large mechanical hysteresis generates a high dissipation during unfolding/refolding cycles, a property that could be essential to protect cellular structures against mechanical shocks~\cite{doi:10.1021/acsnano.8b05721}.
 In this context, we speculate that such shock-absorbing proteins will be characterized by more pronounced asymmetries in entropy production across cyclic force protocols. Applying our method to these proteins would allow us to disentangle the differences in dissipation between folding and unfolding, identifying which step is responsible for most of the dissipation.

 \section{Conclusions}\label{sec: conclusion}
Investigating the out-of-equilibrium behavior of small systems sheds light on their features, properties, and capabilities.
This is especially valuable for living systems, which typically operate away from equilibrium.
To this aim, we have proposed an experimentally accessible model-free method, and related it to entropy production.
We have highlighted the situations where the relation with entropy is tight and scrutinized the reasons behind it.
For stationary systems, the method can be leveraged to provide a spatial profile of dissipation identifying where the system dissipates the most.
Testing the method on nonequilibrium force-spectroscopy measurements of proteins revealed how, by providing a quantitative framework for nonequilibrium thermodynamics in biomolecular transitions, our method is sensitive to subtle mechanical and thermodynamic signatures that govern protein responses to force. This establishes it as a rigorous tool for rationalizing the function of mechanically active proteins.

The method we have proposed relies on the displacements measured at fixed time intervals. In the future, it will also be interesting to explore approaches where time is the random variable being observed to probe equilibrium, as done for example in Refs.~\cite{Alvarez2006,Berezhkovskii2006,Gladrow2019ExperimentalTimes,Neri_2022, Raghu_2025}.

\begin{acknowledgments}
SB wishes to thank Frank Jülicher for suggesting using the difference in displacement distributions to measure irreversibility. The authors also acknowledge inspiring and insightful discussions with 
Izaak Neri, Adarsh Raghu, Lennart Dabelow, Lars Hubatsch, Martina Bulgari, Tan Van Vu, and Felix Ritort. We thank Patricia Paracuellos for expression and purification of the protein constructs, and Sergi Garcia-Manyes for use of the molecular biology equipment. R.T-R. is supported by an EPSRC Horizon Europe Guarantee ERC Starting Grant (EP/Y036085/1), an EPSRC ECR International Collaboration Grant (EP/Y001125/1), and a Royal Society Research Grant (RG/R2/232147).
\end{acknowledgments}

\bibliography{apssamp}

\pagebreak
\widetext
\appendix

 \section{Proof of the entropy bound.}\label{app: proofBound}
 To prove that our measure of irreversibility gives a lower bound to entropy production,  Eq.~\eqref{eq: bound}, we proceed in two steps.
 The first step is to show that considering only the initial and final point of a displacement contains less information about irreversibility than the whole trajectory. The second step demonstrates that focusing on displacements $\ell=x-x'$ and on which class they belong to instead of retaining both $x'$ and $x$, results in a further loss of information.
 We start by introducing the notation $P_{\Delta t}(x', x) = P_{\Delta t}(x|x')P(x')$ to denote the joint probability of starting with a particle in position $x'$ and finding it in position $x$ after a time $\Delta t$.
 Note that the order of the two variables matters, and $P_{\Delta t}(x, x') = P_{\Delta t}(x'|x)P(x)$ denotes the joint probability of starting in $x$ and arriving at $x'$.
 
Starting from Eq.~\eqref{eq: pathEP}, we can marginalize the intermediate points in the trajectories $x'\rightarrow x$. For a Markov process, this yields the joint probabilities as a function of the propagator and the initial distribution $P_{\Delta t}(x', x) = P_{\Delta t}(x|x')P(x')$. Therefore as shown in~\ref{supp: prop},
\begin{eqnarray}
    D_{KL}\left(\mathcal{P}(\mathbf{x}\right) || \mathcal{P}(\widetilde{\mathbf{x}}))
    \geq D_{KL}(P_{\Delta t}(x',x) ||P_{\Delta t}(x,x'))\,, \label{eq: jointCG}
\end{eqnarray}
where equality holds for one-dimensional systems and for systems where the propagator obeys a detailed-balance-like condition $\log[P(x|x')/P(x'|x)] = g(x)-g(x')$ for a generic function $g$.
For the second step, let us re-express the Kullback-Leibler divergence as a sum over the different classes of displacements that we are considering.
For a generic collection of classes  $\mathcal{C}_n$, we have
\begin{align}
    D_{KL}(P_{\Delta t}(x',x) ||P_{\Delta t}(x,x')) = 
     \sum_{C\in \mathcal{C}_n} \iint\limits_{\Omega_C} dx' 
    \, dx \,P_{\Delta t}(x',x) \ln{\frac{P_{\Delta t}(x',x)}{P_{\Delta t}(x,x')}} \,.
     \label{eq: dkl_class}
\end{align}
Each class $C$, is defined by its support in $x'$ and $x$, $\Omega_C$, meaning from which region the displacement starts in and in which region it arrives at.
The time reversal of a displacement entails swapping the initial and final points. The time reversal of displacements belonging to class $C$, then, belong to class $\tilde{C}$ with a support over domain $\Omega_{\tilde{C}}$ where the initial and final domains are swapped (\textit{i.e.}, $x'$ and $x$ are exchanged). 
We restrict ourselves to the case in which the domains are continuous intervals such that they can be written as $x'\in[x'^{C}_{min},x'^{C}_{max}]$ and $x\in[x^{C}_{min},x^{C}_{max}]$, with $x'^{C}_{min}<x'^{C}_{max}$ and $x^{C}_{min}<x^{C}_{max}$. 
The integral over the domain ${\Omega_C}$, can then be written in terms of Heaviside theta functions, yielding 
\begin{align}
    \iint\limits_{\Omega_C} dx' \, dx = \int\limits_{-\infty}^\infty\int\limits_{-\infty}^\infty dx' \, dx \, [\Theta(x'-x'^C_{min})-\Theta(x'-x'^C_{max})][\Theta(x-x^C_{min})-\Theta(x-x^C_{max})]\,.
    \end{align}
     We define the short-hand notation
    \begin{align}
    \theta_{x',x}^C \equiv [\Theta(x'-x'^C_{min})-\Theta(x'-x'^C_{max})][\Theta(x-x^C_{min})-\Theta(x-x^C_{max})]\,,
\end{align}
substitute these into Eq.~\eqref{eq: dkl_class}, and perform the change of variables $x\rightarrow x'+\ell$ with $d\ell=dx$ to get
\begin{align}
    D_{KL}(P_{\Delta t}(x',x) ||P_{\Delta t}(x,x')) = 
     \sum_{C\in \mathcal{C}_n} \int\limits_{-\infty}^\infty\int\limits_{-\infty}^\infty dx' d\ell \; \theta_{x',x'+\ell}^C \,P_{\Delta t}(x',x'+\ell) \ln{\frac{P_{\Delta t}(x',x'+\ell)}{P_{\Delta t}(x'+\ell,x')}} \,.
     \label{eq: dkl_class_ell}
\end{align}
Note $x^{C}_{min},x^{C}_{max}$ are constants, which remain in
$\theta_{x',x'+\ell}^C$, which reads
    \begin{align}
    \theta_{x',x'+\ell}^C = [\Theta(x'-x'^C_{min})-\Theta(x'-x'^C_{max})][\Theta(x'+\ell-x^C_{min})-\Theta(x'+\ell-x^C_{max})]\,,
\end{align}
thus enforcing the same domain after the change of variables.

Focusing on a single $C$ term of the class summation above, we can introduce the theta terms in the log-ratio and apply the log-sum inequality for a $dx'$ integration,
\begin{align}
    &\int\limits_{-\infty}^\infty\int\limits_{-\infty}^\infty dx' d\ell \; \theta_{x',x'+\ell}^C \,P_{\Delta t}(x',x'+\ell) \ln{\frac{\theta_{x',x'+\ell}^C\,P_{\Delta t}(x',x'+\ell)}{\theta_{x',x'+\ell}^C \,P_{\Delta t}(x'+\ell,x')}}\geq \nonumber \\ & \int\limits_{-\infty}^\infty d\ell \int\limits_{-\infty}^\infty dx' \; \theta_{x',x'+\ell}^C \,P_{\Delta t}(x',x'+\ell) \ln{\frac{\int\limits_{-\infty}^\infty dx'\,\theta_{x',x'+\ell}^C\,P_{\Delta t}(x',x'+\ell)}{\int\limits_{-\infty}^\infty dx'\,\theta_{x',x'+\ell}^C \,P_{\Delta t}(x'+\ell,x')}}\, \label{eq: dklndef_supp}
\end{align}
where the inequality follows from the log-sum inequality, which states that, for non-negative $a_i$ and $b_i$, $\sum a_i\log (a_i/b_i)\ge a\log (a/b)$, where $a=\sum_i a_i$ and $b=\sum_i b_i$.
The integrands in $x'$ on the RHS of Eq.~\eqref{eq: dklndef_supp}
can be linked to the displacement distributions.
The term appearing in the numerator in the log ratio
coincides with the definition of the joint displacement distributions $q_{\Delta t}(\ell,C)$, as can be shown by carrying out the integral over $x$ 
in Eq.~\eqref{eq: qdefn}:
\begin{align}
    q_{\Delta t}(\ell, C) \equiv \iint\limits_{\Omega_C} dx' 
    \, dx \, \delta\left(\ell-(x-x')\right)P_{\Delta t}(x | x')P(x') =\int\limits_{-\infty}^\infty dx' \; \theta_{x',x'+\ell}^C \,P_{\Delta t}(x',x'+\ell)\,.
\end{align}

The denominator of the log-ratio of Eq.~\eqref{eq: dklndef_supp} is the integral of the time reverse joint probability $P_{\Delta t}(x'+\ell,x')=P_{\Delta t}(x'|x'+\ell)P(x'+\ell)$, but still in the support $\theta_{x',x'+\ell}^C$ of the forward class $C$.
To connect to Eq.~\eqref{eq: dklndef}, we wish to express this integral in terms of the displacements in the time-reverse class $\tilde{C}$. 
Since the domain of the time-reverse displacements $\tilde{C}$ corresponds to swapping the initial and final domains of the original class, we have
$\theta_{x',x'+\ell}^{C}=\theta_{x'+\ell,x'}^{\tilde C}$ (as one can check by direct substitution).
We then have (renaming the dummy integrating variable $x'$ as $x$ for convenience)
\begin{align}
    \int\limits_{-\infty}^\infty dx\,\theta_{x,x+\ell}^C \,P_{\Delta t}(x+\ell,x) 
    = \int\limits_{-\infty}^\infty dx\,\theta_{x+\ell,x}^{\tilde C} \,P_{\Delta t}(x+\ell,x) 
    \\\nonumber = \iint\limits_{\Omega_{\tilde{C}}} dx' 
    \, dx \, P_{\Delta t}(x|x')P(x')\delta (\ell -(x'-x)) = q_{\Delta t}(-\ell, \tilde{C}).
\end{align}
This shows how the RHS of 
 Eq.~\eqref{eq: dklndef_supp} simplifies as
\begin{align}
    \int\limits_{-\infty}^\infty d\ell \int\limits_{-\infty}^\infty dx' \; \theta_{x',x'+\ell}^C \,P_{\Delta t}(x',x'+\ell) \ln{\frac{\int\limits_{-\infty}^\infty dx'\,\theta_{x',x'+\ell}^C\,P_{\Delta t}(x',x'+\ell)}{\int\limits_{-\infty}^\infty dx'\,\theta_{x',x'+\ell}^C \,P_{\Delta t}(x'+\ell,x')}}=\int\limits_{-\infty}^\infty d\ell\;q_{\Delta t}(\ell,C)\ln{\frac{q_{\Delta t}(\ell,C)}{ q_{\Delta t}(-\ell, \tilde{C})}}\,.
\end{align}
Recalling the definition of our measure of irreversibility from
 Eq.~\eqref{eq: dklndef}, the bound expressed in Eq.~\eqref{eq: bound} then follows by applying the log-sum inequality Eq.~\eqref{eq: dklndef_supp} to each class in $\mathcal{C}_n$. 
To summarize the proof and illustrate the bound, the two steps of information processing, starting from the full path probabilities, give
\begin{align}
    \langle \Delta S_{tot} \rangle = k_B D_{KL}\left(\mathcal{P}(\mathbf{x}\right) || \mathcal{P}^R(\widetilde{\mathbf{x}}))
    \geq k_BD_{KL}(P_{\Delta t}(x',x) ||P_{\Delta t}(x,x')) \geq \Sigma_{\Delta t}
(\mathcal{C}_n).
\end{align}
This bound holds for any stationary systems described by an overdamped Langevin process. In the following subsection, we explain how it is generalized to non-stationary cases where the underlying forces are not time-dependent.

\subsection{Time-dependent systems}\label{app: td}
We present here the generalization of our measure of irreversibility to time dependent systems Eq.~\eqref{eq: irr_transient}, and discuss its relation to entropy production.
We consider two types of time-dependent systems: systems subject to a time-constant force profile, relaxing towards a steady-state, and systems that are driven by time-dependent force protocols.
For time-constant profiles, we show how the measure of irreversibility provides a lower bound to entropy production for arbitrary $\Delta t$, while for time-dependent force protocols this holds in the $\Delta t\rightarrow 0$. We discuss the properties of the measure of irreversibility for finite $\Delta t$.

\subsubsection{Relaxation dynamics}\label{app: relax}
Consider the entropy production $\langle \Delta S_{tot} \rangle^t_{t-\Delta t}$ in a time window $\Delta t$, happening between the times $t-\Delta t$ and $t$.
When the forces governing the Langevin dynamics are time-independent, but the system has not relaxed to a steady-state yet, we have a time-dependence in the joint probability due to the starting distribution $P_{\Delta t,t-\Delta t}(x',x)=P_{\Delta t}(x|x')P_{t-\Delta t}(x')$. In analogy to what we showed for time-independent dynamics in Eq.~\eqref{eq: jointCG}, we have 
\begin{align}
    \langle \Delta S_{tot} \rangle^t_{t-\Delta t} \geq k_B D_{KL}(P_{\Delta t,t-\Delta t}(x',x) ||P_{\Delta t,t}(x,x')) = 
     k_B\iint dx' 
    \, dx \,P_{\Delta t}(x|x')P_{t-\Delta t}(x') \ln{\frac{P_{\Delta t}(x|x')P_{t-\Delta t}(x')}{P_{\Delta t}(x'|x)P_t(x)}}\,.
\end{align}
The main difference with Eq.~\eqref{eq: dkl_class} is that for the time reversal 
the trajectories start at time $t$, sampled from $P_t(x)$. 
We take this into account for classifying the displacements where the joint probability $q_{\Delta t, t-\Delta t}(\ell, C)$ of displacements $\ell=x_{t}-x'_{t-\Delta t}$ starting at $t-\Delta t$ in class $C$, and $q_{\Delta t, t}(-\ell, \tilde C)$ for the time-reverse displacements $-\ell=x_t-x'_{t+\Delta t}$ in class $\tilde C$. With this identification, we generalize our measure of irreversibility to time-dependent systems by defining 
\begin{align}
    \Sigma_{\Delta t, t}
(\mathcal{C}_n) \equiv k_B\sum_{C\in\mathcal{C}_n} \int d\ell \, q_{\Delta t, t-\Delta t}(\ell, C) \ln{\frac{q_{\Delta t, t-\Delta t}(\ell, C)}{q_{\Delta t,t}(-\ell, \tilde{C})}}\,.\label{eq: irr_transient_app}
\end{align}
 This is Eq.~\eqref{eq: irr_transient} in the main text. Following the same logic as Appendix~\ref{app: proofBound}, this quantity provides a lower bound to the average entropy production
\begin{align}
     \langle \Delta S_{tot} \rangle^t_{t-\Delta t}\geq\Sigma_{\Delta t, t}
(\mathcal{C}_n) \label{eq: irr_transient_bound}
\end{align}
An additional remark is necessary for the calculation of $\Sigma_{\Delta t}^{trans}$ under transience, modifying its definition Eq.~\eqref{eq: dkltrans}.
The marginal probabilities of belonging to class $C$ defined in Eq.~\eqref{eq: pdefn}, now have a global time $t$ component
\begin{equation}\label{eq: p_transient}
    p_{\Delta t,t}(C) \equiv \iint\limits_{\Omega_C} dx' 
    \, dx \, P_{\Delta t}(x | x')P_t(x')=\int d\ell \, q_{\Delta t, t}(\ell,C)\,.
\end{equation}
$\Sigma_{\Delta t}^{trans}$ is then defined as
\begin{align}
\Sigma_{\Delta t, t}^{trans}\equiv k_B \sum_{C\in\mathcal{C}_n} p_{\Delta t, t-\Delta t}(C) \ln{\frac{p_{\Delta t, t-\Delta t}(C)}{p_{\Delta t, t}(\tilde C)}} \label{eq: trans_transient}
\end{align}
For the stationary case, displacements that under time reversal belonged to the same class (as the bulk classes in Section \ref{sec: squarepot}) canceled out and did not appear in Eq.~\eqref{eq: dkltrans}.
For time-dependent systems, however, the probability of belonging to a class changes with time so that also classes that do not change under time reversal (as the bulk classes) must be accounted for in Eq.~\eqref{eq: trans_transient}.
From Eq.~\eqref{eq: p_transient} and applying the log-sum inequality to Eq.~\eqref{eq: irr_transient_app} (integrating over $d\ell$ for each $C$) we have the bound
\begin{align}
    \Sigma_{\Delta t, t}(\mathcal{C}_n)\geq \Sigma_{\Delta t, t}^{trans}\equiv k_B \sum_{C\in\mathcal{C}_n} p_{\Delta t, t-\Delta t}(C) \ln{\frac{p_{\Delta t, t-\Delta t}(C)}{p_{\Delta t, t}(\tilde C)}}\,. \label{eq: trans_transient_bound}
\end{align}

\subsubsection{Time-dependent driving} \label{supp: shift}

In cases where the forces $F(x_\tau,\lambda(\tau))$ in the Langevin dynamics have a time-dependent component $\lambda$, the time-reverse protocol [$\tilde \lambda(\tau)=\lambda(t-\tau)$, for a total protocol time $t$] must be included in the reverse path $\widetilde{\mathbf{x}}$'s probabilities to define the correct path-wise entropy production
\begin{align}
    \Delta S_{tot}(\mathbf{x}) = k_B \ln{\frac{\mathcal{P}(\mathbf{x},\lambda(\tau))}{\mathcal{P}^R(\widetilde{\mathbf{x}},\lambda(t-\tau))}} 
\end{align} \\
As in Appendix~\ref{app: relax}, we consider the entropy production $\langle \Delta S_{tot} \rangle^t_{t-\Delta t}$ in a time window $\Delta t$, happening between the times $t-\Delta t$ and $t$. Here the propagator component of the joint probability will also have an explicit global time dependence, $P_{\Delta t,t-\Delta t}(x',x)=P_{\Delta t,t-\Delta t}(x|x')P_{t-\Delta t}(x')$, with $P_{\Delta t,t-\Delta t}(x',x)$ marking the evolution for a time $\Delta t$ starting at $t-\Delta t$.
\paragraph*{Infinitesimal time intervals}
For infinitesimally small time intervals $dt$, the explicit time dependence of the propagator plays a secondary role and can be ignored. In order to see this,
consider the entropy produced for an infinitesimal time step $dt$ during $[t-dt,t]$, including protocol reversal from $t$ in the next step $[t,t+dt]$,
\begin{align}\label{eq: dS_def}
    dS_{tot}(x | x')=k_B\ln\frac{P_{dt,t-dt}(x|x')P_{t-dt}(x')}{\widetilde P_{dt, t}(x'|x)P_{t}(x)} 
   \,,
\end{align}
where $dS_{tot}(x | x')$ denotes the entropy produced in a trajectory starting at position $x'$ and arriving at position $x$. $\widetilde P_{dt, t}(x'|x)$ is the infinitesimal propagator for the subsequent step with protocol reversal such that $\tilde \lambda(t+d t)=\lambda(t-dt)$.
To explicitly evaluate this expression, one can make use of the explicit Gaussian expression of the propagator for infinitesimal time steps \cite{Lau07}
\begin{align}
    P_{dt,t-dt}(x|x') \propto \exp\left[-\frac{\left( x-x'-v_{\bar{t}}(\bar{x})dt\right)^2}{4Ddt}-\frac{1}{2}\frac{\partial v_{\bar{t}}(\bar{x})}{\partial x}\right]
\end{align}
where we evaluate the deterministic drift $v_{\tau}(x)=F(x,\lambda(\tau))/\gamma$ at the midpoint $v_{\bar{t}}(\bar{x})$ where $\bar{t}=t-dt/2$, $\bar{x}=(x'+x)/2$, and for simplicity, assume a constant $D$.
Denoting $dx=x-x'$, the ratio in Eq. \eqref{eq: dS_def} becomes
\begin{align}
    \frac{dS_{tot}(x | x')}{k_B}=
    \frac{d x}{D}v_{\bar{t}}(\bar{x})+ \ln{\frac{P_{t-dt}(x')}{P_t(x)}} + \mathcal{O}(dt^{3/2})  \label{eq: dS}
\end{align}
where we have used that the reversed protocol, $\tilde \lambda(t+dt/2)=\lambda(t-dt/2)$.
Note that evaluating the drift at the start of the protocol, $v_{t-dt}(\bar{x})$, yields the same result to leading order, because, from the Langevin equation, $dx$ contains contributions of order $dt$ and $dt^{1/2}$.

If one cannot reverse the protocol, in an infinitesimal time step, comparing the displacements starting at $t-dt$ with the ones starting at $t$ leads to the same result as Eq.~\eqref{eq: dS}, to leading order
    \begin{align}
        \ln{\frac{P_{\Delta t,t-dt}(x|x')P_{t-  dt}(x')}{P_{\Delta t, t}(x'|x)P_{ t}(x)}} = \frac{dx}{D}v_{t}(\bar{x}) + \ln{\frac{P_{t-dt}(x')}{P_{t}(x)}} +\mathcal{O}(dt^{3/2}) \label{eq: shift}\,.
    \end{align}
    Repeating the same steps, as for the relaxation dynamics, shows that the measure of irreversibility introduced in Eq.~\eqref{eq: irr_transient_app} provides a lower bound to the average entropy production also for time-dependent driving, in the limit of infinitesimally small time intervals: $\Sigma_{\Delta t, t}(\mathcal{C}_n)\leq\langle \Delta S_{tot} \rangle^t_{t-\Delta t}$ as $\Delta t\rightarrow0$.
    For larger time intervals, the bound is not valid. This becomes more evident for faster dynamics, such as the fast pulling rates, which feature a strong time-dependence of $v$.
    In Fig.~\ref{fig: fig3}(c), we observed a slight overestimation of the true entropy production for the faster pulling rates measured with $\Delta t = 5$~\si{ms} for the low barrier case. Note that, if one can reverse the protocol in the $q_{\Delta t, t}(-\ell,\tilde C)$ displacements of Eq.~\eqref{eq: irr_transient_app}, the bound will hold for any $\Delta t$, but this is typically not operationally accessible.

 \section{Fluctuation Theorem}\label{app: ft}
As discussed in Appendix \ref{app: proofBound}, in the limit of vanishing $\Delta t$, Eq.~\eqref{eq: jointCG} is an equality. It then follows that 
\begin{align}
    \frac{P_{\Delta t}(x|x')P(x')}{P_{\Delta t}(x'|x)P(x)} 
    = e^{\Delta S_{tot}(x | x')/k_B} = e^{-\Delta S_{tot}(x'|x)/k_B}\,,
\end{align}
where $\Delta S_{tot}(x | x')$ denotes the entropy produced in a trajectory starting at position $x'$ and arriving at position $x$\footnote{In addition to infinitesimally short trajectories, the entropy produced is a function of the initial and final point only, also for one-dimensional trajectories as shown in~\ref{supp: prop}.}.
This allows us to rewrite the probability of a displacement in terms of that of its time reversal and the entropy produced along such displacement.
We can rewrite Eq.~\eqref{eq: qdefn} as
\begin{align}\label{eq: q_S}
    q_{\Delta t}(\ell, C) \equiv \iint\limits_{\Omega_C} dx' 
    \, dx \, \delta\left(\ell-(x-x')\right)P_{\Delta t}(x | x')P(x')=
    \iint\limits_{\Omega_C} dx' 
    \, dx \, \delta\left(\ell-(x-x')\right)P_{\Delta t}(x' | x)P(x)e^{-\Delta S_{tot}(x'|x)/k_B}\,.
\end{align}
Exchanging the role of the starting and final position $x\leftrightarrow x'$ maps the integration domain $\Omega_C$ on to the one of its corresponding time-reversed class $\Omega_{\tilde{C}}$, so that we have 
\begin{align}\label{eq: q_S_tilde}
    q_{\Delta t}(\ell, C) =
    \iint\limits_{\Omega_{\tilde{C}}} dx' 
    \, dx \, \delta\left(\ell+(x-x')\right)P_{\Delta t}(x | x')P(x')e^{-\Delta S_{tot}(x|x')/k_B}\,.
\end{align}
Taking the ratio of this probability with the probability of a time-reversed displacement, we obtain 
\begin{align}
    \frac{q_{\Delta t}(\ell, C)}{q_{\Delta t}(-\ell, \tilde{C})}=
    \langle e^{-\Delta S_{tot}/k_B}|-\ell, \tilde{C}\rangle\,,
\end{align}
which is the conditional average of the exponential of minus the entropy that is produced in a displacement, with the conditioning on the fact that the displacement is $-\ell$ and of class $\tilde{C}$.
Similarly, one has
\begin{align}\label{eq: ft_app}
    \frac{q_{\Delta t}(-\ell, \tilde{C})}{q_{\Delta t}(\ell, C)}=
    \langle e^{-\Delta S_{tot}/k_B}|\ell, C\rangle\,,
\end{align}
which is Eq.~\eqref{eq: ft} in the main text.
This expression can be used to provide an alternative proof that the measure of irreversibility we have introduced provides a lower bound to entropy production,  Eq.~\eqref{eq: bound}.
Applying Jensen inequality for convex functions, we have that
\begin{align}
    \langle e^{-\Delta S_{tot}/k_B}|\ell, C\rangle\ge \exp\left[\langle-\Delta S_{tot}/k_B|\ell, C\rangle\right]
\end{align}
Combining this with Eq.~\eqref{eq: ft_app}, taking the log and changing signs, we have 
\begin{align}
     \langle\Delta S_{tot}/k_B|\ell, C\rangle\ge \ln \frac{q_{\Delta t}(\ell, C)}{q_{\Delta t}(-\ell, \tilde{C})}\,,
\end{align}
from which Eq.~\eqref{eq: bound} follows, upon multiplication by $q_{\Delta t}(\ell, C)$ and integration.

\let\section\normalsection
\makeatletter
\let\@hangfrom@section\normalhangfromsection
\let\@sectioncntformat\normalsectioncntformat
\makeatother
\newpage

\begin{center}
\textbf{\large Supplementary Material}
\end{center}

\setcounter{equation}{0}
\setcounter{figure}{0}
\setcounter{table}{0}
\setcounter{section}{0}
\makeatletter
\renewcommand{\theequation}{S-\Roman{section}.\arabic{equation}}
\renewcommand{\thefigure}{S\arabic{figure}}
\renewcommand{\thesection}{SM-\Roman{section}}

\startcontents[SM]
{\noindent\large\bfseries Contents\par\medskip}
\printcontents[SM]{l}{1}{}

\section{1D Langevin Conservation of Entropy under Marginalization of Propagators}\label{supp: prop}

The joint probability distribution $P_{\Delta t}(x',x)$ is obtained by integrating out the constituent infinitesimal steps of a path $\mathbf{x}=[x_0=x',x_1,...,x_{N-1},x_N=x]$ with $Ndt=\Delta t$, as by the Chapman-Kolmogorov relation
\begin{equation}
    P_{\Delta t}(x',x) = P(x') \int \prod_{j=1}^{N-1} \, dx_j \,\mathcal{P}(\mathbf{x} | x')=P(x') \int \prod_{j=1}^{N-1} \, dx_j \prod_{i=0}^{N-1}P(x_{i+1}|x_{i})
\end{equation}

Equivalently, we have $P_{\Delta t}(x,x')$ for the time-reverse trajectory $\widetilde{\mathbf{x}}=[x,x_{N-1},...,x_{1},x']$. These marginalizations give Eq.~\eqref{eq: jointCG} via the log-sum inequality:
\begin{align}
    \langle \Delta S_{tot} \rangle &= k_B D_{KL}(\mathcal{P}(\mathbf{x}) || \mathcal{P}(\widetilde{\mathbf{x}})) \nonumber\\
    &= k_B \int \mathcal{D}\mathbf{x} \, \mathcal{P}(\mathbf{x}|x')P(x') \ln{\frac{\mathcal{P}(\mathbf{x}|x')P(x')}{\mathcal{P}(\widetilde{\mathbf{x}}|x)P(x)}} \label{eq: 3.1}\\
    &\geq k_B D_{KL}(P_{\Delta t}(x',x) ||P_{\Delta t}(x,x')) \equiv k_B \iint dx' \,dx \, P_{\Delta t}(x|x')P(x') \ln{\frac{P_{\Delta t}(x|x')P(x')}{P_{\Delta t}(x'|x)P(x)}}\,. \label{eq: CGpath}
\end{align}
From above, it is clear the equality will occur when the log ratio of the propagators $P_{\Delta t}(x|x')/P_{\Delta t}(x'|x)$ is equal to that of trajectory probabilities $\mathcal{P}(\mathbf{x}|x')/P(\widetilde{\mathbf{x}}|x)$, \textit{i.e.}, if $x'$ and $x$ are the only relevant degrees of freedom to the total dissipation. As we will show below, this is the case for a one-dimensional Langevin system subject to a time-independent drift and diffusion coefficient: with dynamics $\dot{x}_t=v(x_t)+\sqrt{2D}\xi_t$ but it is no longer true for a time-dependent drift.

Let us start by constructing the ratio of the full path probabilities before marginalizing to the joints. We discretize the path integral in the Stratonovich sense, which makes the discussion of time-reversal more direct.  For an additive noise, the infinitesimal propagator~\cite{Lau07} reads 
\begin{align}
    P(x_{i+1}|x_i)\propto \exp\left[{-\frac{1}{4Ddt}(\Delta x_i-v_i(\bar{x}_i)dt)^2-\frac{1}{2}\nabla v_i(\bar{x}_i)dt}\right]
\end{align}
with $\bar{x}_i=\frac{1}{2}(x_i+x_{i+1})$ the midpoint where the forces $F_i(\bar{x}_i)=\gamma v_i(\bar{x}_i)$ are evaluated, and $\Delta x_i=x_{i+1}-x_i$. In one dimension, a non-conservative force $f_{\rm{0}}$ and a conservative one stemming form a potential $W$ can be written as the derivative of an effective potential $V(x)=W(x)-f_{\rm{0}}x$. Here, we also consider a time-linear force ramp
with rate $r$, so $F_i(\bar{x}_i)=-\nabla V(\bar{x}_i)+r(i+\frac{1}{2})dt$, and the path probabilities
\begin{align}
    \mathcal{P}(\mathbf{x} | x')=\prod_{i=0}^{N-1}P(x_{i+1}|x_{i}) &\propto \exp\left[\sum_{i=0}^{N-1}{-\frac{1}{4Ddt}(\Delta x_i-v_i(\bar{x}_i)dt)^2-\frac{1}{2}\nabla v(\bar{x}_i)dt}\right] \label{eq: forInfProp} \\
    \mathcal{P}(\widetilde{\mathbf{x}}|x)=\prod_{i=0}^{N-1}P(\tilde x_{i+1}|\tilde x_{i}) &\propto \exp\left[\sum_{i=0}^{N-1}{-\frac{1}{4Ddt}(\Delta x_i+v_i(\bar{x}_i)dt)^2-\frac{1}{2}\nabla v(\bar{x}_i)dt}\right] \label{eq: backInfProp}
\end{align}
with $\tilde{x}_i=x_{N-i}$ leading to a sign flip in Eq.~\eqref{eq: backInfProp} to substitute $\Delta x_i$ in terms of the time-forward increments.  We drop the $i$ subscript of $\nabla v_i(x_i)$ because it does not depend on time. The two expressions above differ only in the cross term of infinitesimal displacement with drift in the squared brackets ($\Delta x_iv_i(\bar x_i)$, other terms will cancel inside the log-ratio of Eq.~\eqref{eq: 3.1}). Substituting for $v(\bar{x}_i)$, these terms give
\begin{align}
    \sum_{i=0}^{N-1}{\frac{1}{2D\gamma}\Delta x_i(-\nabla V(\bar{x}_i)+r(i+\frac{1}{2})dt)} &= \frac{1}{2k_BT}(W(x')-W(x) + (x-x')f_0) + \frac{r}{2k_BT}\sum_{i=0}^{N-1}{\Delta x_i(i+\frac{1}{2})dt} \label{eq: crosstermF}\\
    \sum_{i=0}^{N-1}{\frac{-1}{2D\gamma}\Delta x_i(-\nabla V(\bar{x}_i)+r(i+\frac{1}{2})dt)} &= \frac{1}{2k_BT}(W(x)-W(x') + (x'-x)f_0) - \frac{r}{2k_BT}\sum_{i=0}^{N-1}{\Delta x_i(i+\frac{1}{2})dt} \label{eq: crosstermB}
\end{align}
for the time-forward and reverse path probabilities, respectively. The first term in each expression contains a contribution from the conservative and non-conservative forces, which only depend on the initial and final states of the paths.
This shows how, if the force is not changed ($r=0$), the ratio of path probabilities will depend only on the initial and final points. As a consequence, marginalizing over the intermediate steps is trivial, with $\mathcal{P}(\mathbf{x} | x')=P(x|x')$ and $\mathcal{P}(\widetilde{\mathbf{x}}|x_N)=P(x'|x)$, there is no information lost going from Eq.~\eqref{eq: 3.1} to Eq.~\eqref{eq: CGpath}. In this case, we can recover the full entropy production with only information about the $x'$ and $x$ degrees of freedom,
\begin{align}
    \langle \Delta S_{tot} \rangle = k_B D_{KL}(\mathcal{P}(\mathbf{x}) || \mathcal{P}(\widetilde{\mathbf{x}})) 
    = k_B D_{KL}(P_{\Delta t}(x',x) ||P_{\Delta t}(x,x')).
\end{align}

For time-dependent forces, however, we cannot simplify the last ramp-dependent terms, so applying the $\prod_{j=1}^{N-1} \, dx_j$ marginalization to the path propagators will lose information about the dissipation the longer the considered $\Delta t$ is.\\
The log-ratio of the two path probabilities is given by the difference of Eqs.~\eqref{eq: crosstermF} and~\eqref{eq: crosstermB}
\begin{equation}
    \frac{\mathcal{P}(\mathbf{x} | x')}{\mathcal{P}(\widetilde{\mathbf{x}}|x)}=\frac{1}{k_BT}\left[W(x')-W(x)+(x-x')f_0\right]+\frac{r}{k_BT}\left[-\sum_{i=1}^{N-1} x_idt +\left(N-\frac{1}{2}\right)x\,dt-\frac{1}{2}x'dt\right]
\end{equation}
where the last sum is more conveniently represented in the continuum limit $dt\rightarrow0$ ,
\begin{equation}
\frac{\mathcal{P}(\mathbf{x} | x')}{\mathcal{P}(\widetilde{\mathbf{x}}|x)}=
    \frac{1}{k_BT}\left[W(x')-W(x)+(x-x')f_0\right]+\frac{1}{k_BT}\left(rx\Delta t-r\int_0^{\Delta t} x_tdt \right).
\end{equation}
Again, this is equal to $P(x_N | x')/P(x' | x_N)$ only if there is no change in the applied force, $r=0$, or if $\Delta t\to 0$.

\section{Short time expansion of $\Sigma_{\Delta t}^{trans}$}\label{supp: transExpand}
In this section we provide the derivation for the short time approximation Eq.~\eqref{eq: transExpand} of the $\Sigma_{\Delta t}^{trans}$ defined in Eq.~\eqref{eq: dkltrans}.\\
The small $\Delta t$ propagator for a general Langevin process $\dot{x} = v(x_t) + \sqrt{2D}\xi_t$ with deterministic drift $v(x)=\frac{1}{\gamma}(F_0 - \nabla W(x))$, in the Ito discretization is given by~\cite{risken1996fokker}
\begin{equation}
    P_{\Delta t}(x | x') \approx \dfrac{1}{2\sqrt{\pi D \Delta t}} e^{-\left(x-x'-v(x')\Delta t\right)^2/4D\Delta t}. \label{eq: shortprop}
\end{equation}
Assuming that, for short times, displacements which cross $k$ are start/end in its vicinity,
we Taylor expand $P(x')$ and $v(x')$ around $k$. We have $P(x')\approx P(k)+(x'-k)\nabla P(k)$ and $v(x')\approx v(k)+(x'-k)\nabla v(k)$. We keep the 2nd and 1st order terms in each, respectively, because their higher orders of $x'$ do not affect the lowest order term in $\Delta t$ in the final result. 
Expanding the marginal probabilities [see Eqs.~\eqref{eq: pCrossR} and~\eqref{eq: pdefn}] of crossing $k$ in $\Delta t$,

\begin{align}
    p_{\Delta t}(k^\rightarrow) &= \int_{-\infty}^k dx' \int^{\infty}_k dx \, P_{\Delta t}(x | x')P(x') \nonumber \\
    &\approx \dfrac{1}{2\sqrt{\pi D \Delta t}} \int_{-\infty}^k dx' \int^{\infty}_k dx \,  e^{-\left(x-x'-v(k)\Delta t\right)^2/4D\Delta t} (P(k) + (x'-k)\nabla P(k)) \nonumber \\
    &= \sqrt{\frac{D \Delta t}{\pi}}e^{-v(k)^2\Delta t /4D}\left(P(k) - \frac{\Delta t}{2} v(k)\nabla P(k)\right) + \frac{\Delta t}{2}\left(1+\textrm{Erf}\left[v(k)\sqrt{\frac{\Delta t}{4D}}\right]\right)\left(vP(k)-(D+ \frac{\Delta t}{2}v(k)^2)\nabla P(k)\right) \nonumber \\
    &= \sqrt{\frac{D \Delta t}{\pi}}P(k) + \frac{1}{2}(v(k)P(k)-D\nabla P(k))\Delta t+\mathcal{O}(\Delta t^{3/2})\\
    p_{\Delta t}(k^\leftarrow) &= \int^{\infty}_k dx' \int_{-\infty}^k dx \, P_{\Delta t}(x | x')P(x') \nonumber\\
    &= \sqrt{\frac{D \Delta t}{\pi}}P(k) - \frac{1}{2}(v(k)P(k)-D\nabla P(k))\Delta t+\mathcal{O}(\Delta t^{3/2})
\end{align}

Recognizing that $v(k)P(k)-D\nabla P(k)=j(k)$, the probability flux at $k$, we first make the identification $p_{\Delta t}(k^\rightarrow)-p_{\Delta t}(k^\leftarrow)=j\Delta t$. This is generally true for all times. Let's continue with the expansion of the $\Sigma_{\Delta t}^{trans}$ bound for lowest order in $\Delta t$;
\begin{align}
    \Sigma_{\Delta t}^{trans}/\Delta t&=\frac{k_B}{\Delta t}[ p_{\Delta t}(k^\rightarrow)-p_{\Delta t}(k^\leftarrow) ]\ln{\frac{p_{\Delta t}(k^\rightarrow)}{p_{\Delta t}(k^\leftarrow)}}\\ &\approx j(k) k_B\ln{\frac{1+\frac{j(k)}{2P(k)}\sqrt\frac{\pi\Delta t}{D}}{1-\frac{j(k)}{2P(k)}\sqrt\frac{\pi\Delta t}{D}}} = \frac{j(k)^2}{P(k)}k_B\sqrt{\frac{\pi \Delta t}{D}} + \mathcal{O}(\Delta t^{3/2}) \\
    &= \dot\sigma(k) \sqrt{\pi D\Delta t} + \mathcal{O}(\Delta t^{3/2})
\end{align}
where the density of dissipation rate $\dot\sigma(k)=k_B\,j(k)^2/DP(k)$.

\section{NESS case study: Driven particle in square potential on a ring}\label{supp: pSS_NESS}
\begin{figure}
    \centering
    \includegraphics[width=0.85\linewidth]{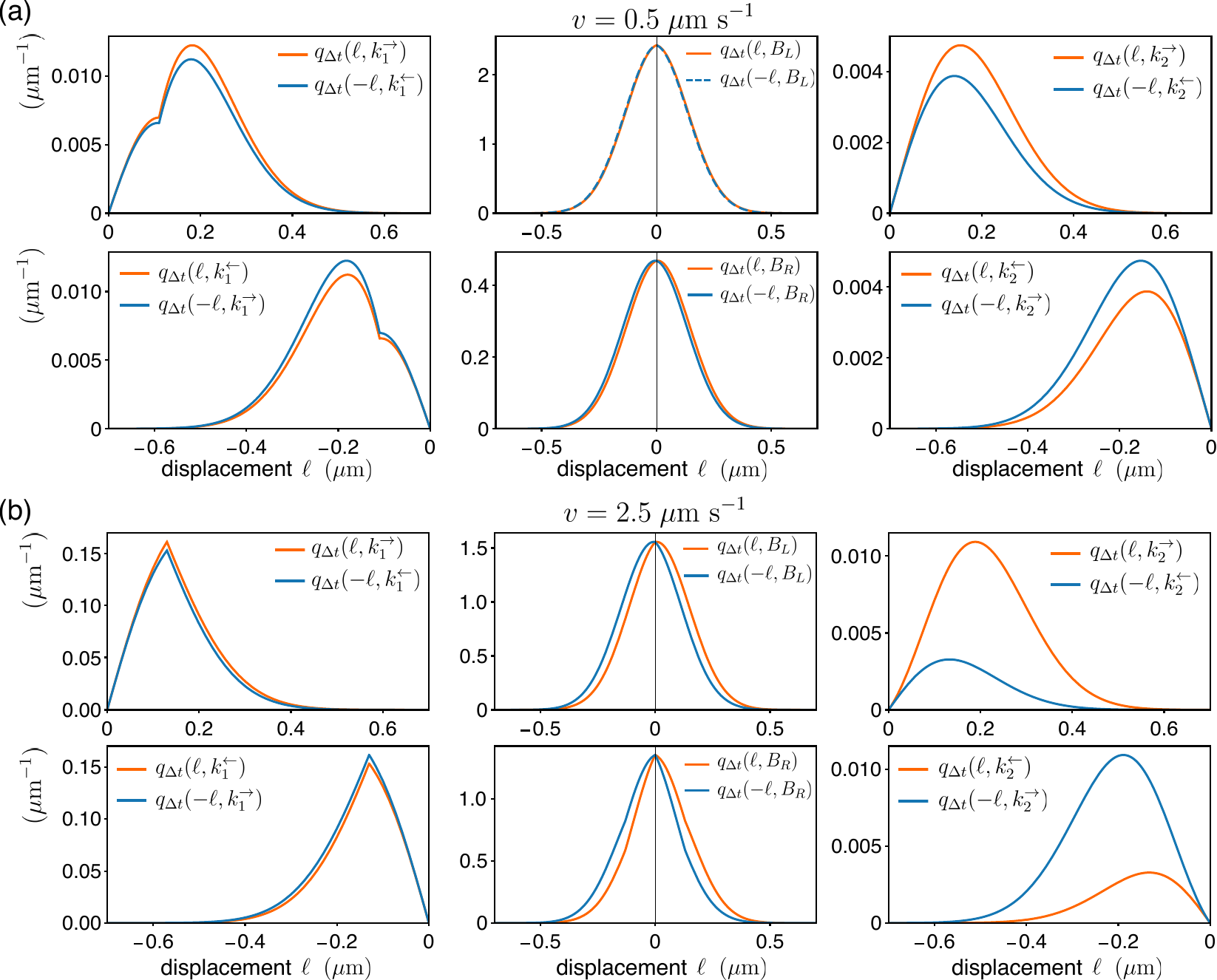}
    \caption{Displacement distributions $q_{\Delta t}(\ell,C)$ (orange) of the $\mathcal{C}_6(k_1,k_2)=\{k_1^\rightarrow,k_1^\leftarrow,B_L,B_R,k_2^\rightarrow,k_2^\leftarrow\}$ classification, for the near-equilibrium (a) and far-from-equilibrium (b) driving in the periodic square potential analyzed in Fig.~\ref{fig: fig2}. Their respective distributions under time-reversal $q_{\Delta t}(-\ell,\tilde C)$ are shown in blue. $k_1=0.11$ for $v=0.5~\si{\micro\meter\per\second}$ and $k_1=-0.13$ for $v=2.5~\si{\micro\meter\per\second}$, the placements offering the tightest bound $\Sigma_{\Delta t}(\mathcal{C}_6)$ to $\langle\Delta S_{tot}\rangle$.}
    \label{fig: 6displacements}
\end{figure}
The steady-state distribution for the NESS of a driven Langevin particle in a periodic square potential studied in Section~\ref{sec: squarepot} is given by
\begin{equation}
    P(x) = 
        \begin{cases}
           P^-(x)\equiv\dfrac{j}{v} \dfrac{1+e^{vL/D}+(\Gamma-1)e^{v(x+L)/D}}{1+e^{vL/D}}, & \text{for } -L \leq x < 0 \\[1em]
            P^+(x)\equiv\dfrac{j}{\Gamma v} \dfrac{\Gamma(1+e^{vL/D})-(\Gamma-1)e^{vx/D}}{1+e^{vL/D}}, & \text{for } 0 \leq x < L.
        \end{cases}
\end{equation}
with 
$
    \dfrac{P(0^-)}{P(0^+)} = \dfrac{P(-L)}{P(L)} = e^{\Delta W / k_B T} \equiv \Gamma, 
$ parameterizing the height $\Delta W$ of the barriers as in \cite{Bo21}
($\Gamma=8$ in Fig.~\ref{fig: fig2}). The probability flux is
\begin{equation}
    j 
    = \dfrac{\Gamma v^2 }{2 \Gamma L v + D(\Gamma-1)^2\tanh{\frac{vL}{2D}}}\,. \label{eq: J-cstF}
\end{equation}
When $j$ is substituted in $P(x)$, the $v$ and $D$ outside of the exponential arguments can be eliminated in such a way that the full distribution becomes a function of the length $D/v=l$. 
\begin{align}
    P^-(x)&=\dfrac{1+e^{L/l}+(\Gamma-1)e^{(x+L)/l}}{2L\Gamma (1+e^{L/l})+L(\Gamma-1)^2(e^{L/l}-1)} \\
    P^+(x)&=\dfrac{1}{\Gamma}\cdot\dfrac{\Gamma(1+e^{L/l})-(\Gamma-1)e^{x/l}}{2L\Gamma (1+e^{L/l})+l(\Gamma-1)^2(e^{L/l}-1)}
\end{align}
The total average entropy production rate in this NESS has an analytical form, found through integration of the density of entropy production rate $\dot\sigma (x)$ [Eq.~\eqref{eq: sigmadot}, visualized in Fig.~\ref{fig: fig2}(c)],
\begin{equation}
    \frac{\langle S_{tot} \rangle}{\Delta t} = k_B\frac{v^2}{D} \frac{2 \Gamma L v}{2 \Gamma L v + D(\Gamma -1)^2\tanh{\frac{v L}{2D}}}, \label{eq: eprNESS} 
\end{equation}
where we have factored out the usual $v^2/D$ term. Interestingly, we note that here $\langle S_{tot} \rangle/\Delta t=k_B\,j\,2L\,v/D$.

Now an example on the calculation of a displacement distribution $q(\ell,k^\rightarrow)$, where the displacement boundary is placed at $k=0$. Following Eq.~\eqref{eq: qCrossR},
\begin{align}
    q_{\Delta t}(\ell,0^\rightarrow) &= \int\limits_{-\infty}^{k=0} dx' \int\limits_{k=0}^{\infty} dx \, \delta(\ell-(x-x'))P_{\Delta t}(x | x')P(x') \\
    &= \Theta(\ell)\int\limits_{-\ell}^{0} dx' \, \,P_{\Delta t}(x'+\ell | x')P^-(x')
\end{align}
Here, $k$ is at the potential barrier, so we just have to use the propagator $P_{\Delta t}(x'+\ell | x')$ for driven diffusion across a step-potential as given in~\cite{Uhl21}. For $P_{\Delta t}(x | x')$ with $x>0$ and $x'<0$,
\begin{multline}
    P_{\Delta t}(x|x')= \dfrac{1-\alpha}{2\sqrt{\pi D t}}e^{-(x-x'-vt)^2/4Dt} + \dfrac{v}{4D}(\alpha-\alpha^2)\left[\text{Erf}\left(\frac{vt\alpha-x+x'}{2\sqrt{Dt}}\right)+1\right] e^{v(1-\alpha)(x-x')/2D-v^2t(1-\alpha^2)/4D} \label{eq: propdendil}
\end{multline}
with $\alpha=(\Gamma-1)/(\Gamma+1)$.

Integrating the above, the end result plotted in Fig.~\ref{fig: fig1}(b) is
\begin{align}
    q_{\Delta t}(\ell,0^\rightarrow)=\frac{j}{2 v^2 (1 + \Gamma) (1 + e^{\frac{v L}{D}})}  \cdot \left( -D e^{\frac{v (L - \ell)}{D}} (-1 + \Gamma) + v \ell + e^{\frac{v L}{D}} \left( D (-1 + \Gamma) + v \ell \right) \right) \nonumber \\ \cdot \left( \frac{2 e^{-\frac{(\ell - v \Delta t)^2}{4 D \Delta t}}}{\sqrt{\pi} \sqrt{D \Delta t}} + \frac{e^{\frac{v (\ell + \Gamma \ell - v \Gamma \Delta t)}{D (1 + \Gamma)^2}} v (-1 + \Gamma) \left( 1 + \erf\left( \frac{-\ell + \frac{v (-1 + \Gamma) \Delta t}{1 + \Gamma}}{2 \sqrt{D \Delta t}} \right) \right)}{D (1 + \Gamma)} \right)
\end{align}

Lastly, we present in Fig.~\ref{fig: 6displacements} plots for all the displacement distributions in the $\mathcal{C}_6(k_1,k_2)$ classification, sketched in Fig.~\ref{fig: fig2}(a), for the NESS system of a driven particle in a periodic square potential near ($v=0.5~\si{\micro\meter\per\second}$) and far ($v=2.5~\si{\micro\meter\per\second}$) for equilibrium. For the location of partitions delimiting the left $B_L$ and right $B_R$ bulk regions, $k_2=\pm L$ (the periodic boundary), whereas $k_1$ is optimized to give the best $\Sigma_{\Delta t}(\mathcal{C}_6)$ bound to $\langle \Delta S_{tot}\rangle$, given by the maxima of the solid red and blue curves in Fig.~\ref{fig: fig2}(e).
\\
It is remarkable that for $v=0.5~\si{\micro\meter\per\second}$ there is such an improvement in entropy inference over the  $\mathcal{C}_3(k_1)$ classification (see again Fig.~\ref{fig: fig2}(d) and~\ref{fig: fig2}(e)), owing to the separation of one bulk term into two, despite the difference between their distributions under time reversal being visibly almost indistinguishable. On the other hand, for $v=2.5~\si{\micro\meter\per\second}$ we perceive a much greater difference under time reversal of the displacement distributions. This makes sense, as the entropy production rate of these NESS, per Eq.~\eqref{eq: eprNESS}, are $0.12\,k_B\,\si{\per\second}$ for $v=0.5~\si{\micro\meter\per\second}$ and $5.02\,k_B\,\si{\per\second}$ for $v=2.5~\si{\micro\meter\per\second}$.

\section{Empirical estimation of Kullback-Leibler Divergences}\label{supp: NN}
Calculating KL-divergences naively from empirical histograms incurs systematic biases~\cite{grandpre24}, difficult to correct for with limited sample sizes, and suffers from a strong dependence on binning size. 
Therefore, we choose to calculate the $D_{KL}$'s of displacement distributions using a nearest-neighbors (NN) technique, which avoids binning~\cite{Wang09}. This must be adapted to the estimation of the mixed discrete and continuous $\Sigma_{\Delta t}(\mathcal{C}_n)$, outlined here.

We exemplify the procedure on one integral term for the displacement class $C\in\mathcal{C}_n$ in the summation of Eq.~\eqref{eq: dklndef}, which can be written in terms of the $D_{KL}$ of the conditional distributions (with respect to just the continuous variable $\ell$):
\begin{align}
    \int d\ell \, q_{\Delta t}(\ell, C) \ln{\frac{q_{\Delta t}(\ell, C)}{q_{\Delta t}(-\ell, \tilde{C})}} = p_{\Delta t}(C)\left[D_{KL}\left(q_{\Delta t}(\ell| C)\big|\big|q_{\Delta t}(-\ell| \tilde{C})\right) + \ln{\frac{p_{\Delta t}(C)}{p_{\Delta t}(\tilde{C})}} \right], \label{eq: integralTerm}
\end{align}
where the conditional probability density $q_{\Delta t}(\ell| C)\equiv q_{\Delta t}(\ell, C)/p_{\Delta t}(C)$.

For a total number of tracked particles $N$, we consider the empirical probabilities of displacement classes $C$ and $\tilde{C}$, $\hat p_{\Delta t}(C) \equiv N_{C}/N$ and $\hat p_{\Delta t}(\tilde{C}) \equiv N_{\tilde{C}}/N$, such that $\langle\hat p_{\Delta t}(C)\rangle=p_{\Delta t}(C)$,$\langle\hat p_{\Delta t}(\tilde{C})\rangle=p_{\Delta t}(\tilde{C})$.
Crucially, this method does not require estimating the empirical densities $\hat q_{\Delta t}(\ell | C)$ and $\hat q_{\Delta t}(\ell | \tilde{C})$, but estimates their $D_{KL}$ based on nearest neighbors assignments across the set of displacement measurements belonging to class $C$ and $\tilde{C}$.
For each element $\ell_j$ of the empirical set of displacement measurements belonging to class $C$, $\{\ell_j\}_{j=1}^{N_C}$,  we identify the element
that has the closest value in the set, $\ell_k$, (\textit{i.e.}, its nearest neighbor, the element with the shortest distance $\rho(j)=|\ell_j-\ell_k|$).
We then turn to the
the set of reversed displacements of class $\tilde{C}$, $\{-\tilde{\ell}_j\}_{j=1}^{N_{\tilde{C}}}$, and identify the nearest neighbor there, $\tilde{\ell}_{k'}$, which has the shortest distance from $\ell_j$, $\nu(j)=|\ell_j-\tilde{\ell}_{k'}|$.
The empirical $D_{KL}$ is then obtained as
\begin{align}
    \hat D_{KL}\left(q_{\Delta t}(\ell| C)\big|\big|q_{\Delta t}(-\ell| \tilde{C})\right) = \frac{1}{N_{C}} \sum_{j=1}^{N_{C}}\left[\log{\frac{\nu(j)}{\rho(j)}}\right]+\ln{\frac{N_{\tilde{C}}}{N_{C}-1}} \, .
\end{align}
Substituting the above into Eq.~\eqref{eq: integralTerm}, and the marginal probability estimates, we have our estimator
\begin{align}
    \hat\Sigma_{\Delta t}(\mathcal{C}_n) \equiv \sum_{C\in\mathcal{C}_n}^n \frac{N_{C}}{N}\left[\frac{1}{N_{C}} \sum_{j=1}^{N_{C}}\left[\log{\frac{\nu(j)}{\rho(j)}}\right] +\ln{\frac{N_{C}}{N_{C}-1}} \right].
\end{align}

\begin{figure}
    \includegraphics[width=0.65\textwidth]{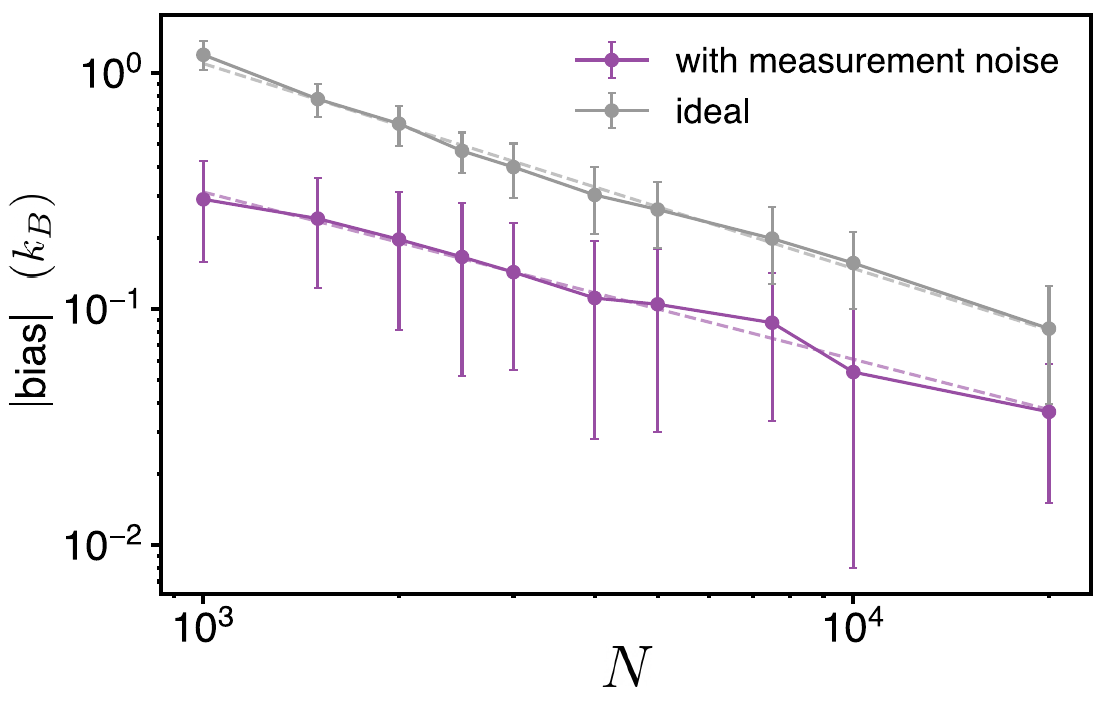}
    \caption{Mean bias of total irreversibility $\Delta\Sigma_{t_\textrm{f}}$ of $10~\si{\pico\newton\per\second}$ force ramps in the high symmetric potential (as in Fig.~\ref{fig: fig3}) using the $\hat\Sigma_{\Delta t}(\mathcal{C}_4)$ nearest-neighbours estimator on $N$ trajectories. Data taken from subsets of $N=10^5$, with the highest $N$ considered in the plot of 20000 (mean of 5 subsets). The error bars, asymmetric due to the log scale, are the standard deviation of the total irreversibility of subsets. Note our definition of the bias Eq.~\eqref{eq: bias} takes an absolute value, with this estimator typically underestimating the true $\Sigma_{\Delta t}(\mathcal{C}_4)$.
    }
    \label{fig: bias}
\end{figure}

In practice, to reduce the variance of the estimator, it is often convenient to use instead of the nearest neighbor the z-th nearest neighbor~\cite{PerezCruz08KDE, Wang09}.
However, this must be carefully balanced, as using too large $z$ induces biases if there are not sufficient samples.
In the analysis of the non-equilibrium protein extension trajectories in the main text, we choose the number $z_C$ based on the amount of data available in each summation term evaluation above, and choose $z_C=\min{(N_{C}^{1/3},N_{\tilde{C}}^{1/3},10)}$.
Our estimator is then 
\begin{align}
    \hat\Sigma_{\Delta t}(\mathcal{C}_n) \equiv \sum_{C\in\mathcal{C}_n}^n \frac{N_{C}}{N}\left[\frac{1}{N_{C}} \sum_{j=1}^{N_{C}}\left[\log{\frac{\nu_{z_C}(j)}{\rho_{z_C}(j)}}\right] +\ln{\frac{N_{C}}{N_{C}-1}} \right]\,,
\end{align}
where $\rho_{z_C}(j)$ is the distance between the $j$-th element of class $C$, $\ell_j$ and its $z_C$-th nearest neighbor, and $\nu_{z_C}(j)$ is defined analogously.

We treat differently the cases in which we have too few displacements belonging to a class. When there are fewer than 3 displacements belonging to either class, instead of computing
$\hat\Sigma_{\Delta t}(\mathcal{C}_4)$, we compute $\hat\Sigma_{\Delta t}^{trans}=\sum_C \hat p_{\Delta t}(C)\ln\frac{\hat p_{\Delta t}(C)}{\hat p_{\Delta t}(\tilde C)}$ if $N_C>0$ and $N_{\tilde C}>0$, otherwise, $\hat\Sigma_{\Delta t}^{trans}=0$

An example implementation is provided on our Github~\cite{github}.

In Fig.~\ref{fig: bias}, we exemplify the behavior of this bias of the estimator as a function of the total number of trajectories $N$ used in the $\hat\Sigma_{\Delta t}(\mathcal{C}_4)$ estimator, for the total measured irreversibility $\Delta\Sigma_{t_\textrm{f}}$ in the $10~\si{\pico\newton\per\second}$ ramp on the high barrier symmetric potential (ie. the final value of the gray curves in the lower inset of Fig.~\ref{fig: fig3}(c)). With a total of $N=10^5$ simulated trajectories, we use $\Delta\Sigma_{t_\textrm{f}}(\mathcal{C}_4,N=10^5)$ as a reference for the bias of its data's $10^5/N$ subsets, so
\begin{align}
    \textrm{bias}(N) = \Delta\Sigma_{t_\textrm{f}}(\mathcal{C}_4,N)-\Delta\Sigma_{t_\textrm{f}}(\mathcal{C}_4,10^5) \label{eq: bias}
\end{align}
In Fig.~\ref{fig: bias} it is plotted, as in Fig.~\ref{fig: fig3}, with added measurement noise to the positions (experimental scenario) and without (ideal scenario). Their respective scalings, taken from the linear fit (dashed lines) on the log-log plot are $N^{-0.71}$ and $N^{-0.87}$. The bias is, in fact, negative so the absolute value is taken for the plot.

\subsection{Finite statistics and noise on the temporal estimate of irreversibility}\label{app: noise}
\begin{figure}
\includegraphics[width=0.8\textwidth]{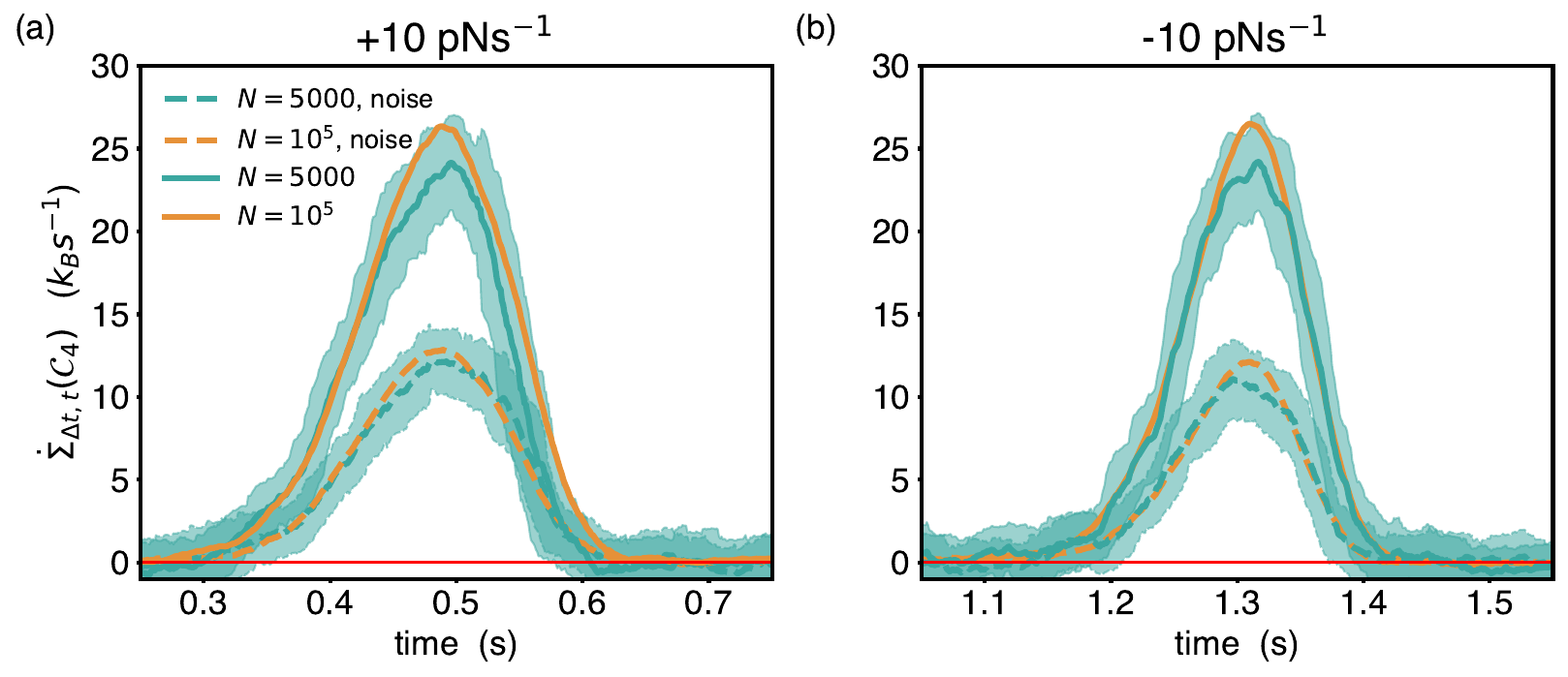}
\caption{\label{fig: width} For the $+10~\si{\pico\newton\per\second}$ unfolding (a) and $-10~\si{\pico\newton\per\second}$ refolding (b) irreversibility peaks of the simulated ramps on the Mutant-like potential [Fig.~\ref{fig: fig4}(f)], we compare the inference for high (orange) or low (blue) amounts of data, and adding (dashed) or without (solid) measurement noise.}
\end{figure}
In Fig.~\ref{fig: fig3} for the symmetric and Fig.~\ref{fig: SI_asym} for the asymmetric potentials, we showed that an estimate of temporal irreversibility at low sample size with measurement noise retains the same features as the ideal estimate (many samples, no added noise). Here in Fig.~\ref{fig: width}, we show that it is the instrumental noise which has the greatest impact on the height of the peaks relative to the best ($N=10^5$ solid orange) estimate. Reducing the number of samples by a factor of 20 slightly decreases the measured irreversibility, but follows the trend of the curve with many samples. Again, this is preserved even for the case of added noise, which still captures the difference in peak widths between unfolding Fig.~\ref{fig: width}(a) and refolding Fig.~\ref{fig: width}(b).


\section{Milestoning estimator of dissipation}\label{app: milestoning}
Milestoning has emerged as a powerful coarse-graining tool in stochastic thermodynamics 
\cite{hartich_emergent_2021, hartich_emergent_2021}. 
In this Section, we compare the milestoning approach of Ref.~\cite{Blom24} to our estimator $\Sigma_{\Delta t}(\mathcal{C}_6)$ for the driven particle in a square potential on a ring discussed in Section~\ref{sec: squarepot} in an empirical setting, with simulated trajectories. We apply the estimators on the raw Markovian trajectory data, 
measured at a coarse frame rate of $\Delta t=10~\si{\milli\second}$ as in Fig.~\ref{fig: fig2}. 
\begin{figure}[h]
    \includegraphics[width=0.95\textwidth]{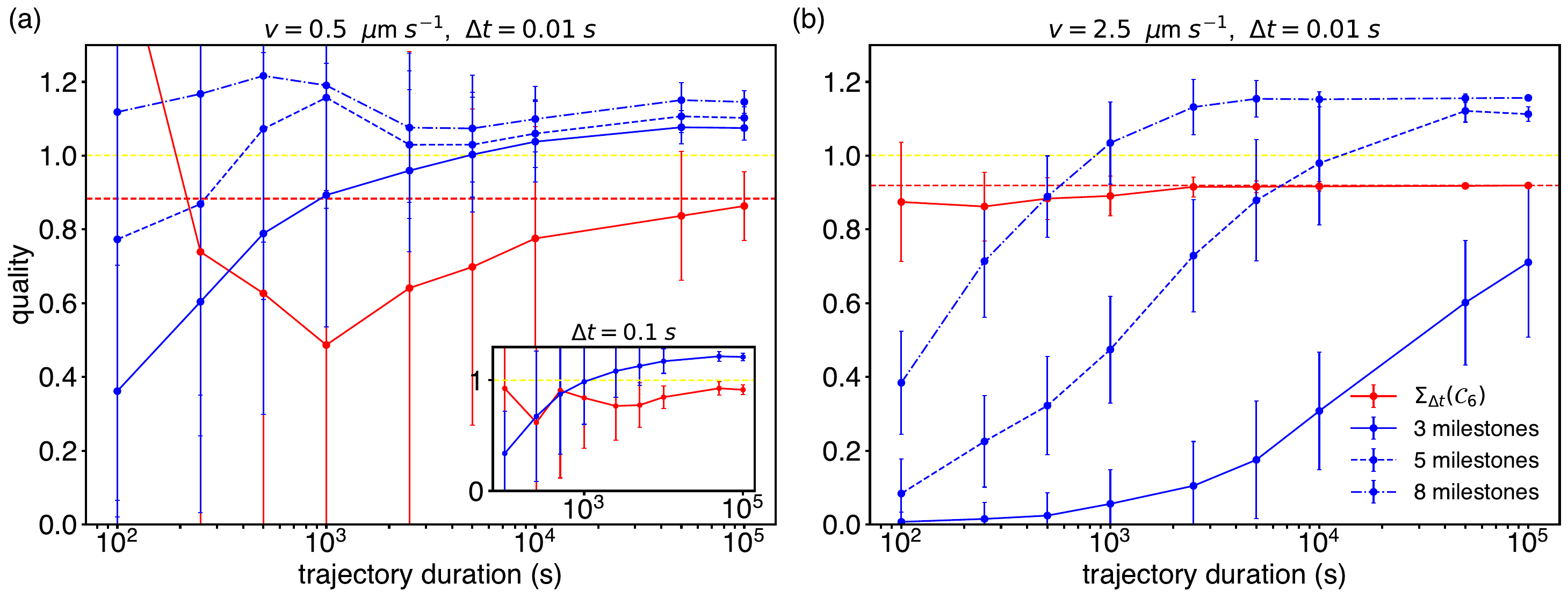}
    \caption{Estimates of entropy production for the square potential on a ring near (a) and far (b) from equilibrium, plotted as a fraction of the true total entropy production rate, their quality. Red points correspond to our displacement-class estimator $\Sigma_{\Delta t}(\mathcal{C}_6)/\langle\Delta S_{tot}\rangle$ using the nearest-neighbors method, with the ground truth given by the horizontal red dashed line. Blue curves correspond to the milestoning estimator $\dot S_{ms}/\langle \dot S_{tot}\rangle$ for different numbers of equally spaced milestones on the ring. Error bars are the standard deviation of 30 estimates of independent trajectories, sampled every $\Delta t=10~\si{\milli\second}$. Inset of (a) is for a sampling window of $\Delta t=100~\si{\milli\second}$, showing better convergence for $\Sigma_{\Delta t}(\mathcal{C}_6)$. Simulation parameters as in Fig.~\ref{fig: fig2}, using the optimal $k_1$ boundaries for $\Sigma_{\Delta t}(\mathcal{C}_6)$ found in Fig.~\ref{fig: fig2}(e).
    }
    \label{fig: milestoning}
\end{figure}
Milestones are placed as virtual boundaries along the ring, which we space equally. When the observed trajectory crosses a milestone $j$, from any direction, if the current state is a different milestone $i$, the present milestoned state of the trajectory switches. Normalizing over all recorded milestone transitions one obtains empirically the jump probabilities $p_{ij}$. For these first-order transition statistics we apply the estimator~\cite{martinez2019inferring, Blom24}
\begin{equation}
    \dot S_{ms} \equiv \frac{1}{\langle\tau\rangle} \sum_{i, j} p_{ij}\ln{\frac{p_{ij}}{p_{ji}}} = \langle \dot S_{tot}\rangle, \label{eq: milestone}
\end{equation}
with $\langle\tau\rangle$ the average time spent waiting in a state for a transition, and where the second equality is guaranteed by the absence of hidden cycles in this case study~\cite{Harunari22, van2022thermodynamic, hartich_emergent_2021}. If a transition is not observed due to insufficient statistics, and either $p_{ij}$ or $p_{ji}$ is measured to be zero,  the corresponding term in the summation above is ignored. Note the $\Delta t$ we use is not so high that the recorded trajectory will cross two milestones in a frame, so all $p_{ij}$ correspond to transitions between adjacent milestones.

Long trajectories are required to obtain significant statistics from a milestoning, which requires waiting to observe sufficient state changes in the clockwise and anticlockwise directions. This can be partially mitigated by including more milestones, at the cost of overlap with the $\Delta t$ timescale. In this comparison, we perform each ($\Sigma_{\Delta t}(\mathcal{C}_6)$ and $\dot S_{ms}$) inference on a single trajectory, studied at different durations. 30 different trajectories are analyzed and their mean estimate divided by the total entropy production--their quality--is plotted, for the near and far from equilibrium cases in Fig.~\ref{fig: milestoning}.

\begin{figure}
    \includegraphics[width=0.9\textwidth]{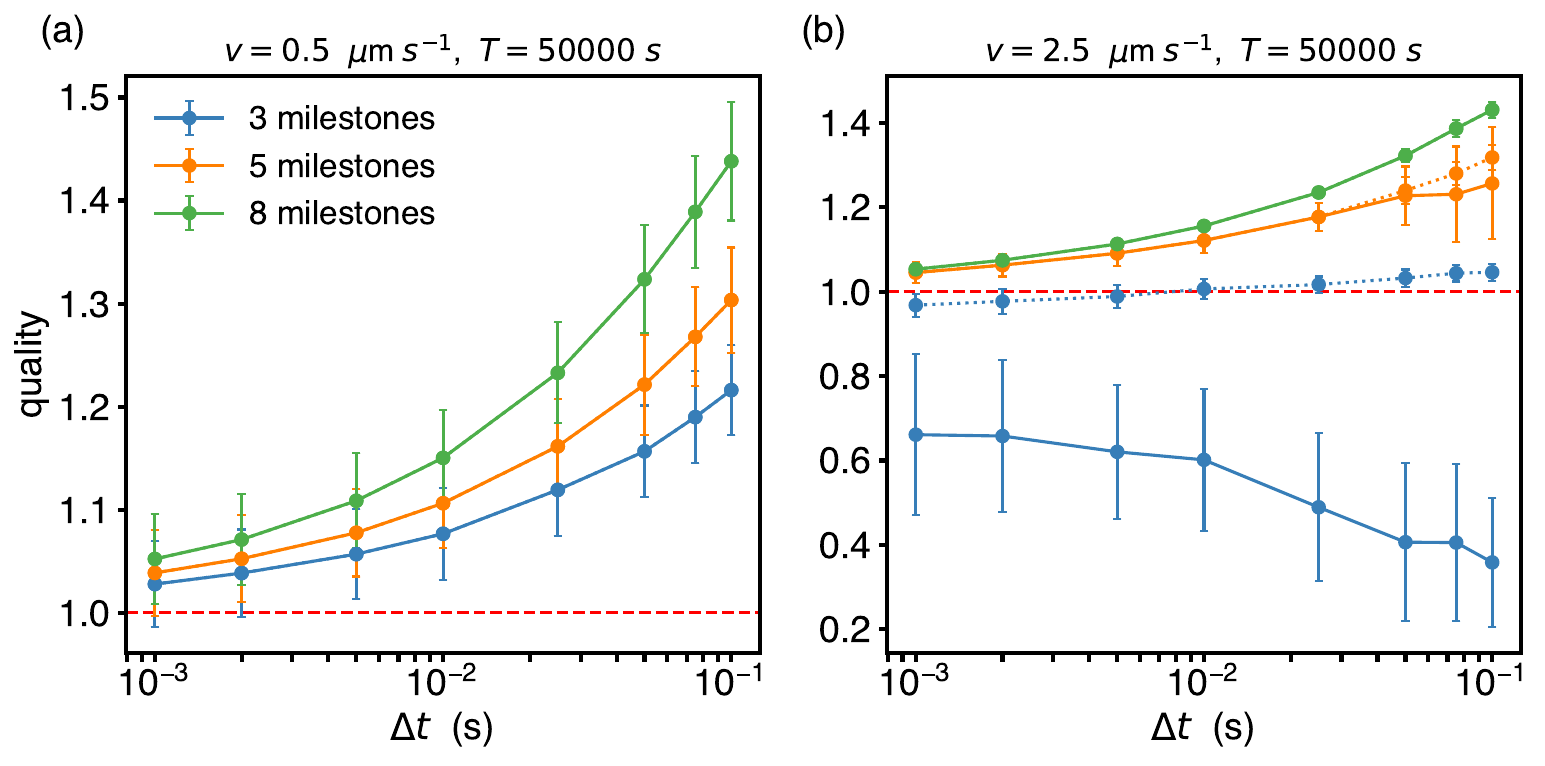}
    \caption{Quality of the milestoning estimator $\dot S_{ms}/\langle \dot S_{tot}\rangle$ as a function of the sampling resolution $\Delta t$. Estimates are of the square potential on a ring near (a) and far (b) from equilibrium, using the mean and standard deviation of 30 independent trajectories, as in Fig.~\ref{fig: milestoning}, for a fixed trajectory duration $T$ of $5\times10^{4}~\si{\second}$. The solid curves are obtained using our default treatment of null reverse jump measurements (no entropy), while the dotted curves substitute $p_{ji}=0$ for $p_{ji}=1/N$ in Eq.~\ref{eq: milestone}, for $N$ total jumps. The difference between the curves is visible only in panel (b).}
    \label{fig: milestoning_dt}
\end{figure}

Both estimators are very noisy in the close-to-equilibrium $v=0.5$ case. In general, it is difficult in these regimes to obtain a statistically satisfying measurement of a deviation from equilibrium at short times, when the dynamics appear so reversible. As a reference, $\langle \Delta S_{tot}\rangle=0.0012\,k_B$ for $\Delta t=10~\si{\milli\second}$ here, $\sim42$ times lower than for $v=2.5$. 
Nonetheless, $\Sigma_{\Delta t}(\mathcal{C}_6)$ approaches the theoretical value for long trajectories, while the milestone estimators plateau higher than $\langle\dot S_{tot}\rangle$. This systematic overshoot suggests the true sequence of milestone states is not being recorded, which we believe is due to the finite $\Delta t$. That is, while transitions following the direction of the deterministic drift will almost always be registered, one may well miss transitions that sojourn into the milestone in the other direction during the limited resolution time window. As a result, $p_{ij}$ with the flux direction is overestimated, and opposite jumps are underestimated, resulting in an overall positive bias.
Considering trajectories sampled every $\Delta t=100~\si{\milli\second}$ (same data but subsampled by a factor 10) supports this reasoning, as we see an increase in the overestimation shown in the inset of Fig.~\ref{fig: milestoning}(a). $\Sigma_{\Delta t}(\mathcal{C}_6)$ is, instead, converging faster, fluctuating less, and approaching a tighter bound.\\ For $v=2.5$, $\Sigma_{\Delta t}(\mathcal{C}_6)$ converges relatively quickly and with less uncertainty to its ground truth, a lower bound to $\langle\Delta S_{tot}\rangle$. Irreversibility is higher here and $\dot S_{ms}$ requires longer trajectories, as the shorter ones severely underestimate $\langle\dot S_{tot}\rangle$ due to missing observations of jumps against the drift direction. This was also previously reported in the case of the waiting-time distribution estimator for networks~\cite{Fritz_2025}.

To verify that the systematic bias in the milestoning estimator is from finite $\Delta t$ sampling, we measure the quality $\dot S_{ms}/\langle \dot S_{tot}\rangle$ for a fixed trajectory duration ($5\times10^{4}~\si{\second}$) as a function of $\Delta t$, given by the solid lines in Fig.~\ref{fig: milestoning_dt}. Indeed, we see that for any number of milestones in both steady states, the inference is much improved upon lowering $\Delta t$. For this system, this is the opposite trend compared to our displacement-class method, which benefits from larger $\Delta t$. We note as well that the quality curves do not seem to approach 1 exactly as $\Delta t\rightarrow0$. This is likely due to an inherent bias in estimators using log-ratios of empirical probabilities at finite sample sizes, which can be corrected if one knows the ground truth jump probabilities~\cite{grandpre24}. 

Alternatively, one could deal with cases where $p_{ji}=0$ by imposing $p_{ji}=\epsilon$, where $\epsilon$ is a conservative estimate of the reverse jump probability. This is shown by the dotted curves in Fig.~\ref{fig: milestoning_dt} for $\epsilon=1/N$, with $N$ the total number of jumps. This correction has no effect in the near-equilibrium case where reverse jumps are always measured, but far from equilibrium it provides a noticeable correction for low numbers of milestones, which improves the 3-milestones case.

\section{Hierarchy of irreversibility and (un)folding times on the symmetric double well}\label{sec: symmC3}
\begin{figure}
    \includegraphics[width=0.8\textwidth]{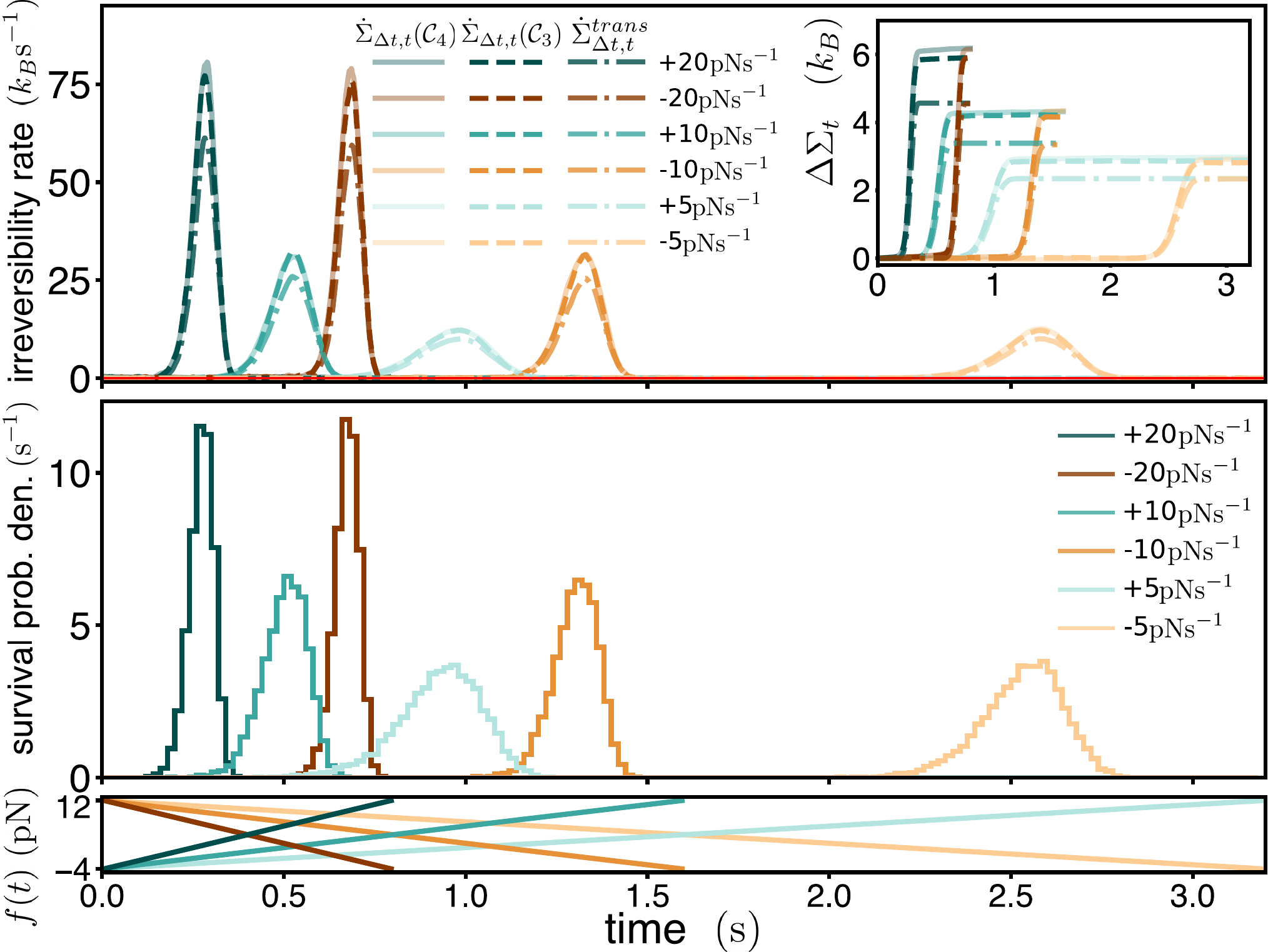}
    \caption{\textit{Top}: using 50000 trajectories simulated on the high barrier symmetric double-well potential of Fig.~\ref{fig: fig3}(a), irreversibility rate estimates for different classifications of displacements $\mathcal{C}_4$ and $\mathcal{C}_3$, and the displacement length-agnostic estimator $\Sigma_{\Delta t}^{trans}$. Inset shows the cumulative time integral of the rates. \textit{Middle}: respective survival (well-crossing) probability densities in each ramp protocol. \textit{Bottom}: force-time ramp protocols.
    }
    \label{fig: sigma3}
\end{figure}

In this Section, we compare the amount of irreversibility captured by partitioning space into four classes:  $\mathcal{C}_4=\{k^\rightarrow,k^\leftarrow,B_F,B_U\}$, as in Fig.~\ref{fig: fig3}(a), to $\mathcal{C}_3$ where $B_F$ and $B_U$ are merged, and to $\Sigma^{trans}$.
We focus on the high barrier symmetric double-well potential, without added measurement noise, shown in the top panel of Fig.~\ref{fig: sigma3}. For all measurements, we place the boundary $k$ in the same location, such that it maximizes the total $\Delta\Sigma_{t_\textrm{f}}^{trans}$ as demonstrated in~\ref{sec: movek}. The $\Sigma(\mathcal{C}_4)$ curves are the same as the ideal case in Fig.~\ref{fig: fig3}(c). We see that $\Delta\Sigma_{t}(\mathcal{C}_4)\geq\Delta\Sigma_{t}(\mathcal{C}_3)\geq\Delta\Sigma_{t}^{trans}$ as expected, so we choose $\mathcal{C}_4$.

In the middle panel of Fig.~\ref{fig: sigma3}, we show the survival probability densities in time for each force ramp. These are measured as the first instance when the particle crosses a point $k+2$ or $k-2$ for the positive (unfolding) ramps and the negative (refolding) ramps, respectively. There is a direct relation between peaks in the irreversibility rate and these events of hopping between the wells. Note that, although not present here, bulk (non-crossing events) displacements make significant contributions to the irreversibility in the experiments, especially the wild-type, and are not captured by these distributions. 

\section{Talin R3's folding dynamics are well-described by a one-dimensional diffusion}\label{app: committor}
It is pertinent to ask, given the staggering complexity of protein structure, about the extent to which we are able to measure a relevant proportion of the total dissipation in the experimental scenario. That is, if it is likely for there to be additional degrees of freedom, apart from the measured end-to-end molecular extension, that are significantly dissipative throughout the (un)folding force protocols of the Talin R3 protein domains. This may arise, for instance, from additional dimensionality in the reaction coordinate energy landscape and non-Markovianity. To this end, a test of the reaction coordinate's suitability as a one-dimensional diffusive process has been developed and experimentally demonstrated in~\cite{neupane_transition-path_2015, neupane_protein_2016}, which we apply in this appendix to long-time constant force trajectories of the R3$^{{\rm{WT}}}$ and R3$^{{\rm{IVVI}}}$ proteins.

\begin{figure}
    \centering
    \includegraphics[width=0.97\linewidth]{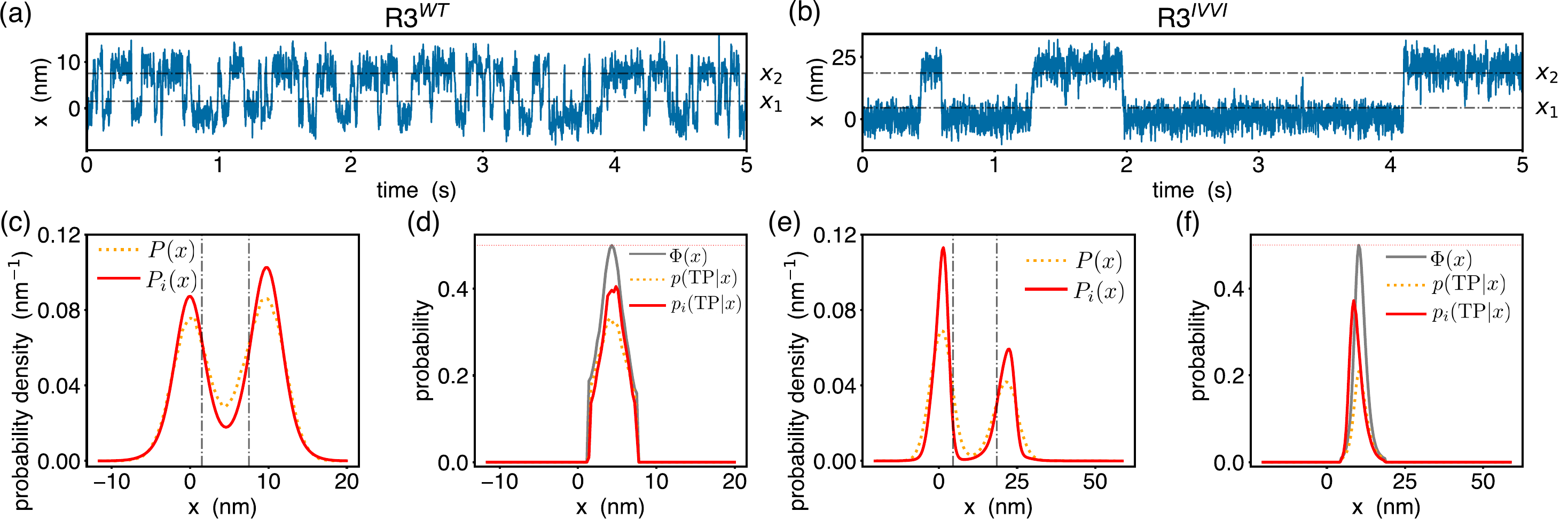}
    \caption{Snippet of constant-force trajectories of the R3$^{{\rm{WT}}}$ (a) and R3$^{{\rm{IVVI}}}$ (b) proteins' extension held at 5.7 and 8.5 pN, respectively, near coexistence. Total duration of the experiments was 1238~\si{\second} and 22591~\si{\second} for R3$^{{\rm{WT}}}$ and R3$^{{\rm{IVVI}}}$, respectively. (c) and (e): equilibrium extension distribution (dotted) before and intrinsic distribution (solid) after deconvolution. (d) and (f): committor function (grey) and transition path probability given some extension before (dotted) and after (red) deconvolution; horizontal dotted line at the expected peak of 0.5. The transition region is delimited by $x_1=1.5(4.5)$nm and $x_2=7.5(18.5)$nm for R3$^{{\rm{WT}}}$(R3$^{{\rm{IVVI}}}$), as shown by the dash-dot lines.}
    \label{fig: committor}
\end{figure}

We identify a transition (also termed reaction) region $x_1 <x<x_2$ between the equilibrium probability $P(x)$ maxima of folded and unfolded states, as demonstrated in the sample traces of Fig~\ref{fig: committor}(a) and (b) and the distributions in (c) and (e). Our results, as in~\cite{neupane_transition-path_2015, neupane_protein_2016}, are agnostic to the precise location of $x_1$ and $x_2$. There are two parts to the test; the first, based on committor analysis. The committor $\phi_2(x)$ gives the probability that a particle starting at an $x$ within the transition region will exit through $x_2$ instead of $x_1$, and vice versa for $\phi_1(x)$ such that $\phi_1(x)=1-\phi_2(x)$. For a 1D diffusion, one should find the function $\Phi(x)=2\phi_2(x)[1-\phi_2(x)]$ to be peaked at a value of $0.5$ near the top of the barrier, as shown by the grey curves in Figs.~\ref{fig: committor}(d) and (f). Note that due to the fast kinetics and narrow transition region of the R3$^{{\rm{WT}}}$, some transitions occur within a single frame of measurement, such that $\phi_2(x)$ is not between 0 and 1, but $\phi_2(x_1)\sim0.1$ and $\phi_2(x_2)\sim0.9$. Nonetheless, $\Phi(x)$ has the expected maximum of $0.5$ at the trough of $P(x)$ -- the top of the energy barrier.

The second part of the test relies on transition path (TP) analysis. TPs are segments of the long constant-force trace that enter through $x_1(x_2)$ and exit through $x_2(x_1)$, which we stitch together. The curve of interest is the conditional probability to be on a transition path given $x$, calculated as
\begin{equation}
    p(\text{TP} | x)=\frac{P(x | \text{TP})p(\text{TP})}{P(x)},
\end{equation}
where $p(\text{TP})$ is the fraction of time spent on TPs. For ideal 1D diffusion, $p(\text{TP} | x)=\Phi(x)$, though it generally suffices to identify a peak at $p(\text{TP} | x)\approx 0.5$. Due to the instrumental noise of the magnetic tweezers, $P(x)$ in the transition region is overestimated, and the $p(\text{TP} | x)$ curves, dotted in Fig.~\ref{fig: committor}(d) and (f), peak at $\sim0.3$ and $0.2$ for R3$^{{\rm{WT}}}$ and R3$^{{\rm{IVVI}}}$, respectively. As in~\cite{neupane_transition-path_2015, neupane_protein_2016}, we remedy this by deconvolving the equilibrium distribution using the method of~\cite{Woodside06}, to obtain the intrinsic extension distribution $P_i(x)$ (red curves Fig.~\ref{fig: committor}(c) and (e)). This procedure is sensitive to parameters such as the number of iterations and the assumed magnitude of measurement noise. Fixing the smoothing noise to a Gaussian with standard deviation of 3~\si{\nano\meter}, as found in Fig.~\ref{fig: expNoise}, we iterate until the peak-to-well height of $P_i(x)$ is consistent with the expected energy barrier obtained from the kinetics in~\ref{supp: expAsymmetry}. It is reassuring that the deconvolution on the R3$^{{\rm{IVVI}}}$ acts to highlight the asymmetry between the folded and unfolded sides of the distribution, with the transition state shifting to the left. Finally, the new probabilities are calculated, after renormalizing for the relative weight of the transition region,
\begin{align}
    p_i(\text{TP} | x)=\frac{P(x | \text{TP})p(\text{TP})}{P_i(x)}\cdot \frac{\int_{x_1}^{x_2}P_i(x)dx}{\int_{x_1}^{x_2}P(x)dx}
\end{align}
These corrected $p_i(\text{TP} | x)$ are shown by the solid red curves in Fig.~\ref{fig: committor}(d) and (f), which are seen to peak around 0.4, an improvement consistent with an effective diffusion in one dimension. We remark that it is not difficult to obtain peaks exactly at 0.5 by slightly tweaking the deconvolution parameters, while still producing sensible $P_i(x)$. In any case, the raw $p(\text{TP} | x)$ measurements peaking around half this value are also well above the unequivocally high-dimensionality cases considered in~\cite{neupane_transition-path_2015, neupane_protein_2016}. We conclude that the projected Talin R3 extension dynamics are consistent with an overdamped diffusion in one dimension, hence the irreversibility measurements are physically meaningful.

\section{Experimental asymmetric rate dependence on force and simulation parameters choice}\label{supp: expAsymmetry}
Here we discuss how equilibrium properties of the wild-type R3$^{\rm{WT}}$ and the mutant R3$^{\rm{IVVI}}$ domains reveal features of their underlying energy landscape, which inform the parameters we can tune to recover their dissipative characteristics in simulation.
We performed equilibrium measurements at different forces to measure how the transition rates between the folded and unfolded states vary with force for the R3$^{\rm{WT}}$ and for the R3$^{\rm{IVVI}}$. 
 In a Bell-Evans framework the slope of the curve is proportional to the distance between the potential minimum and the barrier~\cite{evans1997dynamic} [$x^\dagger_{\rm{u}}$ ($x^\dagger_{\rm{f}}$) from the unfolded (folded) minimum with intrinsic unfolding (folding) rate $k^0_{\rm u}$ ($k^0_{\rm f}$)]:
 \begin{align}\label{eq: bell}
     k_{\rm{f}}(f)=k^0_{\rm{f}}\exp{\left[-\frac{fx^\dagger_{\rm{f}}}{k_BT}\right]}\qquad\qquad
     k_{\rm{u}}(f)=k^0_{\rm{u}}\exp{\left[\frac{fx^\dagger_{\rm{u}}}{k_BT}\right]}
     \,.
 \end{align}

\begin{figure}
\includegraphics[width=0.95\textwidth]{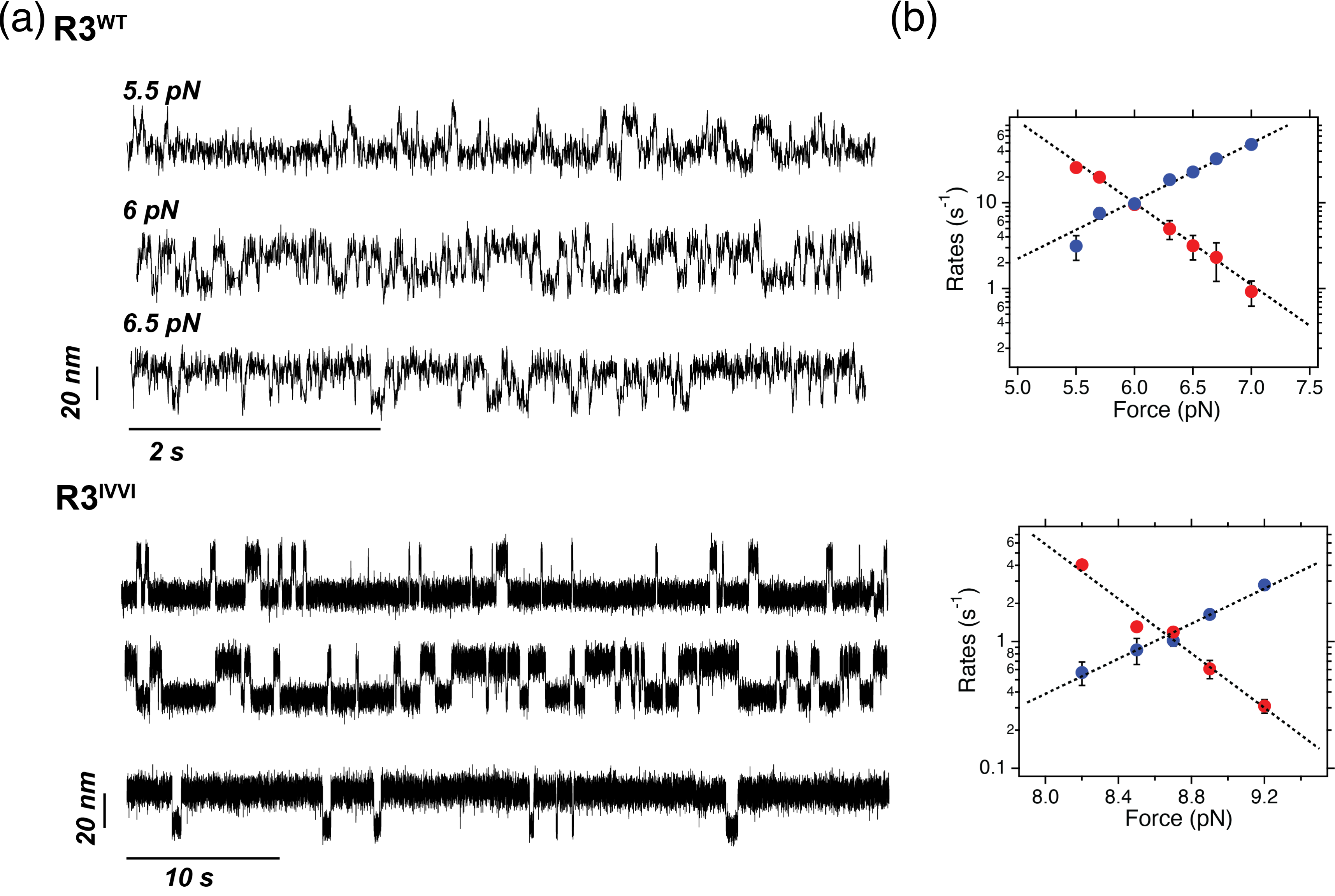}
\caption{\label{fig: constant_force} Equilibrium measurements of talin R3$^{\rm{WT}/\rm{IVVI}}$ folding dynamics as a function of force.
(a) Example trajectories for  R3$^{\rm{WT}}$ (top) and R3$^{\rm{IVVI}}$  (bottom) at different  forces.
(b) Transition rates for unfolding ($k_{\rm{u}}$, blue) and folding ($k_{\rm{f}}$, red) as a function of force $f$. Data is fitted to the Bell-Evans model~\cite{evans1997dynamic} described by Eq.~\eqref{eq: bell}.}
\end{figure}

 For a symmetric potential landscape, the slopes in the log-plot of transition rate vs force should be equal in magnitude and opposite in sign for the unfolding and refolding transitions.
Figure~\ref{fig: constant_force}(b) shows that the slope of the refolding transitions is steeper. Fitting the data, we find for R3$^{\rm{WT}}$ folding: $k^0_{\rm{f}}=(2.37\pm0.25)\times 10^9$ s$^{-1}$, $x^\dagger_{\rm{f}}=10.2\pm1.01$ nm and for unfolding: $k^0_{\rm u}=(1.11\pm0.09)\times 10^{-6}$ s$^{-1}$, $x^\dagger_{\rm{u}}=6.56\pm0.55$ nm. For R3$^{\rm{IVVI}}$, we find, for folding: $k^0_{\rm{f}}=(5.39\pm0.05)\times 10^6$ s$^{-1}$, $x^\dagger_{\rm{f}}=9.04\pm0.31$ nm; and for unfolding: $k^0_{\rm u}=(9.61\pm0.15)\times 10^{-4}$ s$^{-1}$, $x^\dagger_{\rm{u}}=6.37\pm0.29$ nm.
This suggests that the potential minimum corresponding to the unfolded state is further from the barrier than the one of the folded state. 
Therefore, we chose to simulate a `Mutant-like' stochastic molecule on an asymmetric potential landscape, as shown in Fig.~\ref{fig: fig4}.
The barrier height $\Delta U$ parameter is chosen in accordance to previous studies of the R3$^{\rm{IVVI}}$ energy landscape~\cite{Tapia24}, where at a coexistence force ($f=0$ in our simulations) of 8~\si{\pico\newton}, $\Delta U^{IVVI}=15~\si{\pico\newton\nano\meter}=3.65\, k_BT$. We also use Ref.~\cite{Tapia24}  to set the total typical unfolded length of $L=20~\si{\nano\meter}$ and the estimated diffusion coefficient $D=3000~\si{\nano\meter\squared\per\second}$. In~\cite{tapia-rojo_talin_2020} (and shown in Figure~\ref{fig: constant_force}(b)), it was found that the (un)folding rates at the coexistence force were $\sim8$ times faster in the R3$^{\rm{WT}}$ than the R3$^{\rm{IVVI}}$. We assume the curvatures of the well and barrier are the same to approximate the height of the WT-like potential using Kramers theory~\cite{Kramers40}, solving $8e^{-\beta \Delta U^{WT}}=e^{-\beta \Delta U^{IVVI}}$.
We end up with $\Delta U^{WT}=1.56\,k_BT=6.45~\si{\pico\newton\nano\meter}$. 
For the degree of asymmetry of the potential in simulation, we choose 3 times the distance to the barrier from the unfolded state than from the folded.

\section{Force-extension features of non-equilibrium protein pulling experiments}
\begin{figure}
\includegraphics[width=0.8\textwidth]{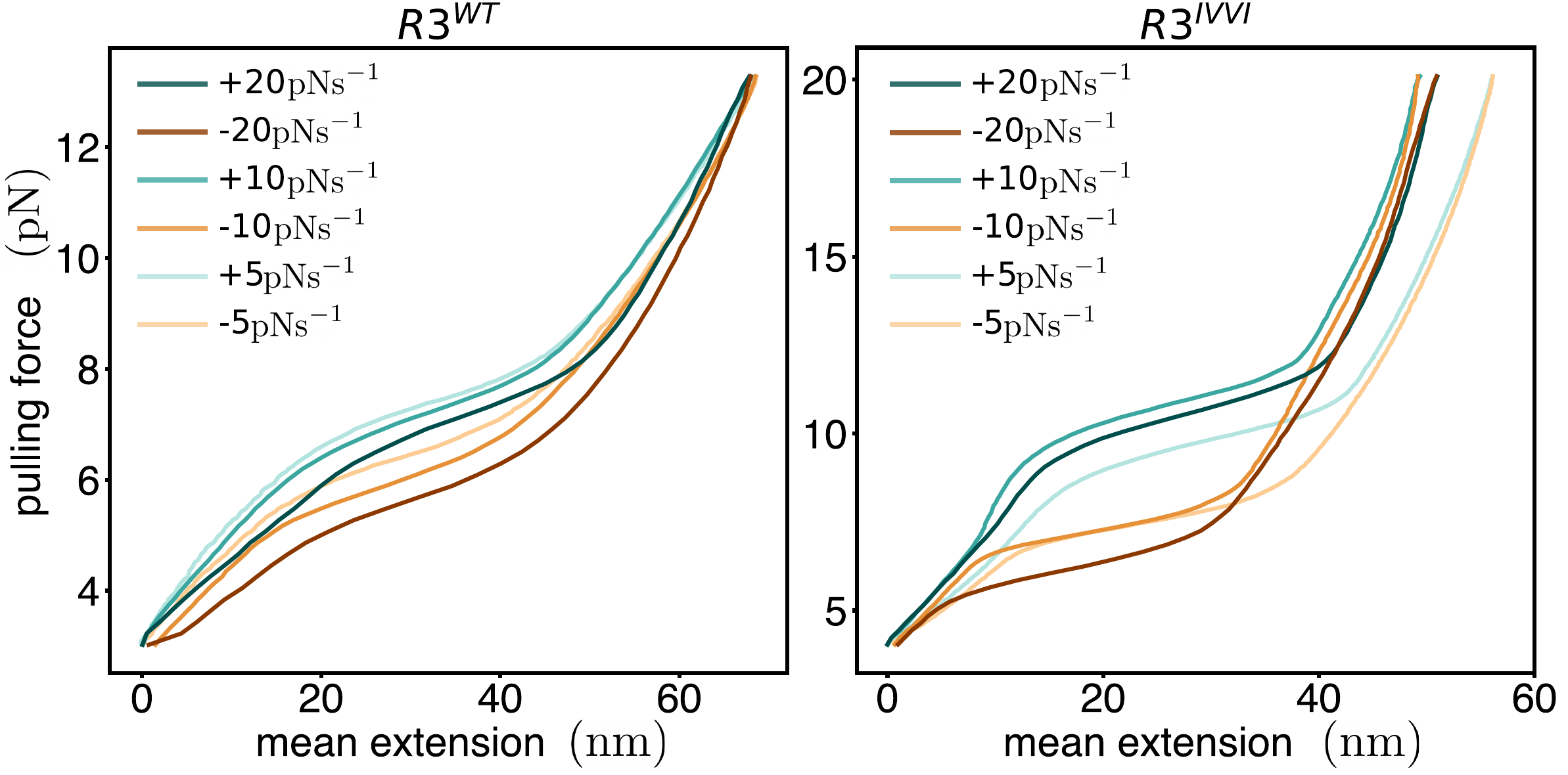}
\caption{\label{fig: exp_extension} At different non-equilibrium pulling rates given in the legends, curves of mean extension with pulling force for the wild-type R3$^{\rm{WT}}$ and mutant R3$^{{\rm{IVVI}}}$ protein constructs.}
\end{figure}

Here we discuss the results of the protein pulling experiments of the wild-type  R3$^{\rm{WT}}$ and mutant R3$^{{\rm{IVVI}}}$ constructs, providing additional details. Figure~\ref{fig: exp_extension} shows the curves of mean extension with pulling force, which display the characteristic hysteresis of cyclic non-equilibrium (un)folding. Here, the lower mechanical stability of the R3$^{\rm{WT}}$ is evident in the lower unfolding force. For this reason, the unfolding force protocols commence at lower forces than for the R3$^{{\rm{IVVI}}}$. We also see indications of higher total irreversibility in the R3$^{{\rm{IVVI}}}$, as the total area contained in the force-extension loop, the total work done divided by temperature, is equal to the total dissipation of the cycle (allowing for the system to relax to an equilibrated state at the end of the refolding ramp). Note as well the variability in final polymeric extensions of the unfolded proteins, due to heterogeneity in the bead anchoring across experiments. When measuring R3$^{{\rm{IVVI}}}$ at $20~\si{\pico\newton\per\second}$ we found two proteins with different final extensions (yet, crucially, near identical transition forces) of $48~\si{\nano\meter}$ and $59~\si{\nano\meter}$, and pool them together in our $\Sigma_{\Delta t}(\mathcal{C}_4)$ irreversibility measurements. In Fig.~\ref{fig: exp_extension}, the R3$^{{\rm{IVVI}}}$ $20~\si{\pico\newton\per\second}$ curve is of this pooled data. As we discuss in ~\ref{sec: crooks}, such different proteins must be analyzed separately when measuring irreversibility via Crooks' Fluctuation Theorem. We also pooled two proteins' data for the R3$^{{\rm{IVVI}}}$ at $5~\si{\pico\newton\per\second}$, but these had very similar force-extension profiles, with final unfolded extensions of $55~\si{\nano\meter}$ and $57~\si{\nano\meter}$.

\begin{figure}
\includegraphics[width=0.95\textwidth]{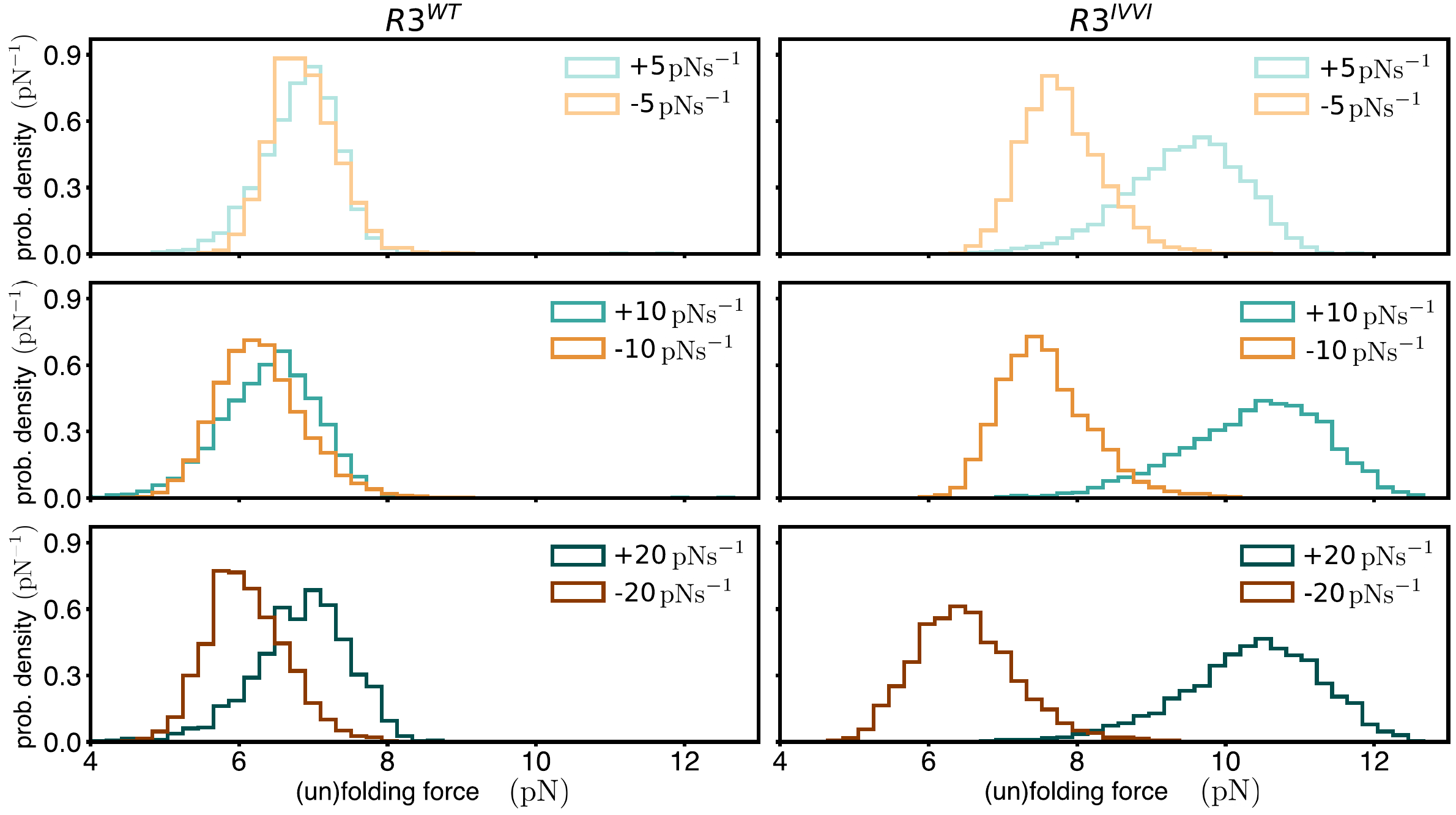}
\caption{\label{fig: exp_escapeF} For different pulling rates of the R3$^{\rm{WT}}$ (left column) and R3$^{{\rm{IVVI}}}$ (right column) protein constructs, histograms (solid lines) of unfolding (positive pulling rate) and refolding (negative pulling rate) forces, at the first recorded event for each ramp iteration.}
\end{figure}

\begin{figure}
\includegraphics[width=0.95\textwidth]{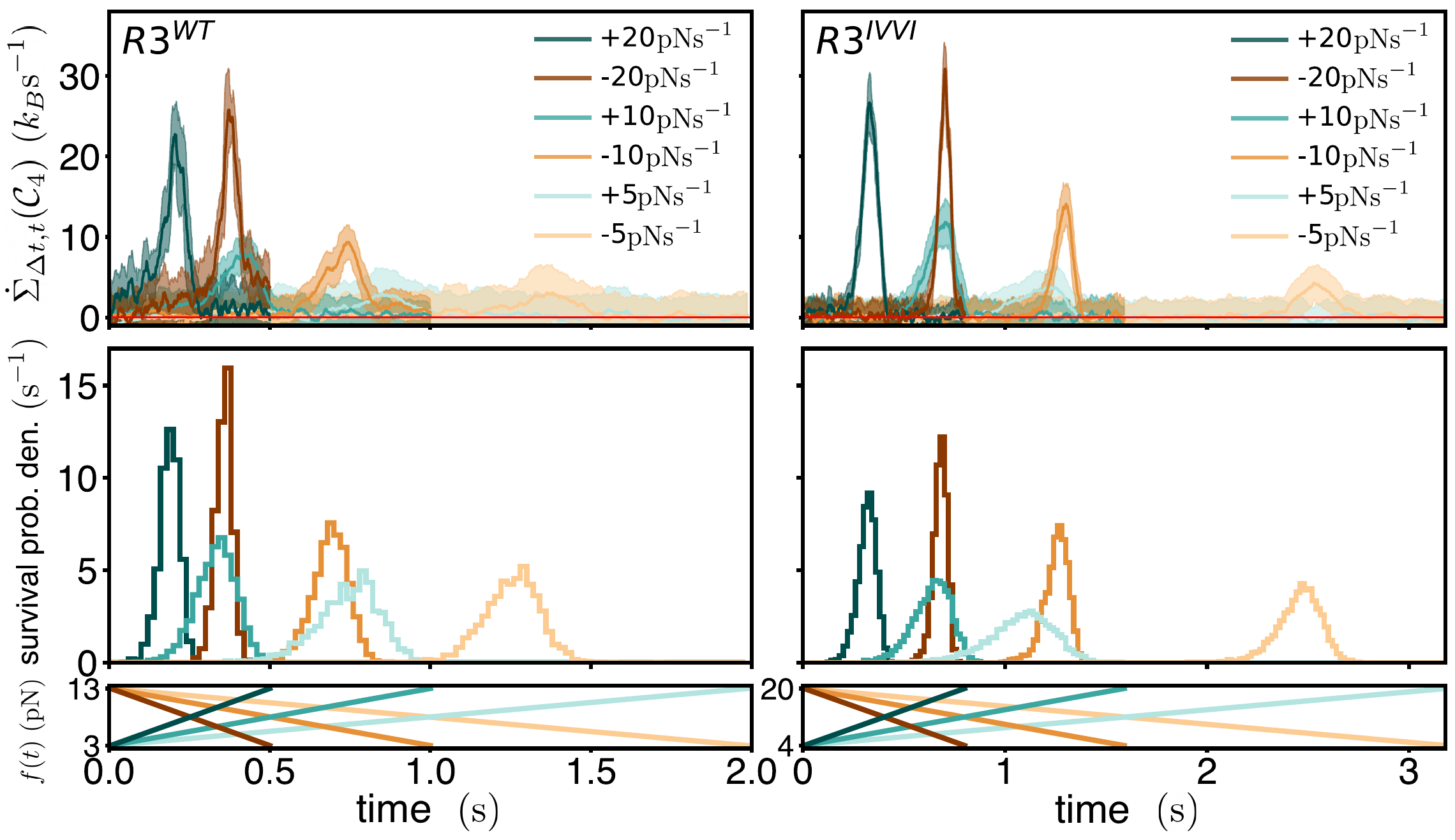}
\caption{\label{fig: exp_escapeTexp} Top row: same curves as in Fig.~\ref{fig: fig4}(c), irreversibility rate $\dot\Sigma_{\Delta t,t}(\mathcal{C}_4)$ measurements of R3$^{{\rm{WT}}}$(left) and R3$^{{\rm{IVVI}}}$ (right). Middle: distributions of (un)folding event times in the (un)folding force-ramp protocols at different pulling speeds. Bottom: pulling force protocol with time.}
\end{figure}

These features are also recovered by the survival probability densities of Fig.~\ref{fig: exp_escapeF}, where the (un)folding force is measured at the first point of crossing $k+2$ and $k-2$ (with $k$ the displacement boundary location for optimal irreversibility measurements) for the unfolding and refolding ramps, respectively. One clearly sees the hysteretic effect increasing with the pulling speeds, as the distributions shift further apart. Furthermore, we see there is a pronounced asymmetry in the shape of the unfolding and folding distributions, especially for R3$^{{\rm{IVVI}}}$. This was also seen and related to asymmetries in the underlying energy landscape in~\cite{alemany2017force}. In Section~\ref{sec: exp} and Fig.~\ref{fig: exp_escapeTexp} we see this is also manifested in the peaks of irreversibility rate and in the total irreversibility of the unfolding and refolding pulling protocols.

Finally, we corroborate the findings of the measurements in simulation illustrated in Fig.~\ref{fig: sigma3}, that peaks in our measured irreversibility rate $\dot\Sigma_{\Delta t,t}(\mathcal{C}_4)$ coincide with these (un)folding events. We plot this in Fig.~\ref{fig: exp_escapeTexp}, converting the force-wise distributions to the time axis. 
\clearpage
\section{Displacement time window sensitivity of experimental irreversibility}

\begin{figure}
\includegraphics[width=0.85\textwidth]{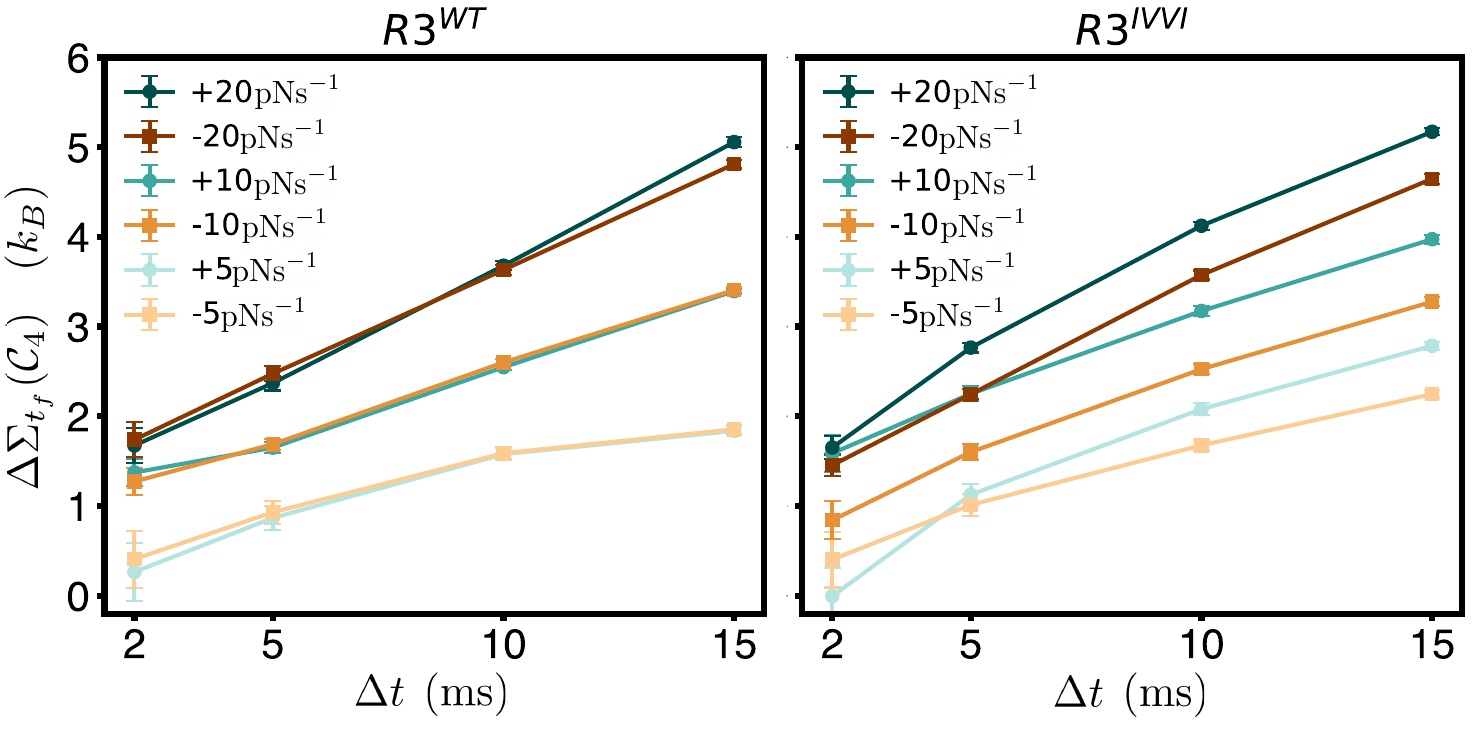}
\caption{\label{fig: dtSigma_exp} Influence of displacement time $\Delta t$ on the total measured irreversibility $\Delta \Sigma_{t_f}(\mathcal{C}_4)$ in each force ramp, for the R3$^{{\rm{WT}}}$(left) and R3$^{{\rm{IVVI}}}$ (right) protein domains. Error bars are given by the accumulated error of the running standard deviation of the irreversibility rate, as in Fig.~\ref{fig: fig4}.}
\end{figure}

We saw in the simulation results of the noisy estimator to measure irreversibility in the symmetric double well potential [Fig.~\ref{fig: fig3}(b)], that there is an important dependence of total irreversibility $\Delta \Sigma_{t_f}(\mathcal{C}_4)$ on the measurement time window $\Delta t$. In Fig.~\ref{fig: dtSigma_exp} we report analogous results on the experimental protein pulls discussed in Section~\ref{sec: exp}. As in the simulations, the curves at $\Delta t=2~\si{\milli\second}$ cross over and have larger errors, indicating that in this regime the estimator is unreliable due to an inability to resolve transitions past the instrumental noise. From $\Delta t=5~\si{\milli\second}$ up, most of the expected trends of dissipation are recovered, so we choose this in our measurements of the main text, as we discussed in~\ref{supp: shift} that smaller $\Delta t$ lead to more faithful reconstructions of the true dissipation. 
It is worth noting, however, that there is still some ambiguity at $\Delta t=5~\si{\milli\second}$ (the $r=\pm 5\si{\pico\newton\per\second}$ points for R3$^{{\rm{IVVI}}}$ are quite close together).

\section{Irreversibility on asymmetric double-well simulations -- comparison with total entropy production}
\begin{figure}
    \includegraphics[width=0.8\textwidth]{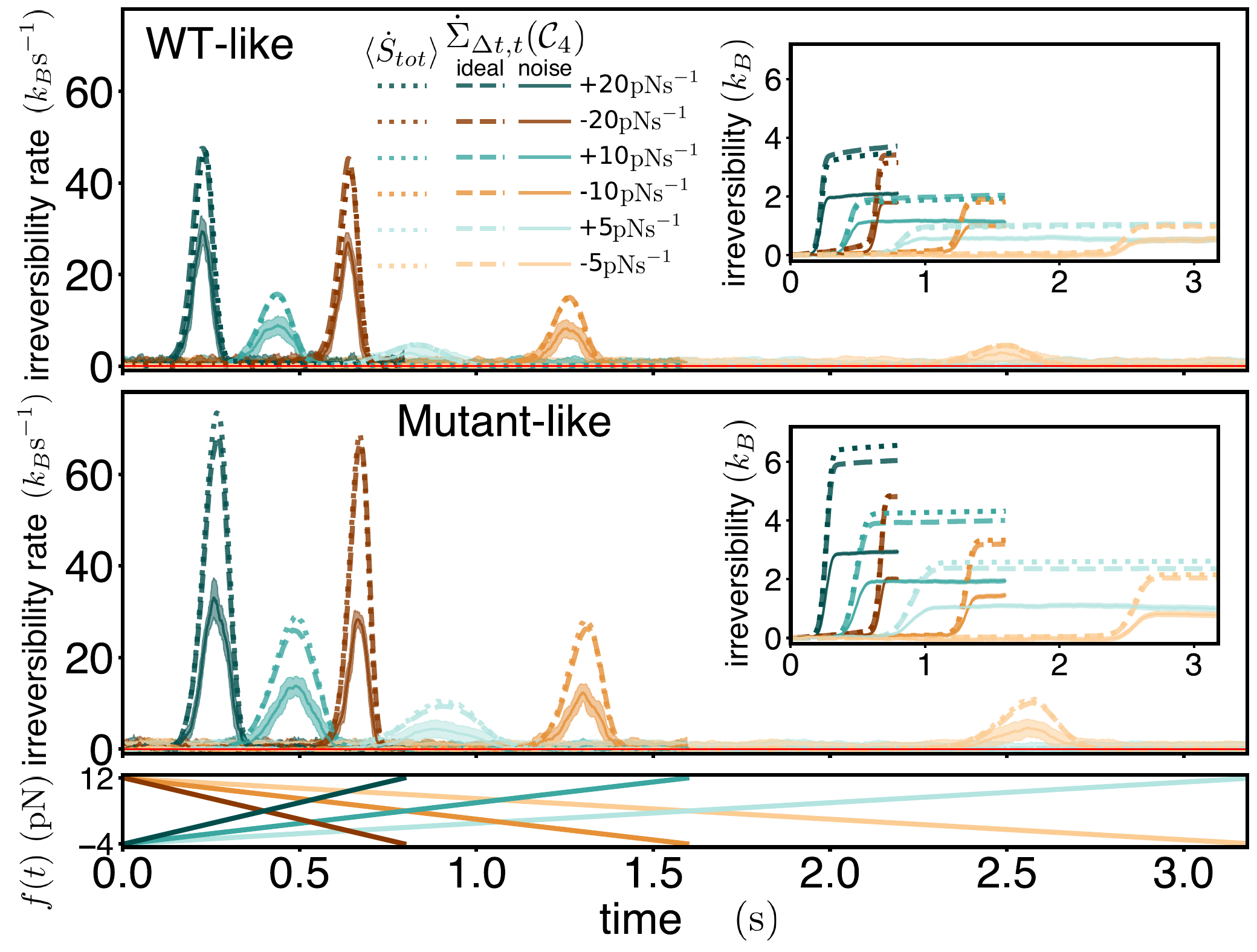}
    \caption{(a) Solid curves: at different force ramp rates $r$ given in the legend, running average of irreversibility rate estimates $\dot{\Sigma}_{\Delta t,t}(\mathcal{C}_4)$ using 5000 simulated trajectories on the asymmetric potentials of Fig.~\ref{fig: fig4}(e), with added noise. The average is done over 70,35,18 points for $|r|=5,10,20$ respectively, and shaded region is the error taken from the running standard deviation. Dashed curves: the same irreversibility estimate without noise and 50000 trajectories. Irreversibilities measured from displacements over $\Delta t=5~\si{\milli \second}$, at a rate of 1~\si{\per\milli\second}. Dotted curves: entropy production rate obtained from averaging Eq.~\ref{eq: thermoEnt}.
    Inset: Cumulative time integral of measured irreversibility and entropy.
    }
    \label{fig: SI_asym}
\end{figure}

To aid the comparison with experiment, we did not include the `ideal' (no added noise) estimated irreversibility and true entropy production curves in the simulation results of the asymmetric potential of Fig.~\ref{fig: fig4}(f). Here, as in Fig.~\ref{fig: fig3}(c), we provide this information in Fig.~\ref{fig: SI_asym}. Once again, we see that the noise impacts the ability to measure the true scale of irreversibility quite significantly. We also see a slight overestimation of $\langle\Delta S_{tot}\rangle$ (top inset) in the WT-like potential, but the trends across non-equilibrium forcings, barrier heights, and between folding and refolding are conserved.

\section{Dependence of irreversibility on displacement boundary location -- experiment and simulation}\label{sec: movek}

\begin{figure}
\includegraphics[width=0.9\textwidth]{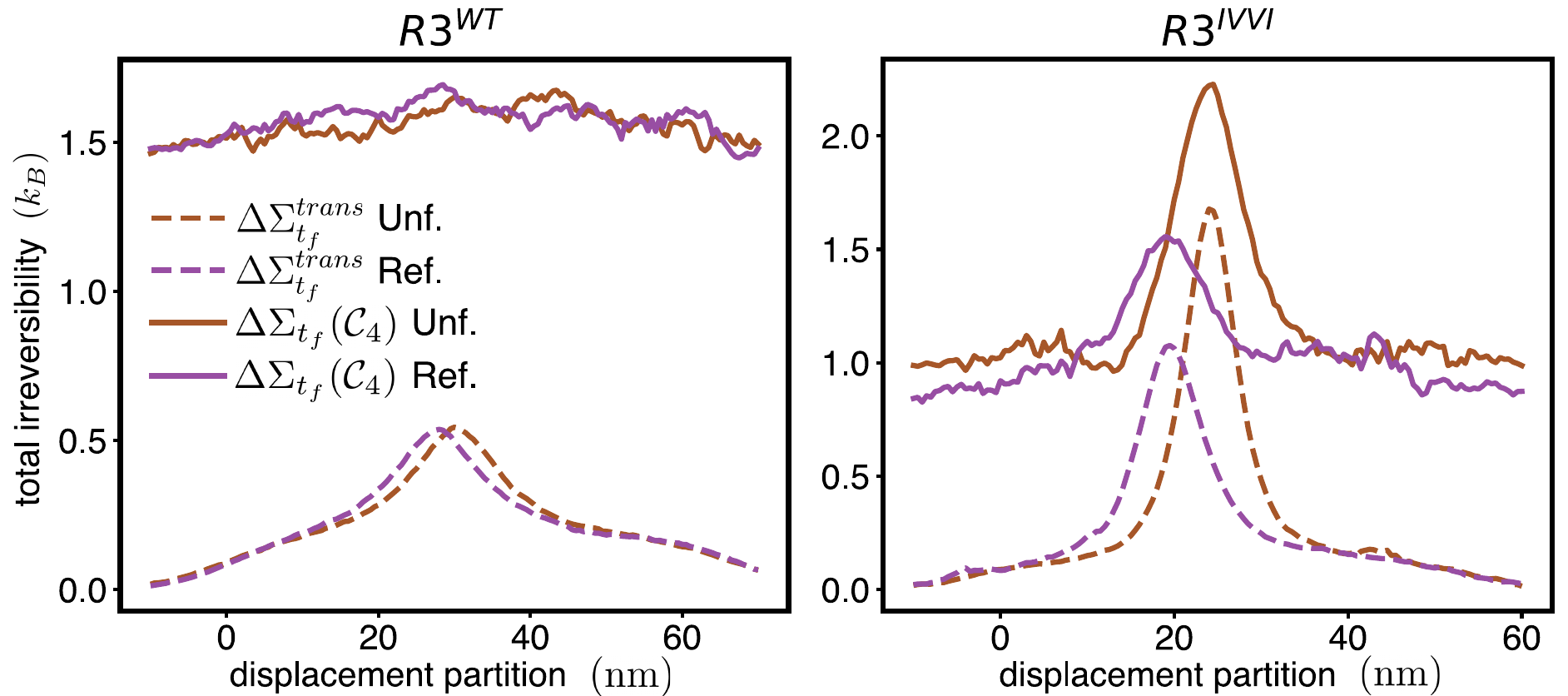}
\caption{\label{fig: vary_k_exp} \textbf{Experiments}.  Total measured irreversibility $\Sigma_{t_f}$ as a function of the displacement partition $k$, using $\Sigma_{\Delta t}(\mathcal{C}_4)$ (solid) and $\Sigma_{\Delta t}^{trans}$ (dashed) on experiments of R3$^{{\rm{WT}}}$(left) and R3$^{{\rm{IVVI}}}$(right) under non-equilibrium force ramps at rates of $r=+10~\si{\pico\newton\per\second}$ (red, Unf.) and $r=-10~\si{\pico\newton\per\second}$ (purple, Ref.).}
\end{figure}

Here we look at how the placement of the boundary $k$ we use to categorize displacements affects the irreversibility inference of the protein experiments and double-well simulations. Recall we decided to keep $k$ fixed for all measurements throughout the protocol. In Fig.~\ref{fig: vary_k_exp}, we report the total measured irreversibility of the $\pm10~\si{\pico\newton\per\second}$ experiments where $k$ is placed at varying positions throughout the protein extension space. For R3$^{{\rm{IVVI}}}$, there are clear peaks of $\Delta\Sigma_{t_f}(\mathcal{C}_4)$ in the transition region (see mean extensions in Fig.~\ref{fig: exp_extension} for reference). Thus, for measurements reported in the main text, we choose $k$ such that $\Delta\Sigma_{t_f}$ is maximized. Here, one can also be guided by the $trans$ classification, which is less computationally intensive to calculate (no nearest neighbors estimation) and has its peaks aligned with $\mathcal{C}_4$. In the more unstable R3$^{{\rm{WT}}}$ system,  $\Delta\Sigma_{t_f}(\mathcal{C}_4)$ is more agnostic to the position of $k$, indicating a greater relevance of processes which are not localized in a particular point (\textit{i.e.} hopping across the barrier). Nonetheless, for consistency with the mutant, we make the decision based on the peaks of $\Delta\Sigma_{t_f}^{trans}$, which, remarkably, are still well resolved.

\begin{figure}
\includegraphics[width=0.85\textwidth]{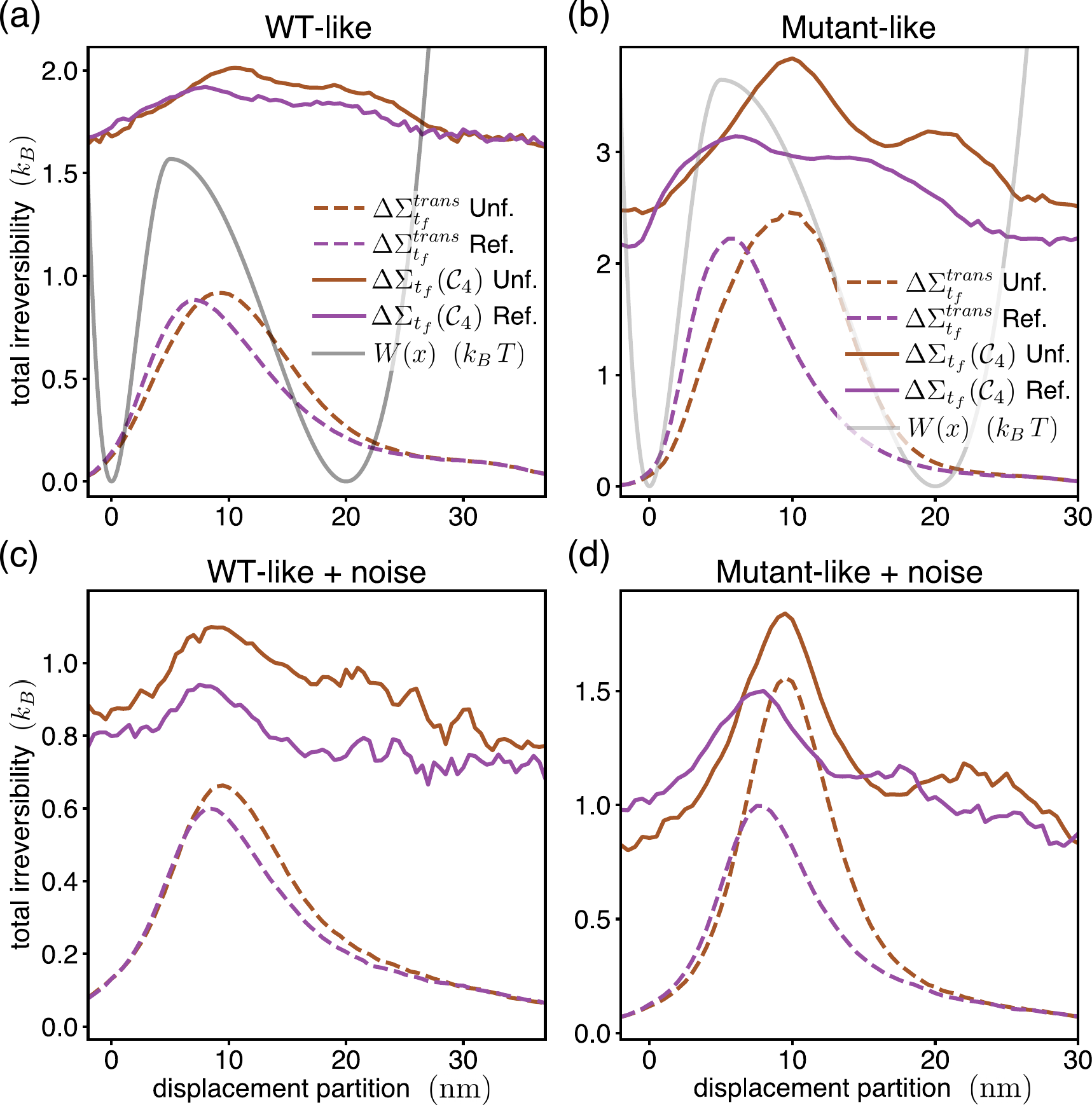}
\caption{\label{fig: vary_k_sim}\textbf{Simulations}. 
Total measured irreversibility $\Sigma_{t_f}$ as a function of the displacement partition $k$, using $\Sigma_{\Delta t}(\mathcal{C}_4)$ (solid) and $\Sigma_{\Delta t}^{trans}$ (dashed) under non-equilibrium force ramps at rates of $r=+10~\si{\pico\newton\per\second}$ (red, Unf.) and $r=-10~\si{\pico\newton\per\second}$ (purple, Ref.).
Results for simulations in a low (a) and high (b) potential barrier shown together with the potential profile $W(x)$. The measurement is made using 10000 trajectories. (c) and (d) are as in (a) and (b) with estimates made using 5000 trajectories and added white Gaussian measurement noise of $\sigma=3~\si{\nano\meter}$}
\end{figure}

This analysis also contributes to our understanding of the simulated protocols, as shown in Fig.~\ref{fig: vary_k_sim}. In the top row, we plot as well the conservative potential $W(x)$ at the coexistence force, so we see that the peak in irreversibility occurs when measuring near the transition point of the barrier. It would be interesting to study how this dependence of dissipation on $k$ relates to equilibrium quantities such as the protein's intrinsic transition state~\cite{hummer_transition_2004, neupane_transition-path_2015}. Adding noise and limiting the amount of data (Fig.~\ref{fig: vary_k_sim}(c) and (d)) places us closer to the experimental measurement regime. This has the effect of `blurring' the difference between the dissipation peaks of unfolding and refolding. Like the experiments, there is a less pronounced maximum in the $\Delta\Sigma_{t_f}(\mathcal{C}_4)$ measurement for the WT-like case, but $\Delta\Sigma_{t_f}^{trans}$ remains comparatively robust.

\section{Comparison to alternative methods to measure irreversibility on protein pulling data}
\subsection{Method of Ref.~\cite{Singh24}}\label{sec: proesmans}
\begin{figure}
\includegraphics[width=\textwidth]{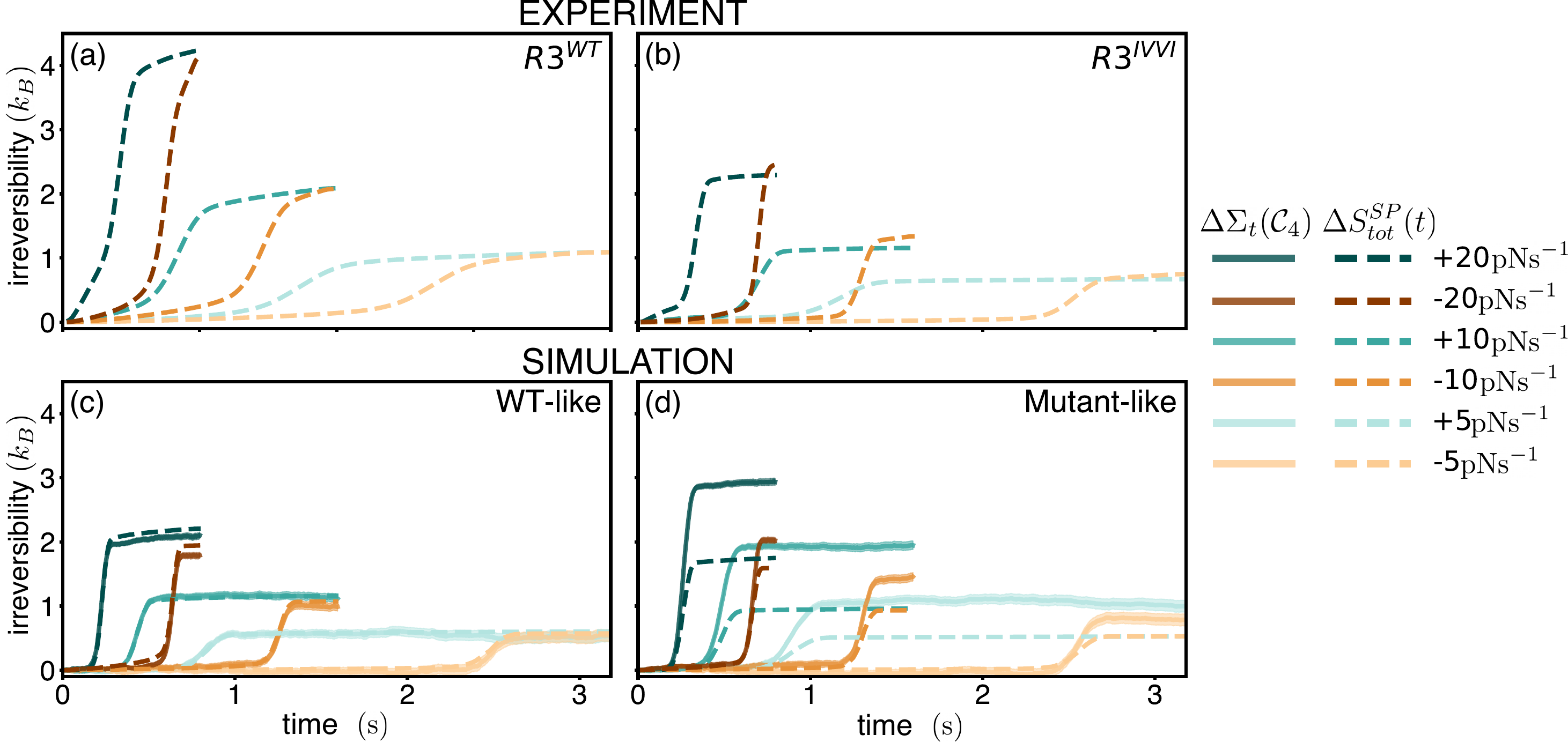}
\caption{\label{fig: proesmans} Bounded measurement of dissipation $S^{SP}_{tot}(t)$ from~\cite{Singh24} on \textit{experimental} non-equilibrium force ramps of the R3$^{{\rm{WT}}}$ (a) and mutant R3$^{{\rm{IVVI}}}$ (b) protein constructs. A constant $D=3000~\si{\nano\meter\squared\per\second}$ is assumed. $S^{SP}_{tot}(t)$ predictions on \textit{simulated} trajectories in a low barrier (WT-like, (c)) and high barrier (mutant-like, (d)) potential, with added measurement noise to the positions (as for those used to make the $\Delta\Sigma_{t_f}(\mathcal{C}_4)$ predictions given by the solid curves).}
\end{figure}

As a first point of comparison to our irreversibility estimator we take the method from Singh and Proesmans' work~\cite{Singh24}, which is applicable as a bound to $\langle\Delta S_{tot}\rangle(t)$. The method has the advantage of not requiring  extensive amount of data, since one only needs the mean $\mu_x(t)$ and variance $\sigma_x^2(t)$ (and their time derivatives $\dot\mu_x(t),\dot \sigma_x^2(t)$) of the positions at each recorded time. This bound $\Delta S^{SP}_{tot}(t)$ is defined as 
\begin{align}
    \langle\Delta S_{tot}\rangle(t)\geq\Delta S^{SP}_{tot}(t) \equiv \frac{k_B}{D}\int^{t}_0 dt' \left[\frac{(\dot\sigma_x^2(t'))^2}{4\sigma_x^2(t')} + (\dot\mu_x(t'))^2 \right] .\label{eq: proesmans}
\end{align}
The method, however, requires precise knowledge of the diffusion coefficient $D$ to use Eq.~\eqref{eq: proesmans}. This is a notoriously difficult measurement in systems with complex energy landscapes and competing timescales such as our non-equilibrium protein pulling experiments. 
Therefore, in our application of this method to our experimental data in Fig.~\ref{fig: proesmans}(a) and (b) we do not show it alongside our estimator $\Delta\Sigma_{t_f}(\mathcal{C}_4)$, and instead focus on whether $\Delta S^{SP}_{tot}(t)$ captures the expected phenomenology. We plug in $D=3000\si{\nano\meter\squared\per\second}$ to Eq.~\eqref{eq: proesmans} as in simulations, and as estimated for R3$^{{\rm{IVVI}}}$ in a previous work~\cite{Tapia24}. We insist there can be heterogeneity in $D$ within one molecule's energy-extension space~\cite{FOSTER20181657}, between proteins of the same sequence\footnote{Different proteins were used for each ramp protocol speed} and, naturally, the R3$^{{\rm{WT}}}$ and R3$^{{\rm{IVVI}}}$. 
$\Delta S^{SP}_{tot}(t)$ captures the expected increasing trend with pulling speeds, but does not capture the difference between the total irreversibilities of unfolding (positive) and refolding (negative) pulling rates in the R3$^{{\rm{IVVI}}}$. In fact, the opposite behavior is suggested, with refolding displaying more entropy production than unfolding for R3$^{{\rm{IVVI}}}$.

The $\Delta S^{SP}_{tot}(t)$ predictions on experimental data suggest that the R3$^{{\rm{WT}}}$ is more dissipative than R3$^{{\rm{IVVI}}}$, as seen comparing Fig.~\ref{fig: proesmans}(a) and (b). This is in disagreement with the force-extension curves (see Fig.~\ref{fig: fig4}(b) and Fig.~\ref{fig: exp_extension}), our findings reported in Fig.~\ref{fig: fig4}(c) and (d), and with the estimates performed using Crooks' fluctuation theorem shown in Fig.~\ref{fig: crooks}.

The results from simulations on the asymmetric potential are given in Fig.~\ref{fig: proesmans}(c) and (d), where we know $D$ and the true dissipation (see Fig.~\ref{fig: SI_asym}). Here, the $\Delta S^{SP}_{tot}$ bound is very similar to our measurement for the low potential barrier (WT-like). It is lower for the high barrier one (Mutant-like), suggesting the incorrect conclusion that the low barrier case is more irreversible. $\Delta S^{SP}_{tot}$ also underestimates the relative difference in dissipation between ramps in the Mutant-like case. Note that while we also added Gaussian noise to the positions to mimic experimental conditions, $\Delta S^{SP}_{tot}$ is quite robust to this and we saw its effect to be minimal compared to our $\Delta\Sigma_{t_f}(\mathcal{C}_4)$.

\subsection{Method of Ref.~\cite{Otsubo22} and short-time TUR}\label{sec: otsubo}
\begin{figure}
\includegraphics[width=\textwidth]{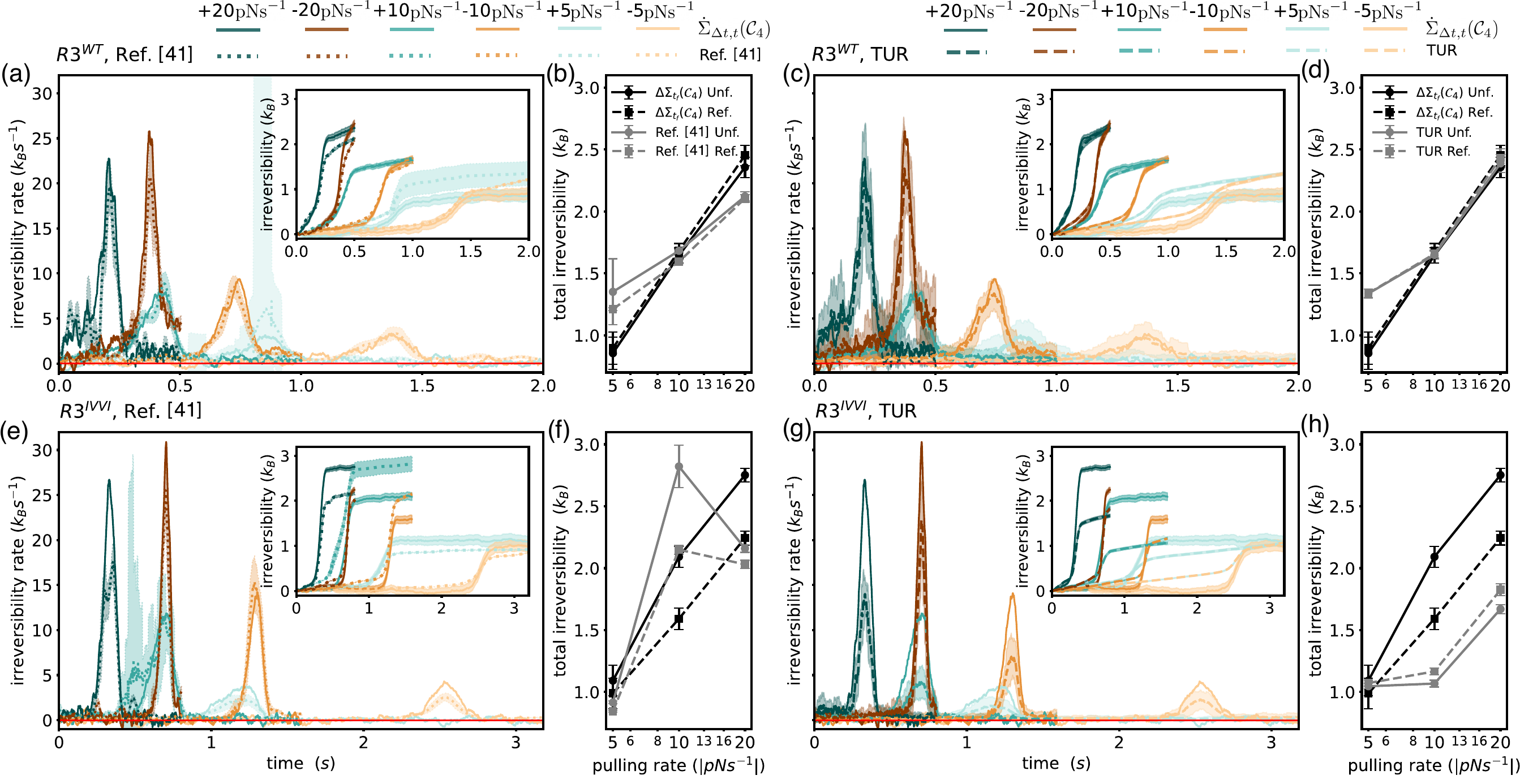}
\caption{\label{fig: otsubo} Irreversibility estimates on the experimental trajectories of non-equilibrium protein pulling in the (a-d) R3$^{{\rm{WT}}}$ and (e-h) R3$^{{\rm{IVVI}}}$ domains. In (a,c,e,g), solid curves show the inference of this work's method $\Delta\Sigma_{t}$, as in Fig.~4(c) in the main text. Colored dotted (a,e) curves with rolling standard deviation error shading correspond to the method of~\cite{Otsubo22}, and colored dashed (c,g) curves are a short-time TUR estimate, using the bulk displacement within time windows $\Delta t=5$ms. Insets display the cumulative irreversibility in time, where the final values at the end of each protocol, the total irreversibility, are shown in the grayscale plots (b,d,f,h) to the right of these. Here, black markers correspond to $\Delta\Sigma_{t_f}$, while the gray are the estimates of~\cite{Otsubo22} in (b, f) and of the TUR in (d,h). Unfolding protocols (positive pulling rate) correspond to solid lines, refolding (negative pulling rate) to dashed.}
\end{figure}
Here, we examine the machine learning (ML) approach of Otsubo et al.~\cite{Otsubo22}, which is based on the TUR in the short-time limit. This TUR is known to saturate to the true entropy production rate using an optimal current proportional to the thermodynamic force. Although it is not often used in practice, we will, in addition to the ML method, apply this bare TUR using displacement currents as in Eq.~\eqref{eq: TUR}, but with respect to the protocol time $t$, with $\ell=\int_{t'=t-\Delta t}^{t'=t} dx_{t'}$. 
Note that while both of these methods should give tight estimates in the $\Delta t\rightarrow0$ limit, as with our $\Sigma_{\Delta t, t}$, this needn't be the case with our experimentally finite resolution, sample size and instrumental noise. We investigate if the expected trends of irreversibility across pulling rates and the two proteins, obtained by $\Sigma_{\Delta t, t}$, and confirmed by the estimates using Crooks' fluctuation theorem in~\ref{sec: crooks}, are recovered. 
We used the ML estimator as suggested in their Github~\cite{githubOtsubo} for nonstationary dynamics, modifying the training iterations to reach convergent test/train curves, and modifying the variance estimator to rectify the bias induced by measurement noise, as described in the supplementary information of~\cite{Otsubo22}. This had the effect of roughly doubling the captured dissipation, with respect to estimating with the biased variance. We also tried this correction on the regular short-time TUR, which did not help. Lastly, to the input of the neural network, we did a linear interpolation in instances of temporal gaps in trajectory data due to instrumental sampling errors.

The results, applying~\cite{Otsubo22} and TUR to our experimental data of proteins under force ramps, are given in Fig.~\ref{fig: otsubo} alongside the $\dot\Sigma_{\Delta t,t}(\mathcal{C}_4)$ estimates of Fig.~\ref{fig: fig4}(c). For the R3$^{{\rm{WT}}}$, the alternative methods [see Fig.~\ref{fig: otsubo}(a-d)] give similar estimates to our own for 20 and 10~\si{\pico\newton\per\second}. For 5~\si{\pico\newton\per\second}, more dissipation is obtained in the times away from the (un)folding jump, as visible in the higher slopes in the insets of Fig.~\ref{fig: otsubo}(a) and (c). This leads to an overall amount of irreversibility which is unexpectedly higher than the 5~\si{\pico\newton\per\second} estimates on the R3$^{{\rm{IVVI}}}$ [see Fig.~\ref{fig: otsubo}(e-h)]. 
The R3$^{{\rm{IVVI}}}$ 5 and 20~\si{\pico\newton\per\second} estimates using Ref.~\cite{Otsubo22} measure less irreversibility than the $\Delta\Sigma_{\Delta t,t}(\mathcal{C}_4)$, seemingly underestimating the degree to which the unfolding and refolding curves should be different.
For 10~\si{\pico\newton\per\second}, Ref.~\cite{Otsubo22} erroneously captures more dissipation than the 20~\si{\pico\newton\per\second} ramps. For all the R3$^{{\rm{IVVI}}}$ ramps, TUR unexpectedly acquires more irreversibility in the refolding than unfolding curves, like the Ref.~\cite{Singh24} method in ~\ref{sec: proesmans}. Both the TUR and Ref.~\cite{Otsubo22} methods also obtain a significant contribution in the times away from the (un)folding events, absent in $\Delta\Sigma_{\Delta t,t}(\mathcal{C}_4)$ for R3$^{{\rm{IVVI}}}$, and also Ref.~\cite{Singh24}. 
Figure~\ref{fig: otsubo}(a) and (e) show that the method of Ref.~\cite{Otsubo22} features a higher uncertainty, for the R3$^{{\rm{WT}}}$ at 5~pN/s [Fig.~\ref{fig: otsubo}(a)] and for the R3$^{{\rm{IVVI}}}$ at 10~pN/s [Fig.~\ref{fig: otsubo}(e)].
In the latter experiment, the method also detects a spurious peak in irreversibility before most of the unfolding events take place.

We remark that the method Ref.~\cite{Otsubo22} relies on choosing specific network architectures and hyperparameters, which may require extensive exploration to be optimized. Our efforts do not rule out that it may be possible to obtain better estimates with this method.

To summarize, while the results on R3$^{{\rm{WT}}}$ are reasonable for all estimators, upon comparison with the more irreversible R3$^{{\rm{IVVI}}}$, our $\Sigma_{\Delta t,t}$ method provides the best recapitulation of the expected dissipation characteristics.

\subsection{Method of Ref.~\cite{Degunther24}}\label{sec: snippet}

Here, we discuss the application of the technique by Degünther et al. for entropic inference based on Markovian event detection~\cite{Degunther24}. 
This technique infers the entropy production of a particular snippet of events, which may not be bounded by the total entropy production (which includes all events). We focus on the unfolding process snippet, defined as an event $\mathcal{F}$: exiting the folded state F at time $\tau$, followed by an event $\mathcal{U}$: entering the unfolded state U at time $(\tau+t)$. The average entropy of unfolding is calculated from the weighted average of the entropy $\Delta S_{unf}(\tau,t)$ of these $\mathcal{F}\rightarrow\mathcal{U}$ events occurring at particular $(\tau, t)$, binned on a 2d histogram as $p_{\mathcal{F}\rightarrow\mathcal{U}}(\tau,t)$. 
\begin{align}
    \langle \Delta S_{unf}(\tau, t)\rangle = \int_0^Td\tau\int^{T-\tau}_0 dt \,p_{\mathcal{F}\rightarrow\mathcal{U}}(\tau,t)\Delta S_{unf}(\tau,t)=k_B\int_0^Td\tau\int^{T-\tau}_0 dt \,p_{\mathcal{F}\rightarrow\mathcal{U}}(\tau,t) \ln \frac{p_{\mathcal{F}\rightarrow\mathcal{U}}(\tau,t)}{p_{\tilde{\mathcal{U}}\rightarrow\tilde{\mathcal{F}}}(\tau,t)} \label{eq: snippet}
\end{align}
The time-reverse probabilities $p_{\tilde{\mathcal{U}}\rightarrow\tilde{\mathcal{F}}}(\tau,t)$ are obtained from the conjugate events in the refolding protocols. This is an important application difference compared to our displacement-based method, which does not require the measurement of reverse protocol experiments to make an inference. Furthermore, as we discuss below, in our regime of limited empirical statistics the result of Eq.~\ref{eq: snippet} is highly dependent on the discretization of $(\tau,t)$ and the definitions of F and U.

\begin{figure}
\includegraphics[width=0.65\textwidth]{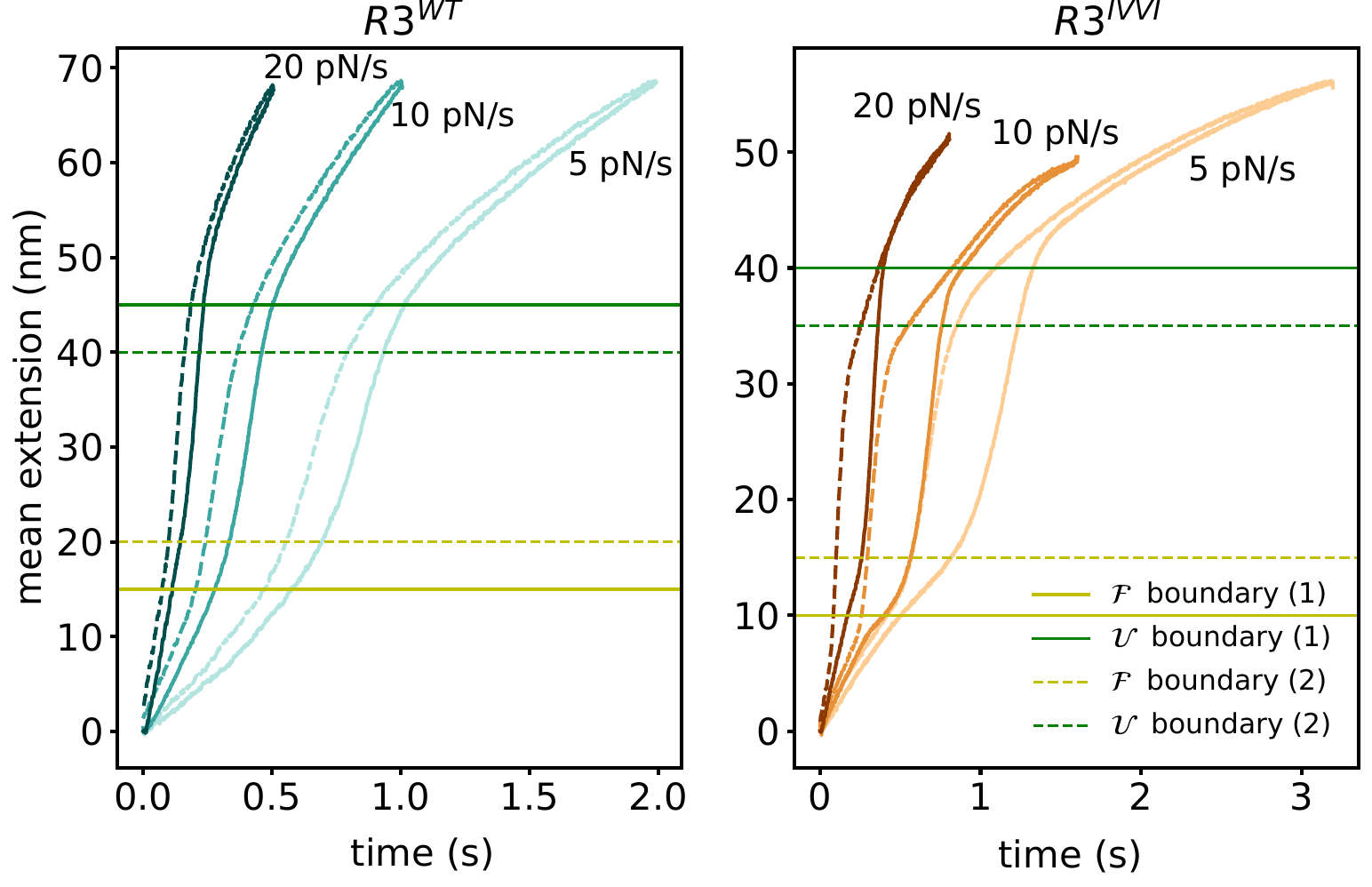}
\caption{Mean extension of the R3$^{{\rm{WT}}}$ and R3$^{{\rm{IVVI}}}$ proteins with protocol duration, for the unfolding ramps (solid) and the refolding (dashed). For the refolding curves, the time axis is reflected with respect to the total protocol duration $T$. The horizontal lines show the two sets (1 and 2) of boundaries where the Markovian events of $\mathcal{F}$ folding and $\mathcal{U}$ unfolding are recorded.}\label{fig: snippetExt}
\end{figure}

To apply the method to our continuous trajectory data, we define the Markovian states F and U as regions enclosed below and above a pair of boundaries in the extension space.
As there is some arbitrariness in choosing these boundaries, we test for each protein two sets of boundary pairs: one further apart (1) and another one closer (2), as shown on the mean extension curves in Fig.~\ref{fig: snippetExt}. Both choices identify unfolded and folded conformations across ramp speeds.

Now we must measure two continuous random variables ($\tau$ – event start, and $t$ – duration between (un)folding events) per ramp, making it quite data-intensive to sufficiently populate the two-dimensional histogram in these variables. Note these should also have non-zero bin probabilities for the backward protocol to avoid as many null measurements as possible \footnote{If either $p_{\mathcal{F}\rightarrow\mathcal{U}}(\tau,t)=0$ or $p_{\tilde{\mathcal{U}}\rightarrow\tilde{\mathcal{F}}}(\tau,t)=0$ for some $\tau,t$ this integrand term in Eq.~\ref{eq: snippet} is chosen to return zero.}. A coarsening of the histogram is indeed recommended by Degünther et al. in answer to insufficient data~\cite{Degunther24}.

\begin{figure}
\includegraphics[width=0.9\textwidth]{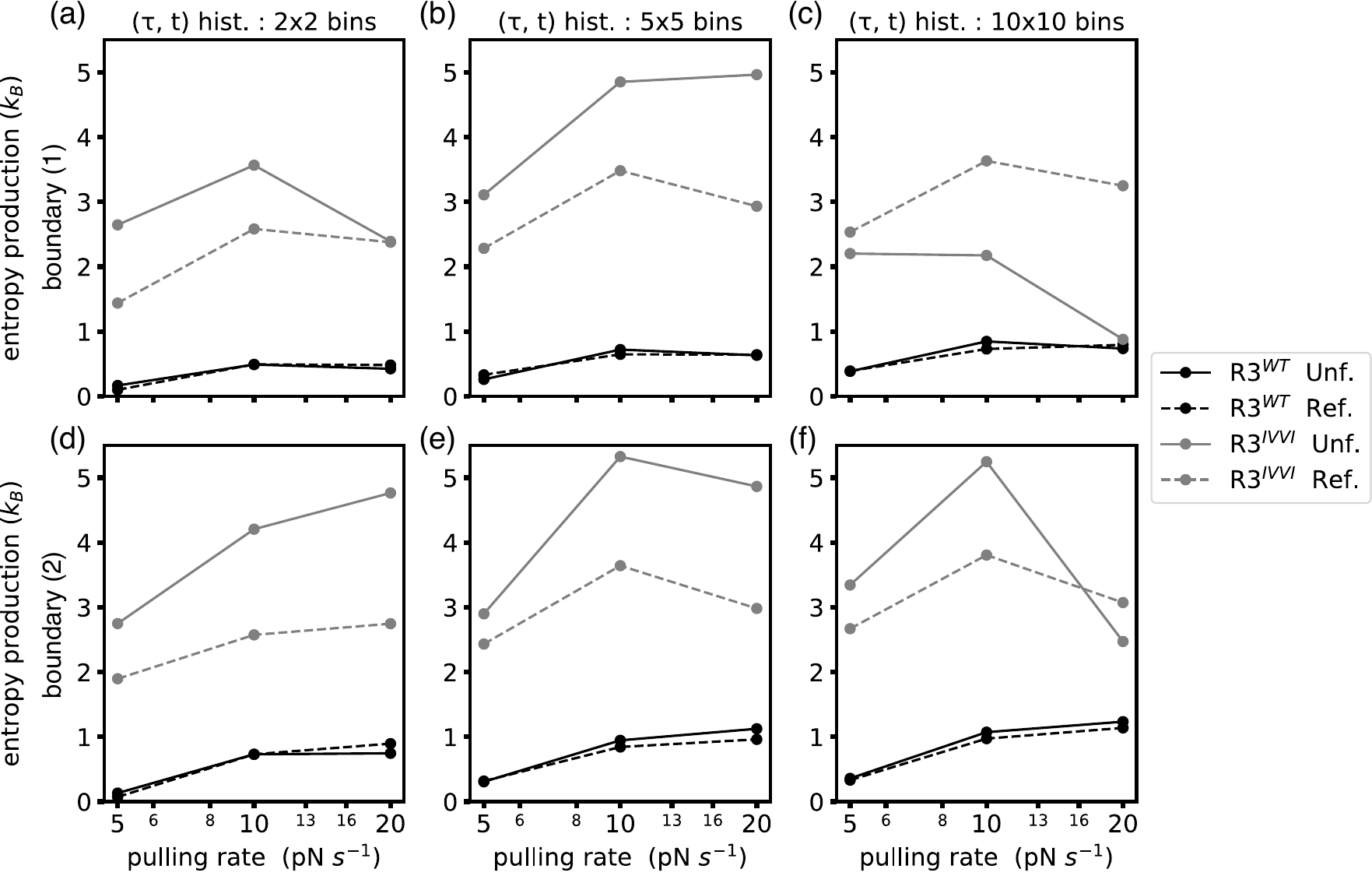}
\caption{Estimates of entropy production for the unfolding (solid) and refolding (dashed) processes of the non-equilibrium pulling of R3$^{{\rm{WT}}}$ (black) and R3$^{{\rm{IVVI}}}$ (grey) proteins, using the method of Ref.~\cite{Degunther24}. In each box, a different set of `hyperparameters' is tested: the different event boundaries ((1) - top row, and (2) - bottom row) shown in Fig.~\ref{fig: snippetExt} and different discretizations (bins, column-wise) of the event-time space.}\label{fig: snippetEnt}
\end{figure}

To compare this approach with our results, we note again that these inferences of the unfolding and refolding process entropies are not necessarily a bound to the total entropy production, as unproductive $\mathcal{F}\rightarrow\tilde{\mathcal{F}}$ sequences, for instance, are not included. However, being the dominant feature of these directed non-equilibrium protocols, they should be expected to obey the same trends as the total irreversibility estimated by our estimate and Crooks'.
In analogy to our total irreversibility plots in Fig.~\ref{fig: fig3}(d), we also measure the entropy of the refolding process (i.e. treating the negative rate refolding ramps as the forward protocol). Our analysis is reported in Fig.~\ref{fig: snippetEnt}, with the unfolding entropies given by solid curves and refolding by the dashed. Additionally, for each U, F boundary definition, we show the results for three different resolutions of the $p_{\mathcal{F}\rightarrow\mathcal{U}}(\tau,t)$ histogram. 

While the estimates of the wild-type (R3$^{{\rm{WT}}}$) are fairly robust, the results for the mutant (R3$^{{\rm{IVVI}}}$) are crucially sensitive to the location of the boundary, binning, and the finite sample size.
The conclusions that can be drawn, such as the trends with the ramp rates, whether unfolding dissipates more than refolding and even if the mutant dissipates more than the wild-type depend on these parameters. This shows 
the difficulties in directly applying the method.
However, we find that the boundary (2) with 2x2 binning provides estimations in line with our displacement-based and Crooks' inferences. This would suggest to err on the side of caution and use the coarsest discretization when data is limited. We leave it for future investigations to determine the optimal location for the event boundaries, as this may also need to be adjusted depending on the protocol speed.

\subsection{Dissipated work via Crooks' Fluctuation Theorem}\label{sec: crooks}
For a non-equilibrium process starting at equilibrium, Crooks' Fluctuation Theorem is a reliable method for measuring equilibrium free energy differences $\Delta F$, provided one has access to the reverse protocol, also starting at equilibrium. It is given by~\cite{crooks1999entropy}
\begin{align}
    \frac{p(W)}{p(-W^R)} = \exp\left(\frac{W-\Delta F}{k_BT}\right),
\end{align}
where $p(W)$ is the distribution of work done during the forward protocol and $W^R$ during the reverse. The quantity in the exponential term is known as the dissipated work. The average dissipated work, which we denote as $W_d$, is bounded by the second law of thermodynamics $W_d\equiv \langle W\rangle-\Delta F\geq0$. When the process in question also \textit{ends} at equilibrium (we have that $\Delta F-\langle\Delta V\rangle=-T\langle\Delta S_{sys}\rangle$ for a change in internal energy $\Delta V$), then $W_d/T=\langle\Delta S_{tot}\rangle$ \cite{Blaber_2023}.  In our non-equilibrium force ramps of proteins, even at high pulling forces, we can assume the system to be practically equilibrated at the end of each ramp. In fact, we have checked in experiment and simulation there is no measurable dissipation during the relaxation of the protein at constant force at the end of the protocol. 

\begin{figure}
\includegraphics[width=0.75\textwidth]{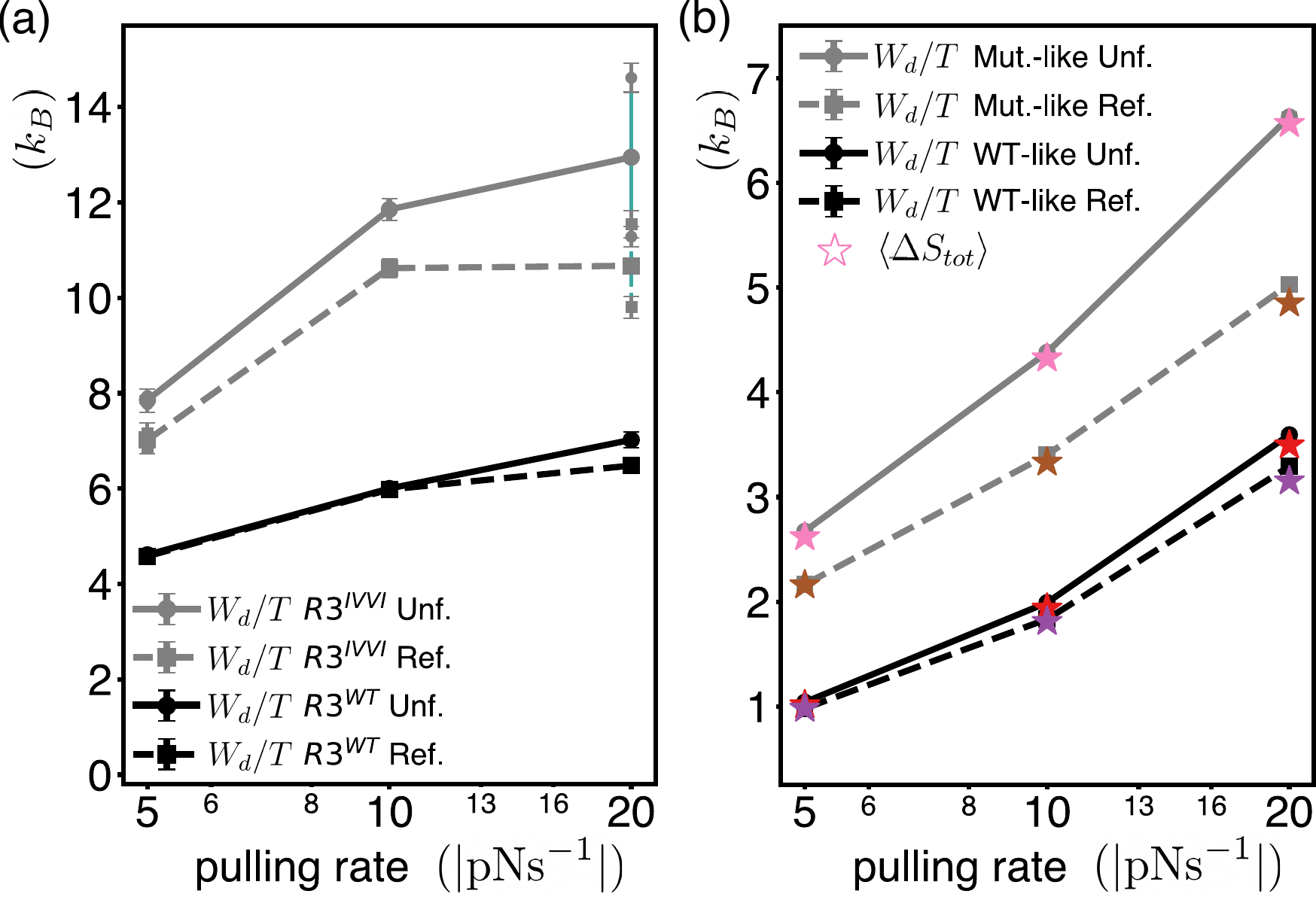}
\caption{\label{fig: crooks} (a) Dissipated work $W_d$ calculated via Crooks' Fluctuation Theorem for protein domains R3$^{{\rm{WT}}}$ (black) and R3$^{{\rm{IVVI}}}$ (gray) under unfolding (positive rate, solid,  circles) and refolding (negative rate, dashed, squares) pulling protocols. For the R3$^{{\rm{IVVI}}}$ at $\pm5$ and $\pm20$~\si{\pico\newton\per\second} the mean of two proteins is used, with their individual $W_d$ shown by the smaller gray markers attached by the cyan vertical lines. Error bars are obtained with jackknife resampling. (b) $W_d$ of the asymmetric double well simulations with 5000 trajectories and added noise, for non-equilibrium driving on a low (WT-like) and high (Mutant-like) barrier. Stars closest to respective markers indicate the true entropy production $\langle\Delta S_{tot}\rangle$.}
\end{figure}

In Fig.~\ref{fig: crooks}, we report our measurements of $W_d$ in experiment and simulation, using Crooks' Fluctuation Theorem for $\Delta F$ by solving $p(W)=p(-W^R)$ via the Bennet Acceptance Ratio method~\cite{BENNETT1976245}. Crucially, with data from a magnetic tweezers setup, $W$ can be directly evaluated from $W=\int \partial_\lambda Vd\lambda=-\int x\,df$, since the control parameter of the protocol $\lambda$ is the pulling force itself~\cite{Alemany_2011}. From the simulation analysis of Fig.~\ref{fig: crooks}(b) we corroborate the equivalence with the total entropy production, given by the stars closest to the respective $W_d$ measurement. Despite using $5000$ trajectories and adding noise to the positions there is good agreement with the theory.

\begin{figure}
    \centering
    \includegraphics[width=0.9\linewidth]{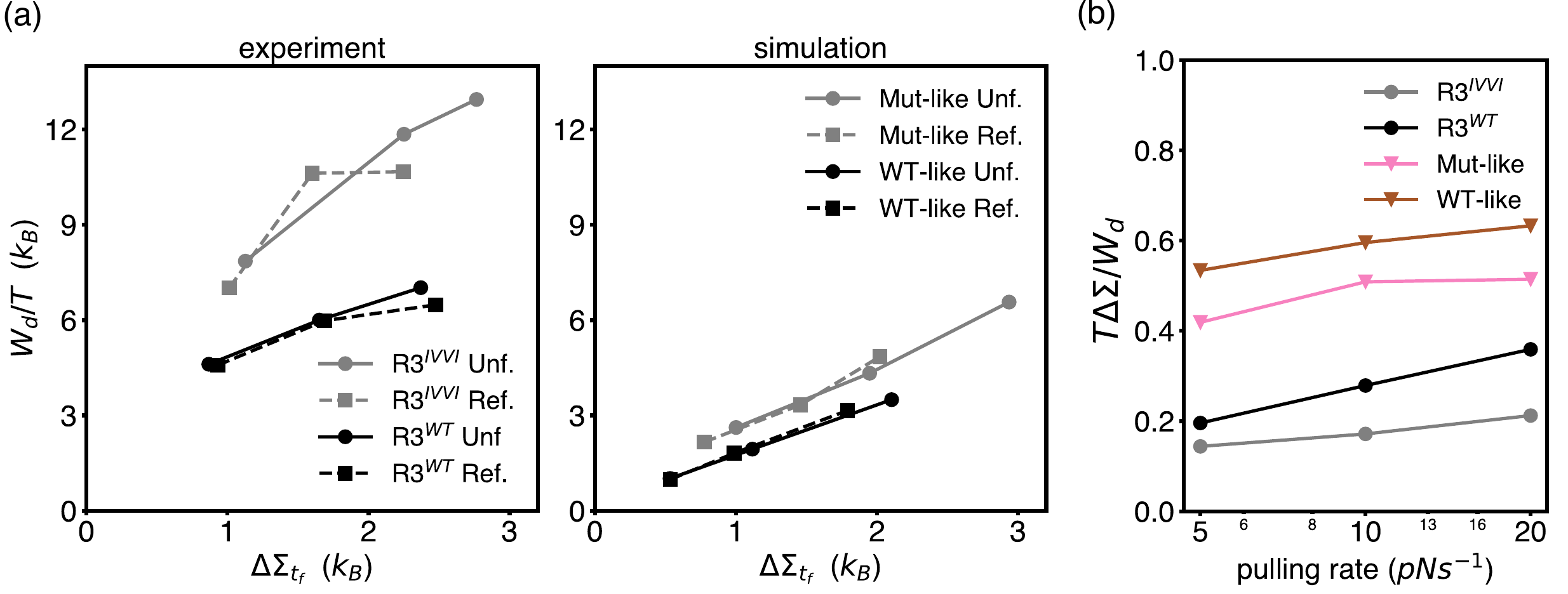}
    \caption{(a) Total dissipated work per unfolding and refolding protocol over temperature, $W_d/T$ obtained using Crooks' Fluctuation Theorem, against estimated irreversibility $\Delta\Sigma_{t_f}$ of the same. The values increase with the speed of pulling (5,10,20 pN/s). (b) Estimated total irreversibility of the complete unfolding/refolding cycle per pulling rate tested, as a fraction of the total dissipated work in the cycle, $T\Delta\Sigma_{t_f}/W_d$. Experiment: The R3$^{{\rm{IVVI}}}$ protein in grey and the R3$^{{\rm{WT}}}$ in black. Simulation: Mutant(IVVI)-like as pink triangles and WT-like as brown triangles.}
    \label{fig: cycle}
\end{figure}

One expects then, that the experimental measurements in Fig.~\ref{fig: crooks}(a) mark values very close to the true dissipation. We note again the use of two different proteins used for the $|r|=5$ and 20~\si{\pico\newton\per\second} in the R3$^{{\rm{IVVI}}}$ experiments. Unlike our $\Sigma_{\Delta t}(\mathcal{C}_4)$ measurement, we cannot pool both datasets to find $p(W)=p(-W^R)$, as distributions of $W$ are very different when the extensions are too. This can be seen in the $W_d$ measurements for $\pm20~\si{\pico\newton\per\second}$ in Fig.~\ref{fig: crooks}(a) stemming from their averages. The trends across pulling rates are in general agreement with those captured by our irreversibility measurement in Fig.~\ref{fig: fig4}(d). Namely, higher dissipation as pulling rate is increased, similar dissipation between folding and refolding ramps in the R3$^{{\rm{WT}}}$, more dissipation overall in the R3$^{{\rm{IVVI}}}$ compared to the R3$^{{\rm{WT}}}$, and a consistently higher dissipation for unfolding over refolding in the R3$^{{\rm{IVVI}}}$.
This strong correlation is evident upon plotting $W_d$ as a function of the irreversibility $\Delta \Sigma_{t_f}$ as shown in Fig.~\ref{fig: cycle}(a).

It is also insightful to compare
the ratio of the average total irreversibility $\Delta\Sigma$ in an unfolding-refolding cycle to the dissipated work $W_d/T$ in Fig.~\ref{fig: cycle}(b). In experiments, this ratio ranges between 14 and 36\%, while in simulations between 42 and 63\%. 
We note, however, that these ratios are sensitive to the magnitude of the chosen
$\Delta t$, since $\Delta\Sigma$ is a bound to entropy production only in the $\Delta t\to 0$ limit.
The trends observed in Fig.~\ref{fig: cycle}(b) may be rationalized by noticing
 the increase in the ratio for faster pulling rates, where $\Delta t$ increases relative to the timescale of pulling. Similarly, the wild-type, which has faster hopping dynamics, returns higher ratios.
\\

\section{Experimental set-up}\label{supp: setup}
\begin{figure}
    \includegraphics[width=0.9\textwidth]{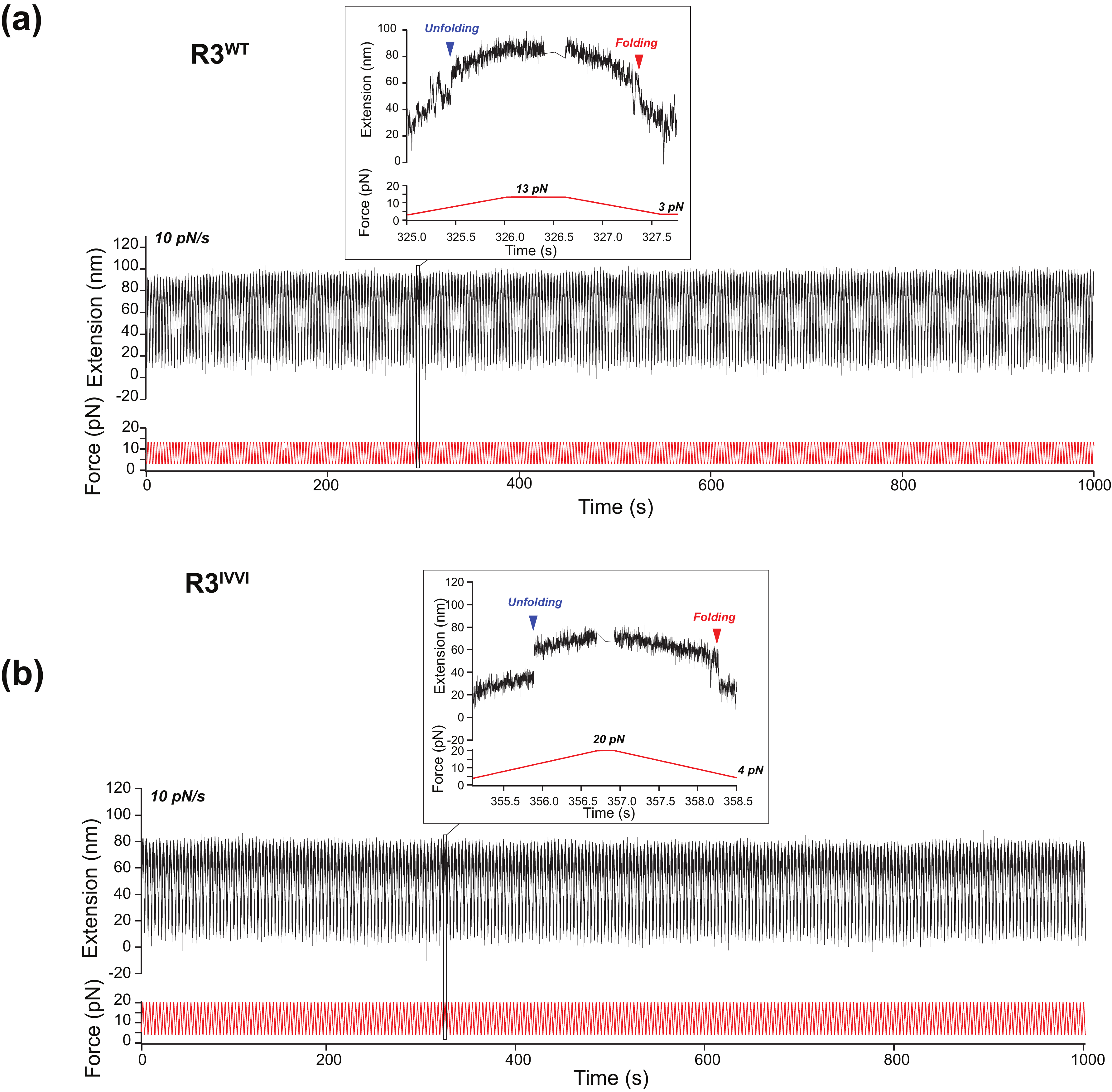}
    \caption{Exemplar magnetic tweezers recordings for cyclic unfolding/refolding pulses at a pulling rate of $\pm$10 pN/s for R3$^{\rm{WT}}$ (a) and R3$^{\rm{IVVI}}$ (b). The extension of the protein is measured as a function of time under a cyclic force ramp protocol, whereby the force is linearly increased/decreased between two force values ($F_{\rm{0}}$ and $F_{\rm{f}}$), ensuring that the protein is folded at $F_{\rm{0}}$, and unfolded at $F_{\rm{f}}$ ($F_{\rm{0}}$=3 pN, $F_{\rm{f}}$=13 pN for R3$^{\rm{WT}}$ and   $F_{\rm{0}}$=4 pN, $F_{\rm{f}}$=20 pN for R3$^{\rm{IVVI}}$).    (Insets) Detail of a single unfolding-refolding cycle, highlighting the unfolding (blue arrow) and refolding (red arrow) events captured as a discrete increase or decrease of the molecular extension. Raw trace is shown, measured at a frame rate of $\sim$ 1 kHz.}  \label{fig: ramps_SI}
\end{figure}

Protein expression and purification were conducted following established molecular biology protocols as described in~\cite{tapia-rojo_single-molecule_2024}.

The single-molecule experiments were conducted on a home-made magnetic tweezers setup as described in~\cite{TAPIAROJO202483}. Briefly, the instrument is based on a custom-made inverted microscope, using a magnetic tape head that enables the application of calibrated forces between 0 and 40 pN with a bandwidth of 100 kHz. The inverted microscope uses a red light-emitting diode (625 nm) as the light source, which is collimated and condensed into a 160 mm oil-immersion objective  mounted on a high-precision piezo-focus scanner. The image is collected with a complementary metal-oxide-semiconductor camera. 

Single-molecule experiments were conducted in custom-made fluid chambers assembled as two cover slides (40x24 mm and 22x22 mm) sandwiched between a laser-cut parafilm pattern. The bottom glasses are silanized for amine functionalization after cleaning and plasma-activation, as described in~\cite{tapia-rojo_single-molecule_2024}. Top glasses are immersed in a repel silane solution to functionalize them with hydrophobic groups. After assembly, the chambers are incubated in a 0.001\% glutaraldehyde solution for 1 h, followed by incubation with 0.002\% amine-polystyrene beads (2.5 $\mu$m) for 20 min, and a 20 $\mu$g ml$^{-1}$ solution of HaloTag amine (O4) ligand  overnight. Finally, chambers are passivated with a sulfhydril-blocked bovine serum albumin (BSA) buffer (20 mM Tris-HCL, 150 mM NaCl, 1\% BSA, pH 7.4) for $>3$ h. The R3$^{\rm{WT}}$ or R3$^{{\rm{IVVI}}}$ protein constructs are freshly diluted in phosphate-buffered saline at $\sim$2 nM and incubated in the fluid chamber for $\sim$30 min to achieve HaloTag conjugation. Finally, streptavidin-coated superparamagnetic beads are flowed into the fluid chamber to bind the biotinylated C-term AviTag. All experiments are conducted on Tris-HCl buffer with 150 mM NaCl and 10 mM ascorbate (pH 7.4) on a temperature-controlled room at 20$\pm$1$^\circ$C.

Single-protein tethers were identified by conducting an initial force-ramp to observe an extension event of $\sim$20 nm, followed by a short constant force pulse to fingerprint the reversible folding dynamics. Cyclic force-ramp experiments at 5, 10, and 20 pN/s (as exemplified in Fig.~\ref{fig: ramps_SI}) were programmed and saved in a JSON file format, enabling independent analysis of each pulse~\cite{Tapia24}.

\begin{figure}
    \includegraphics[width=\textwidth]{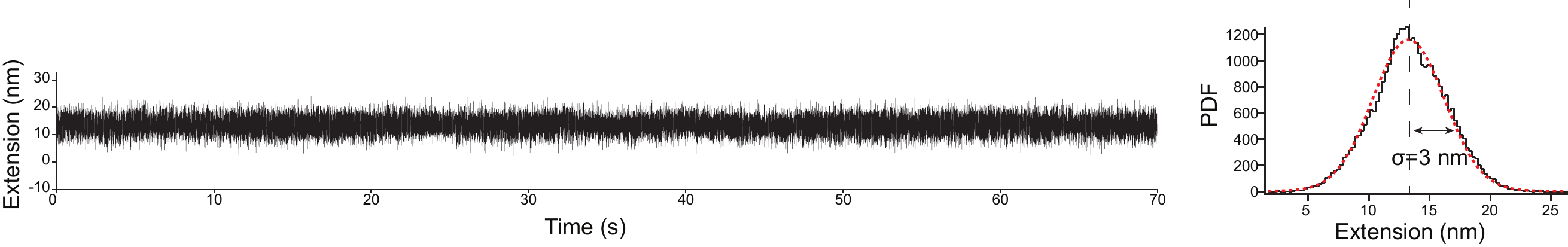}
    \caption{Measuring the imaging noise of the magnetic tweezers setup from the trace of displacement of two fixed reference beads. The fluctuations of this displacement indicate a Gaussian noise with standard deviation of $3~\si{\nano\meter}$.}  \label{fig: expNoise}
\end{figure}

To measure the intrinsic instrumental noise, we perform a measurement without any protein. Two amine-polystyrene beads are anchored to the bottom glass and the apparent relative displacement between them, physically constant, is measured. The fluctuations of this measurement indicate the noise, which we find to be Gaussian-like with a standard deviation of $3~\si{\nano\meter}$, as shown in Fig.~\ref{fig: expNoise}.

\end{document}